\documentclass[amsmath,trackchanges]{aastex702}

\usepackage{longtable}
\newcommand*{\Euclid}{\textit{Euclid} }

\newcommand*{\Euclidc}{\textit{Euclid}, }
\newcommand{\sfont}[1]{{\scriptscriptstyle\rm #1}}
\newcommand{\IE}{\ensuremath{I_\sfont{E}}\,}
\newcommand{\YE}{\ensuremath{Y_\sfont{E}}\,}
\newcommand{\JE}{\ensuremath{J_\sfont{E}}\,}
\newcommand{\HE}{\ensuremath{H_\sfont{E}}\,}

\begin{document}

\title{Euclid DR1 Legacy Science on Ultracool Dwarfs:  Search, Classify and Analyze L-, T- \& Y-type Objects}

\author[gname=Carlos,sname=Dominguez-Tagle]{C. Dominguez-Tagle}
\affiliation{Instituto de Astrofisica de Canarias, La Laguna, Spain}
\affiliation{Universidad de La Laguna, Dpto. Astrofisica, La Laguna, Spain}
\email{carlos.dominguez@iac.es}

\author[gname=Sara,sname=Muñoz Torres]{S. Muñoz Torres}
\affiliation{Instituto de Astrofisica de Canarias, La Laguna, Spain}
\affiliation{Universidad de La Laguna, Dpto. Astrofisica, La Laguna, Spain}
\email{sara.munoz@iac.es}

\author[gname=Nafise,sname=Sedighi]{N. Sedighi}
\affiliation{Instituto de Astrofisica de Canarias, La Laguna, Spain}
\affiliation{Universidad de La Laguna, Dpto. Astrofisica, La Laguna, Spain}
\email{nafise.sedighi@iac.es}

\author[gname=Marusa,sname={\v Z}erjal]{M. {\v Z}erjal}
\affiliation{Instituto de Astrofisica de Canarias, La Laguna, Spain}
\affiliation{Universidad de La Laguna, Dpto. Astrofisica, La Laguna, Spain}
\email{maruska.zerjal@iac.es}

\author[gname=Styliani,sname=Tsilia]{S. Tsilia}
\affiliation{Instituto de Astrofisica de Canarias, La Laguna, Spain}
\affiliation{Universidad de La Laguna, Dpto. Astrofisica, La Laguna, Spain}
\email{stella@iac.es}

\author[gname=Pedro,sname=Mas-Buitrago]{P. Mas-Buitrago}
\affiliation{Centro de Astrobiologia (INTA-CSIC),Villanueva de la Cañada, Madrid, Spain}
\email{pmas@cab.inta-csic.es}

\author[gname=Eduardo,sname=Martin]{E. L. Martin}
\affiliation{Instituto de Astrofisica de Canarias, La Laguna, Spain}
\affiliation{Universidad de La Laguna, Dpto. Astrofisica, La Laguna, Spain}
\email{ege@iac.es}

\author[gname=Victor,sname=Bejar]{V. J. S. B{\' e}jar}
\affiliation{Instituto de Astrofisica de Canarias, La Laguna, Spain}
\affiliation{Universidad de La Laguna, Dpto. Astrofisica, La Laguna, Spain}
\email{victor.bejar@iac.es}

\author[gname=Nicolas,sname=Lodieu]{N. Lodieu}
\affiliation{Instituto de Astrofisica de Canarias, La Laguna, Spain}
\affiliation{Universidad de La Laguna, Dpto. Astrofisica, La Laguna, Spain}
\email{nlodieu@iac.es}

\author[gname=Khalid,sname=Barkaoui]{K. Barkaoui}
\affiliation{Instituto de Astrofisica de Canarias, La Laguna, Spain}
\affiliation{Universidad de La Laguna, Dpto. Astrofisica, La Laguna, Spain}
\email{khalid.barkaoui@iac.es}

\author[gname=Laura,sname=Rodriguez-Torres]{L. Rodríguez-Torres}
\affiliation{Universidad de La Laguna, Dpto. Astrofisica, La Laguna, Spain}
\email{.}

\author[gname=Adam,sname=Burgasser]{A.~Burgasser}
\affiliation{Department of Astronomy \& Astrophysics, UC San Diego, USA}
\email{aburgasser@ucsd.edu}


\begin{abstract}

\Euclid mission Data Release 1 includes  near infrared spectra for about 250 million objects. These spectra are the result of a single visit to the fields of the Wide Survey, plus 8 to 10 visits to the Deep Fields and over 20 visits for auxiliary and calibration fields. Using all the released spectra from these fields, we search for  L-,  T- and Y-type ultracool dwarfs (UCDs), derive their spectral-type classification, spectro-photometric distances, and measure the spectral indices of water, methane and other molecular bands. We measure the pseudo equivalent widths of the alkali lines present in the red-grism spectra. From the fields that include spectra obtained with the blue grism, we also investigate their quality and the main spectral features in this spectral range. We also show how the water and methane spectral indices correlate well with spectral types, and that the pseudo equivalent width of KI line (1.253~$\mu$m) increases from mid-L dwarfs to T6, and then, sharply decreases down to late-T dwarfs. We find from our estimations of spectro-photometric distances, that L and T dwarfs distribute between 25 and 450 pc. With all these results we present a catalog of more than 760 spectroscopically-confirmed UCDs, including 400 T dwarfs and 600 new discoveries. We discuss peculiar objects, including several binary candidates and possible subdwarfs. We search for Y dwarfs and present the  \Euclid  images and  spectra of two known Y dwarfs.  And we build new standard templates based on \Euclid spectra for all UCD spectral types.

\end{abstract}





\section{\label{sc:Intro}Introduction}

  Ultracool dwarfs (UCDs) have spectral types M7 and later \citep{Kirkpatrick95}, which correspond to effective temperatures below 2700~K and, currently, down to $\sim$~250~K \citep{Luhman14, Luhman24}. 
Their spectral classification includes:  late-M dwarfs (M7 and later, \citealt{Kirkpatrick97},  L-type \citep{ege97,ege98,ege99,Kirkpatrick99,Kirkpatrick00},  T dwarfs \citep{Burgasser02, Geballe02, Burgasser06}, and  Y dwarfs
\citep{Cushing11,Luhman2011,Liu2011,Kirkpatrick11}.
  In the literature, UCDs are already very numerous for M and early-L types. The number of published objects start to decrease for the mid-L dwarfs, being low in the so called ``L/T transition" (i.e., L8 to T4).   
The numbers for every spectral subtype in this transition region are still low to address all the  questions on their formation and evolution \citep[see, for example,][]{Gizis2024}. For the mid and late T dwarfs, the numbers can surpass 100 per subtype, and the total number of Y dwarfs is currently increasing over 50 (see, for example SIMLPE\footnote{simple-bd-archive.org} or MocaDb\footnote{MocaDb.ca}). 

The observation strategy of \Euclid mission \citep{EuclidSkyOverview}, doing image and spectroscopy simultaneously, 
has opened a new strategy to search for and study UCDs. The first data set released in 2025,
Q1 data \citep{Q1} allowed us \citep{Dominguez2025} to develop a specific method to search for late type dwarfs directly in the near infrared (NIR) spectra observed by \Euclidc without any photometric pre-selection. We found 11 new late-L and T dwarfs, classified them and demonstrated the feasibility of the spectral analysis with  \Euclid data. Complementary works
carried  photometric and  statistical methods for the search in Q1 data \cite[][Vitas  et al. accepted]{Marusa2025, Anjana2026}.
Two others works also contributed to the search and analysis of UCDs in Q1 data  \citep{ Kiwy2026, USMILE}. As well,  \cite{Tu2025} conducted a systematic search over 40 thousand JWST NIR spectra, also without previous photometric pre-selection, and found 8 new T dwarfs. Other teams, using different data sources are also fostering the study of the cooler dwarfs, mainly  \cite{Schneider2026}, who have published a recent catalog including over 600 T dwarfs, many of them pending for spectroscopic confirmation. As well, the very recent work of  \cite{Tu2026} adds over 100 T dwarfs with the very low resolution spectrophotometry from  SPHEREx \citep{SPHEREx}.
Still, augmenting these numbers is crucial to improve their statistics and to study the diversity in their physical properties or, even the diversity in the viewing angle at which we observe them. Improving the spectroscopic study is required to constrain the current theoretical models and to set telluric-free standard templates. This can only be done with large samples of spectra, and \Euclid telluric-free data is perfect for that purpose in the NIR.
The planned survey strategy of \Euclid \citep{Scaramella-EP1} with  repetitive visits to some fields is  very convenient to go deeper and also for variability studies and parallax measurements.

\Euclid Data Release 1 (DR1; Aussel et al. submitted; Tereno et al. submitted)\footnote{Officially known as DR1-Foundation}
 includes  images and NIR spectra of millions of sources. It revisits the fields observed in Q1 and 
releases data in a part of the Wide Survey that is already 30 times larger than Q1. Using these data, in this paper we focus on the  study of UCDs based on their spectral features. We investigate how \Euclid DR1 wide and deep observations allow a detailed analysis on the NIR spectral range provided by NISP.  
The estimation of effective temperatures using the updated ATMO model grid, developed by Sedighi et al. (in prep.), based on the ATMO2020 models \citep{ATMO}, will be presented in a separate paper. 

As we will see in Sect.~\ref{sc:data}, this paper includes DR1 data with two different depths, therefore we have organized the sections, the analysis and the figures in such a way that we can show the best from both. We present a comparison between them when the depth has an impact on the results. The paper is organized as follows:
Sect.~\ref{sc:data} describes \Euclid DR1 data split in the different fields;
Sect.~\ref{sc:q1} describes the improvement in quality with respect to Q1 data;
Sect.~\ref{sc:search} describes the search for L, T \& Y dwarfs; 
Sect.~\ref{sc:spt} describes our classification by spectral type and presents the {\it Euclid} templates;
Sect.~\ref{sc:dist} presents the estimation of spectrophotometric distances;
Sect.~\ref{sc:anal} describes the spectral analysis 
in two subsections, one for the deep observations and the other for the Wide Survey;
Sect.~\ref{sc:notes} presents a detailed  analysis for some interesting dwarfs: Ys, possible subdwarfs or binaries; 
 Sect.~\ref{sc:concl} presents our conclusions,
and Appendix~\ref{sc:catal} describes the full catalog of UCDs.

\begin{figure}[ht!]
\plottwo{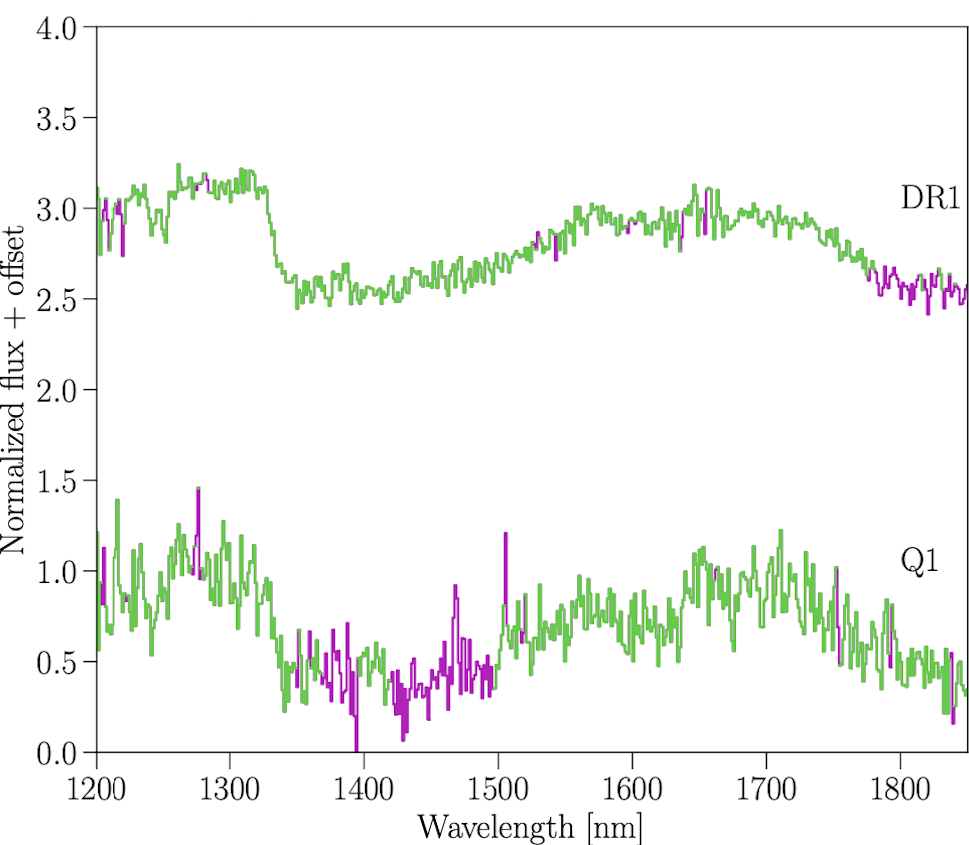}{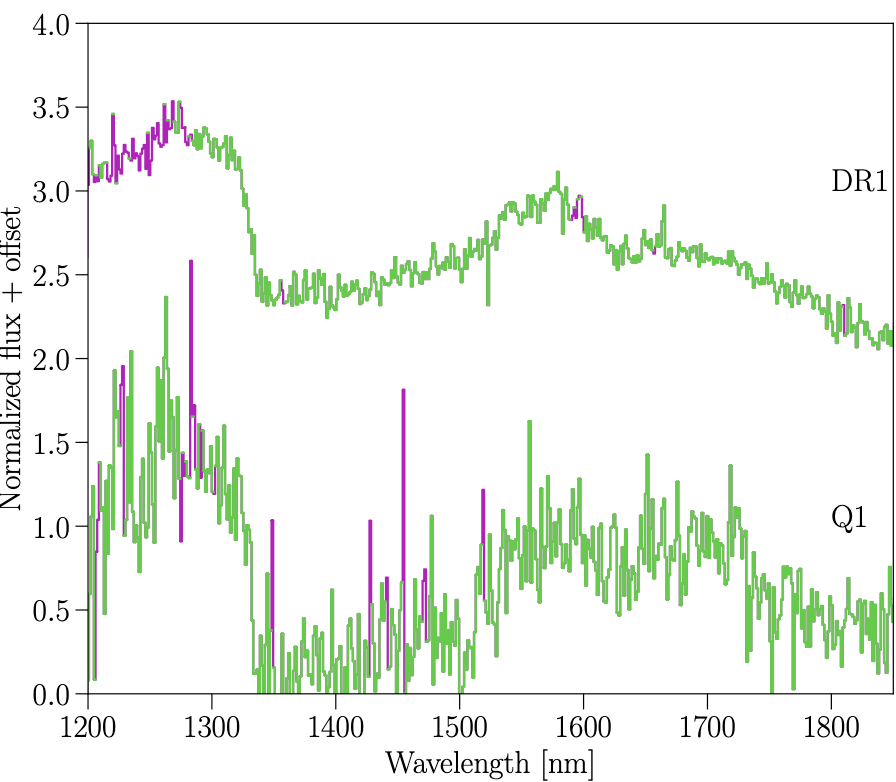}
\plottwo{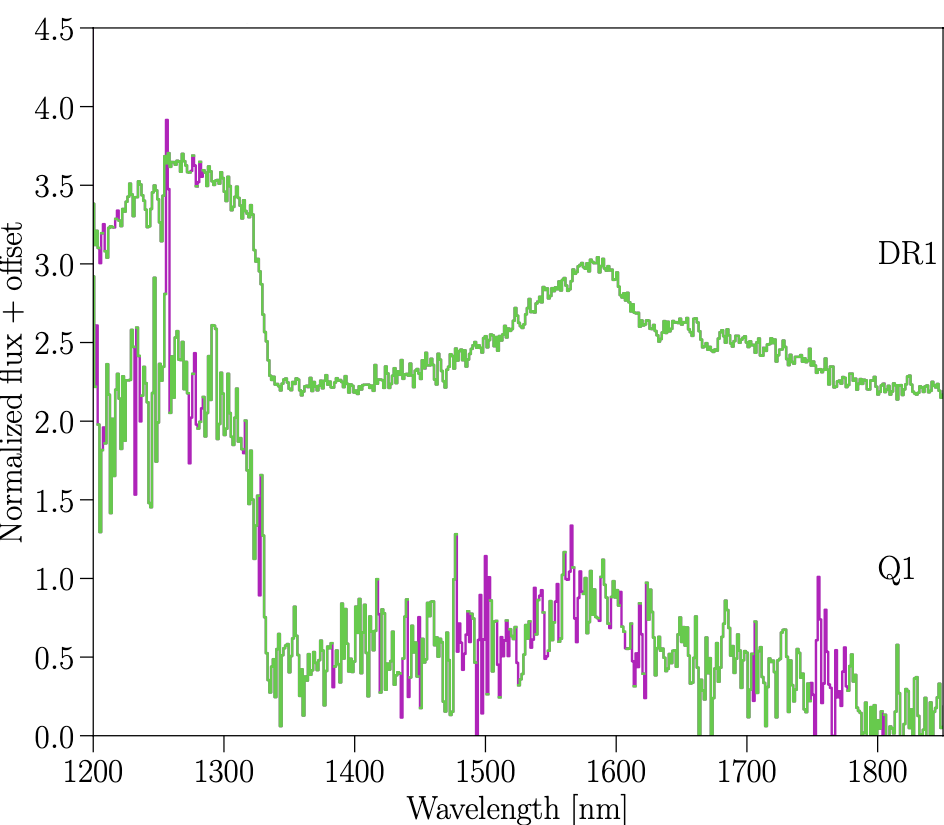}{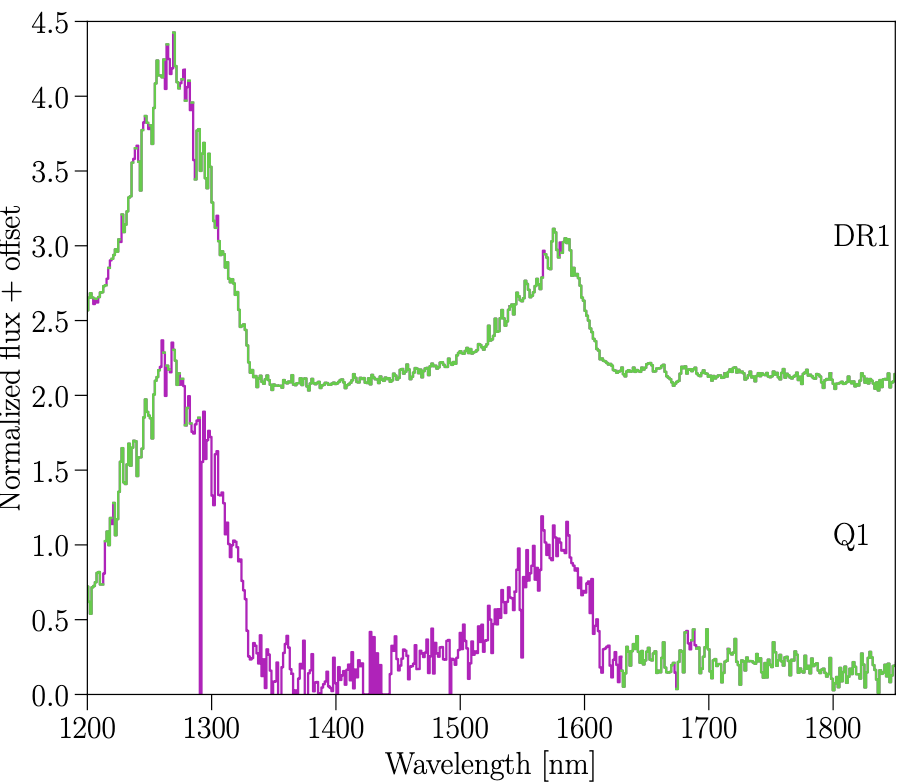}
\caption{ Four examples of spectra from Q1 and DR1 for the same object. The green lines represent the spectra with \texttt{NDITH}~$>$~2 and  the  magenta lines  indicate the values with  \texttt{QUALITY} $<$~0.7.  }
\label{fig:q1dr1}
\end{figure}

\section{\label{sc:data}Data}

\Euclid data includes visible and NIR images, as well as NIR spectra, from its two instruments VIS \citep{EuclidSkyVIS} and The Near-Infrared Spectrometer and Photometer  \citep[NISP;][]{EuclidSkyNISP}. The observations are conducted in the Reference Observing Sequence  \cite[ROS;][]{Scaramella-EP1}. Each ROS includes four images in the four bands (\IE,  \YE, \JE, and \HE) and four spectra by performing a dithering pattern.  NISP has two operational red grisms and one blue. They cover the spectral ranges of 1206-1892~nm and 930-1366~nm, respectively, and have an overlap region of about 166~nm. The wavelength dispersion of the blue grism (1.23~nm/pix) is slightly lower than that for the red grism (1.34~nm/pix). 
 The difference in the number of grisms for the blue and the red, translates into different observational sequences. The red-grism observations in the ROS include four dithering positions, two for each red grism changing its angle by four degrees. This strategy has demonstrated to overcome the cross-contamination in slitless spectroscopy \citep{Copin2026, Passalacqua2026}. The observations with the blue grism, in contrast, are limited to only two angles, reducing the effectiveness of the decontamination strategy.
The \Euclid pipeline reduces the data and extracts the spectra  \citep[see][]{Copin2026,LeBrun2026}. The output 1D spectra files include wavelength, flux density, the flux density variance, along with some flags for  \texttt{QUALITY}, \texttt{NDITH}, and  \texttt{MASK}\footnote{\url{https://euclid.esac.esa.int/dr/q1/dpdd/}}.

\Euclid mission survey \citep{Scaramella-EP1}  is observing the \Euclid Deep Fields (EDF) in several epochs. DR1 (Aussel et al. submitted;  Tereno et al. submitted) includes 8 visits of EDF-N and 10 of EDF-S. As each visit or ROS takes images and spectra in 4 dithering positions, the number of individual spectra available to be coadded can exceed 30. From these two fields, only EDF-N includes blue-grism spectra.
The third field, EDF-F has one visit only, with a mean of only three observations per combined spectra (similar to Q1 data). In contrast, it includes blue-grism spectra.
DR1 also incorporates deep data for the auxiliary and calibration fields with more than three visits (Tereno et al. submitted). The auxiliary fields include the COSMOS region \citep{COSMOS2007}, with up to 29 visits and blue-grism spectra. The large volume of data in DR1, comes from the Wide Survey, which covers an area over 2100 deg$^2$ and is observed in one visit and using only the red grism. The Wide Survey covers regions in the north and south hemispheres at high galactic latitudes.

In this paper, we collectively refer to the EDF and all fields with more than one visit (including auxiliary and calibration fields) as “deep observations”, to distinguish them from the Wide Survey fields, which have only a single visit.

\section{\label{sc:q1}Revisiting the Q1 UCDs}

A single-visit data from Q1 allowed us to find new UCDs, classify and analyze them  \citep{Dominguez2025,Marusa2025}. Many objects were left out owing to insufficient data on their spectra or poor signal-to-noise ratio (SNR). The new spectra to be released by DR1 for these regions overcomes those limitations as the spectra is the result of averaging a large number of single spectra. 

As DR1 data became available we searched for the spectroscopically confirmed dwarfs (M7 to T7) from our previous works with Q1. The object ids changed from one catalog to the other, so we cross-matched their positions in the sky with the DR1 catalog of spectra, in a radius of three arcsecond. We also cross-matched our list of unpublished candidates and found matches for over 90\% of them. Together we created a list of close to 600 dwarfs spanning from M7 to T7 spectral types. We reclassified them (see Sect.~\ref{sc:spt}), using their DR1 spectra, which allowed to reduce the uncertainties. The full list of these UCDs is included in Appendix~\ref{sc:Q1600}.  
 Fig.~\ref{fig:q1dr1} shows a comparison between Q1 and DR1 spectra for two UCDs. The improvement in SNR, owing the larger number of averaged  spectra is evident. The UCDs are named by an alias, which composed by the first 8 digits of the \Euclid  ID.
 We used DR1 data of these UCDs to tune the search algorithm described in Sect.~\ref{sc:search} and for the creation of the templates described in Sect.~\ref{sc:template}. For the rest of the analysis in the following sections of this paper, we focus only on the dwarfs cooler than L3.

\begin{table}[h!]
\caption{\label{tab:regions}  Spectral index search regions.}
\centering

\begin{tabular}{lcc}
\hline\hline
Region & L4-T2 & T3-Y1 \\
 & (nm) & (nm) \\
\hline
P$J$  &  (1270.4 - 1310.6) & (1243.6 - 1283.8)  \\
P$H$  &  (1559.8 - 1640.2) &  (1559.8 - 1600.0)  \\
$JH$  &  (1349.5 - 1429.9) &  (1349.5 - 1470.1)  \\
$HK$  &  (1819.8 - 1846.6) &  (1649.6 - 1810.4)  \\
\hline
\end{tabular}
\end{table}


\section{\label{sc:search}Spectral index search} 

We searched for mid-L to Y dwarfs by looking for the deep and broad absorption features of H$_2$O and CH$_4 $, included in the spectral range of NISP. We modified the search algorithm used in our previous work  \citep{Dominguez2025}, to run directly in the Datalabs platform  \citep{DataLabs},  made available by ESA. The number of spectra released in the EDF  doubled from Q1 to DR1 and, for the Wide Survey surpasses 240 million.
The search algorithm from  \cite{Dominguez2025} works in sequential steps:  it filters first the data to eliminate unreliable values using the flags \texttt{QUALITY}, \texttt{NDITH}, and  \texttt{MASK}, described in Sect.~\ref{sc:data}. Then, it computes index ratios from four spectral regions. Using the UCDs confirmed in Q1, we optimized these regions to search for mid- and late-L, T, and  Y dwarfs. 
The regions are labeled as P$J$ and P$H$ for the peak regions and $JH$ and $HK$ for the deepest region of the broad absorption bands between two consecutive bands ($J$, $H$, and $K$).  Their ranges are listed in Table~\ref{tab:regions}.
All the wavelength range can be employed owing to the fact that  \Euclid observations are free from telluric absorption bands.
The search algorithm computes three ratios using the integrated fluxes:  P$J$-$JH$,  P$H$-$JH$, and P$H$-$HK$. 
In the second step, the algorithm convolves the spectra with a gaussian profile and computes the ratios again. This allows to filter unreliable values that escaped from the flags mentioned before.
The  L-type candidates are selected by setting values larger than 0.39 for the three ratios in both steps. For the T-type candidates it uses a minimum of 0.6 for the first step and 0.3 for the second one. These values were  calibrated using the DR1 spectra of the  UCDs described in Sect.~\ref{sc:q1}. We did a visual confirmation of the spectra for the selected candidates. We also looked to their \Euclid images in the four bands and we found some objects with a close companion. These objects and their images are discussed in Sect.~\ref{sc:notes}.

\subsection{\label{sc:searchedf}UCDs found in the fields with deep observations} 

Up to date, running the search algorithm in the  8.3 million spectra from the Deep Fields, we have found $> 50$~L dwarfs and $> 50$~T dwarfs. Despite a search focused to mid and late  L dwarfs, the results include a dozen of early L dwarfs. After verifying previous references to the UCDs in the literature, it turned that we found, only for the EDF, more than 70 UCDs not published before.  Table~\ref{tab:numeritos} summarizes the exact numbers. In the auxiliary and calibration fields, we found a dozen more dwarfs. Their numbers are also included in the table. 

\begin{figure}[ht!]
\centering
\includegraphics[width=0.6\hsize]{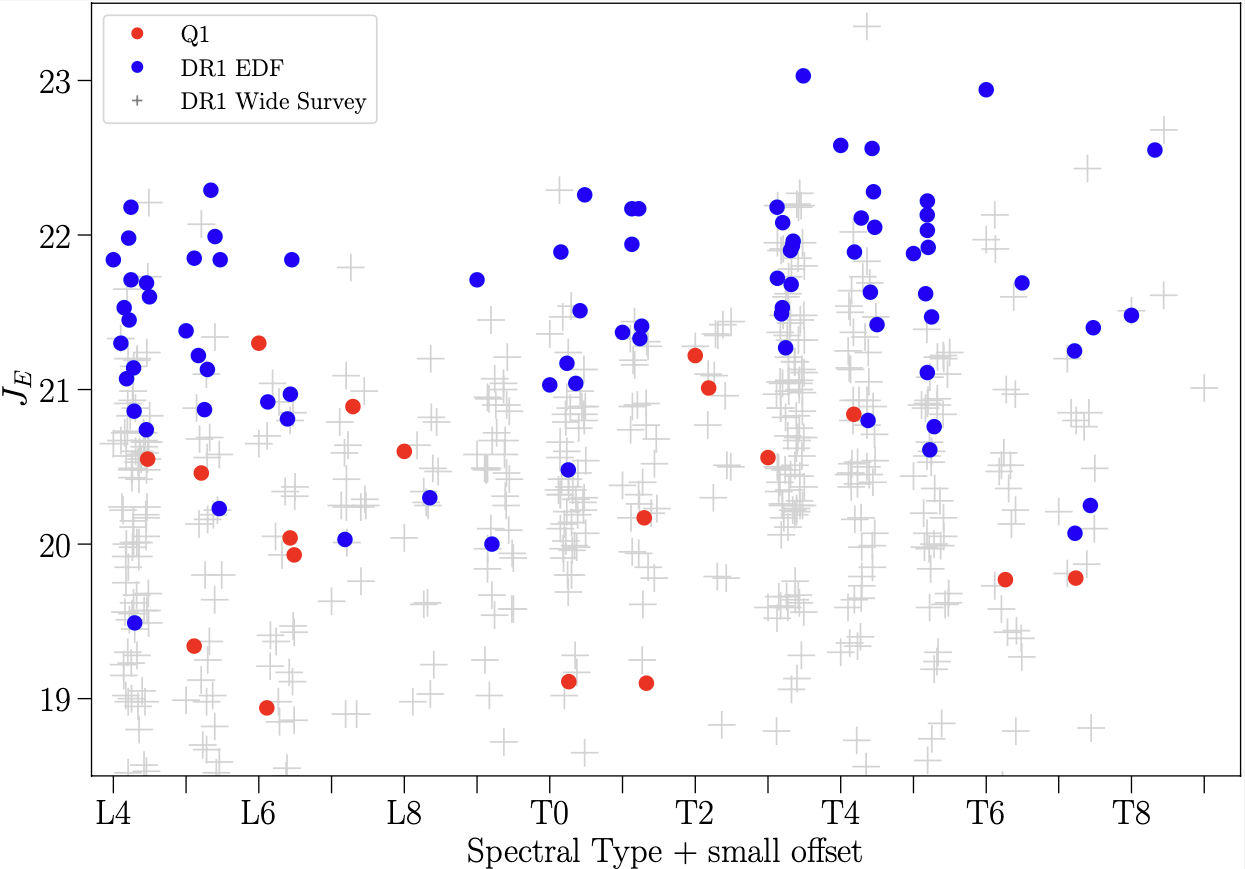}
\caption{ \JE magnitude vs spectral type for Q1 (red circles) and DR1. DR1 data is represented for the EDF (blue circles, to compare with Q1) and for Wide Survey (grey plus).  The spectral classification is estimated in Sect.~\ref{sc:spt}.  }
\label{fig:q1dr1magj}
\end{figure}

We can see in Fig.~\ref{fig:q1dr1magj} the \JE magnitude vs. the spectral classification, which is estimated in Sect.~\ref{sc:spt}. Most of objects found in DR1 EDF (i.e., same fields than Q1) are one magnitude deeper. The faintest dwarf is E66068698 a T3 with a \JE AB magnitude of 23.0. 
It makes us to foresee the discovery of even fainter UCDs in future data releases.  

We have computed two \Euclid colors:  \JE -  \HE  and \IE -  \YE, and plotted them as a function of the spectral type. The left panels in Fig.~\ref{fig:color}  shows the result for the deep observations.  The plot for  \JE -  \HE follows a well defined sequence. There are objects that have clear red or blue colors. In the latter case these plots include the diamond symbol  for  the possible subdwarfs found during the spectral classification (Sect.~\ref{sc:spt}).
They are analyzed in Sect.~\ref{sc:notes}.

\subsection{\label{sc:searchws}UCDs found in the Wide Survey} 

As we can see in the third row of Table~\ref{tab:numeritos}, much larger numbers apply for the search in the 240~million spectra of the Wide Survey. Up to date, running the search algorithm in these fields, we have found more than 260 L dwarfs and more than 340 T dwarfs. From both groups, more than 500 UCDs have not been published before. Fig.~\ref{fig:q1dr1magj} includes these objects (represented as a grey plus) and we can see that the bulk of objects are, as in Q1, found below 21.5 in the \JE AB magnitude, with exception of some dwarfs, which can exceed 22. The faintest object, however is E22929186, a T4 with a  of 23.3 \JE AB magnitude.

Table~\ref{tab:find} lists all the UCDs discovered in both searches, the deep observations and the Wide Survey. Their \Euclid photometry is included, as well as their spectral types and distance estimations described in Sections~\ref{sc:spt} and \ref{sc:dist}, respectively.

Fig.~\ref{fig:colorws} shows the  \JE -  \HE  and \IE -  \YE colors vs.  spectral type (Sect.~\ref{sc:spt}) for the Wide Survey. As in the previous case, the plot for  \JE -  \HE follows a well defined sequence, and there are a group of red objects in the L/T transition. They are analyzed in detail in Sect.~\ref{sc:notes}, as well as the possible subdwarfs.  The  late-T dwarfs present a bluer color in the \IE -  \YE  plot, owing to a detection in  \IE   below $10\sigma$.
\subsection{\label{sc:searchy}The search for Y dwarfs} 

We started the search for Y dwarfs by cross matching the known objects from the literature to the regions observed  by \Euclid DR1. We found two of them: the Y0 dwarf WISEA J030237.52-581740.3  \citep{Tinney2018} and the Y1 dwarf WISE J035000.32-565830.2 \citep{Kirkpatrick2012}  by searching in a radius of 10 arcsecond around the coordinates corrected by the proper motions estimated by WISE \citep{WISE}. The search included reviewing the VIS and NIR images for all the possible candidates, estimating their \JE~-~W2 color, and reviewing the \Euclid spectra.  Both dwarfs were found more than two arcsecond away from where they were expected, which could be explained by the errors in the proper motion estimations.
These two Y dwarfs  are described in detail in Sect.~\ref{sc:ys}. As they fall in the region Wide Survey (i.e., one visit only), their spectra has a very low SNR, which makes it very difficult to be retrieved by our spectral index search in these fields.

We continued the search by using the spectral index for late T dwarfs (column 3 of Table~\ref{tab:find}), particularly in the fields with deep observations. In order to asses the possibilities of success, we compared the  simulated \JE AB magnitude from \citep{Leggett2025}  with those for the  two Y dwarfs (see Sect.~\ref{sc:ys}). As other Y0 dwarfs could be as bright as 20.8 in  \JE AB magnitude, we inspected the VIS and NIR images of the possible candidates retrieved by the spectral index search, giving zero confirmations. 
 From what we saw in Fig.~\ref{fig:q1dr1magj}, one-visit \Euclid observations could retrieve the spectra of Y0 dwarfs. Even the brightest Y1 dwarf (\JE = 22.9) seems to be at or beyond the depth limit for DR1. 
 The multiple-visits in the deep observations, which will continue to gain depth throughout the \Euclid nominal mission, keeps open the possibility of finding new Y0 dwarfs in future data releases.

\begin{table}
\caption{Number of UCDs found and discovered per field. }
\centering

\begin{tabular}{lrrrrr}
\hline\hline
Field & early-L dwarf & L4-L9 dwarf & T dwarf & Total & Discovered\\
\hline

EDF                    &  18   &  50  & 71    & {\bf139}   & $>$60  \\
Aux. and cal.       & 1    &  10   &  2      &  {\bf13}   & $>$10  \\
Wide Survey       &  41  &  229  & 341  & {\bf611}  & $>$530  \\
\hline
All  fields              & 60   &  289  & 414  & {\bf763}  &  $>$600 \\
\hline
\end{tabular}
\label{tab:numeritos}
\tablecomments{EDF and auxiliary and calibration form together what we call "the deep observations". The last column gives the number of UCDs from the Total that are not found in the literature. }
\end{table}

\begin{figure}[ht!]
\plottwo{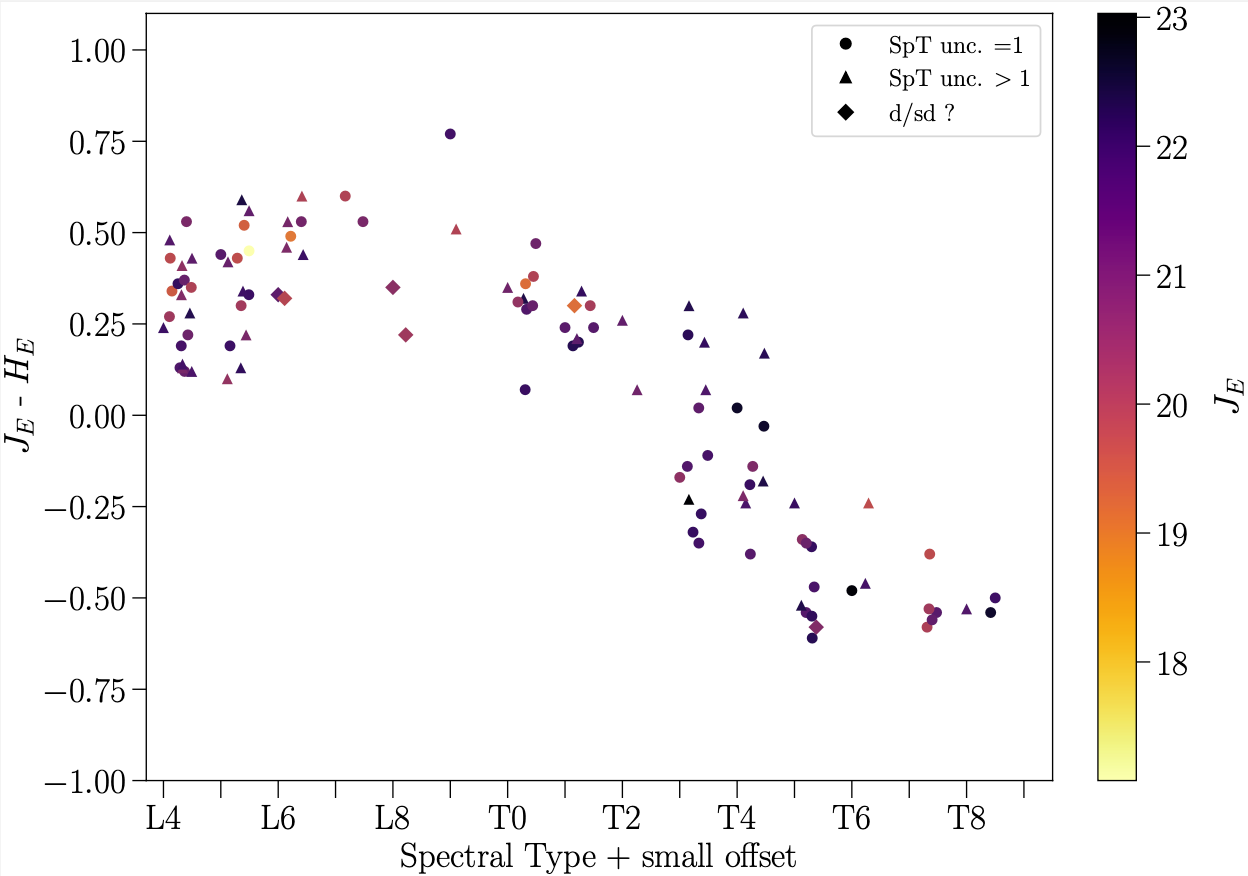}{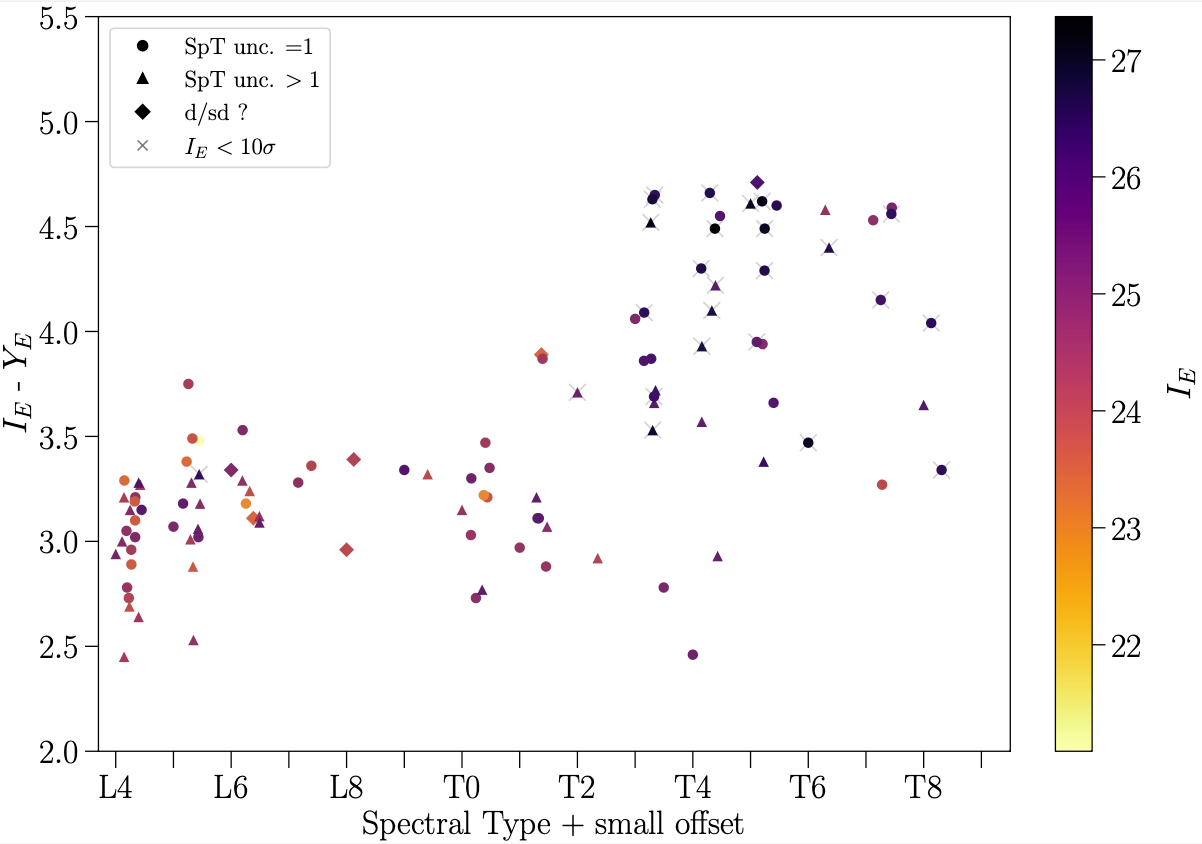}
\caption{  \JE -  \HE (left) and \IE -  \YE (right) color vs. spectral type for the UCDs (circles) in the deep observations.  The symbol color follows the \JE and \IE AB magnitudes, respectively. The triangles represent the spectral classification with an uncertainty larger than 1. The 
diamonds represent the possible subdwarfs. A grey cross in the  \IE -  \YE (right) color plot, indicates that the SNR in \IE magnitude is lower than 10. }
\label{fig:color}
\end{figure}

\begin{figure}[ht!]
\plottwo{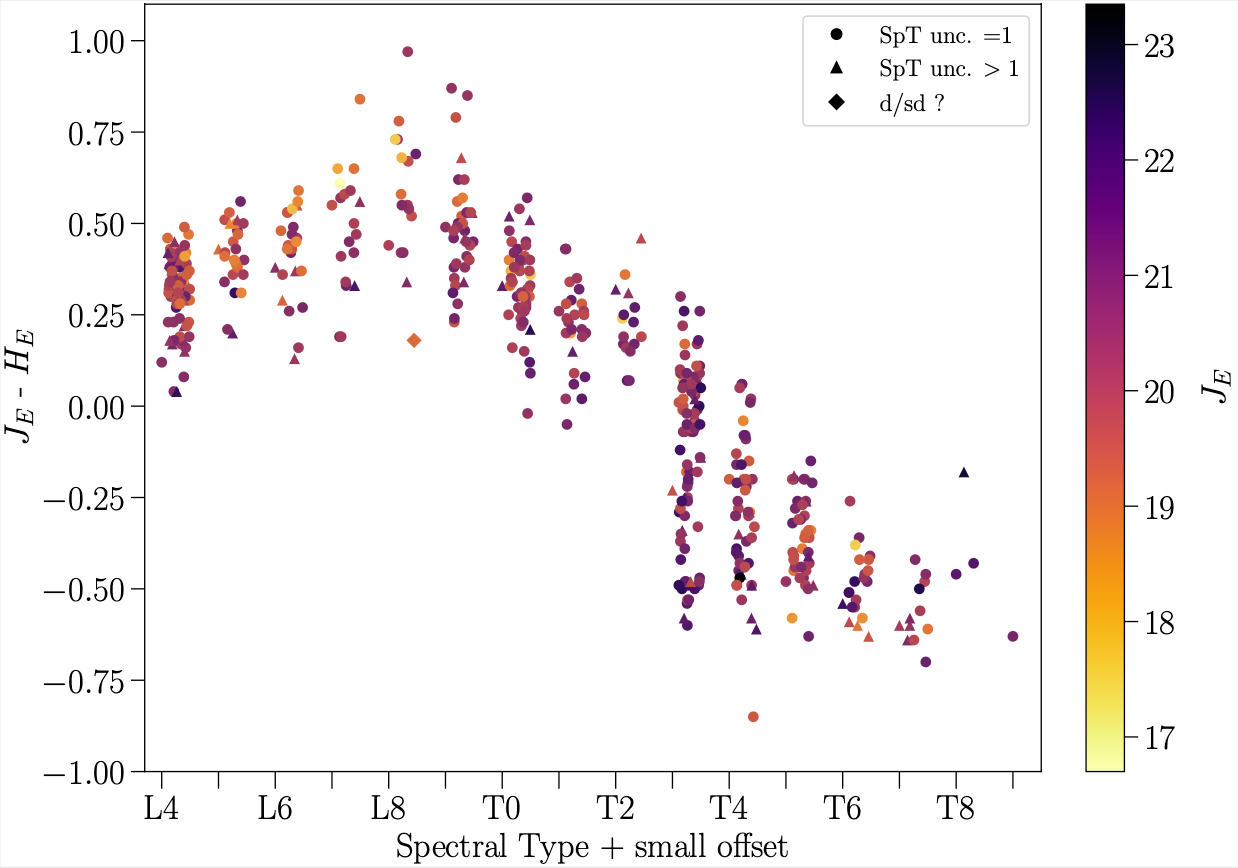}{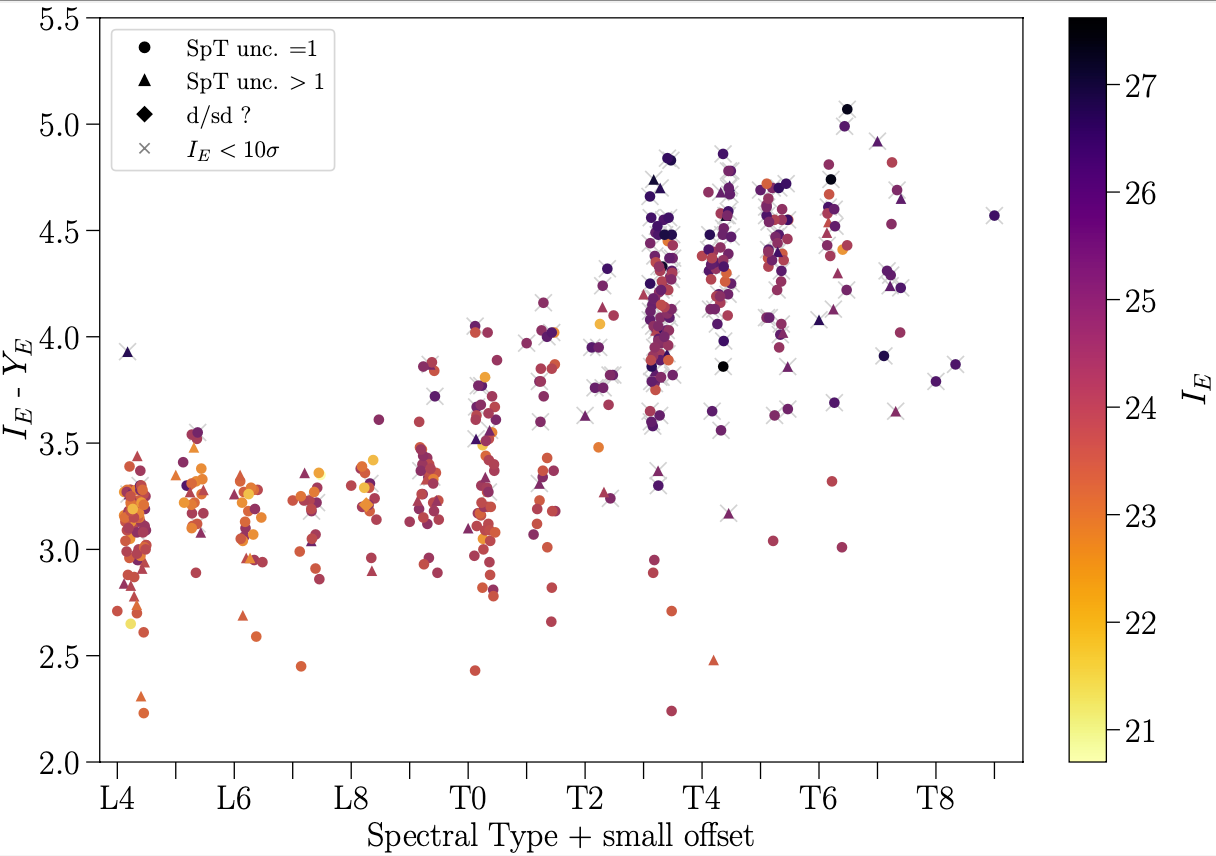}
\caption{  \JE -  \HE (left) and \IE -  \YE (right) color vs. spectral type for the UCDs (circles) in the Wide Survey. The symbol color follows the \JE and \IE AB magnitudes, respectively. The triangles represent the spectral classification with an uncertainty larger than 1. The 
diamonds represent the possible subdwarfs. A grey cross in the  \IE -  \YE (right) color plot, indicates that the SNR in \IE magnitude is lower than 10. }
\label{fig:colorws}
\end{figure}

\begin{longtable}{llrrlrrrrr}
\caption{ UCDs discovered by this work using the spectral index search. It includes first spectroscopic confirmations. } \\
\hline\hline
Alias   &  Designation   \texttt{Euclid DR1 +}   &     R.A. (J2000)   &    Dec. (J2000)   &   SpT       &   \IE    &  \YE  & \JE  & \HE &  SP Dist. \\
    &         &      &       &      &     (AB)   &  (AB)    &  (AB)    &  (AB) &  (pc)   \\ 
\hline
\endfirsthead
\multicolumn{10}{@{}l}{\ldots continued}\\ \hline
Alias   &  Designation    \texttt{Euclid DR1 +}  &     R.A. (J2000)   &    Dec. (J2000)   &   SpT       &   \IE    &  \YE  & \JE  & \HE &  SP Dist. \\
    &        &      &       &      &     (AB)   &  (AB)    &  (AB)    &  (AB) &  (pc)   \\ 
\hline
\endhead
\hline
\multicolumn{10}{r@{}}{continued \ldots}\\
\endfoot
\endlastfoot
E85298142 & J003407.155+031424.35 & 8.5298143 & 3.2400971 & T4: & 26.3 & 21.6 & 21.2 & 21.8 & 140.3 \\
E10654771 & J004237.145+042848.62 & 10.6547715 & 4.4801731 & T4 & 26.5 & 22.1 & 21.7 & 21.9 & 155.9 \\
E11146411 & J004435.138+044015.66 & 11.1464119 & 4.6710163 & L4 & 23.6 & 20.4 & 19.8 & 19.4 & 139.1 \\
E11521966 & J004605.271-555050.34 & 11.5219660 & -55.8473174 & L9p & 23.9 & 20.8 & 20.3 & 19.9 & 95.6 \\
E11742418 & J004658.180-525010.66 & 11.7424184 & -52.8362950 & L6 & 23.1 & 19.8 & 19.1 & 18.6 & 74.1 \\
E12027617 & J004806.628+044511.56 & 12.0276178 & 4.7532110 & T3 & 25.6 & 21.1 & 20.7 & 20.8 & 106.6 \\
E12113518 & J004827.244+061052.36 & 12.1135186 & 6.1812103 & L5p & ... & ... & ... & ... & ... \\
E12132150 & J004831.716-521741.82 & 12.1321504 & -52.2949498 & L2: & 23.3 & 21.0 & 20.8 & 20.8 & 345.6 \\
\end{longtable}


\section{\label{sc:spt}Spectral type (SpT) classification}

We classified the spectra combining the two methods described in  \cite{Dominguez2025}: a $\chi ^2$ minimization in the full spectral range and  a residual minimization over short telluric-free wavelength ranges. The classification is based on the red-grism spectra only, owing to the limited number of good blue-grism spectra and the calibration issues discussed in Sect.~\ref{sc:analdo}.
We use as a reference the standard templates observed with SpeX \citep{spex} and compiled in SPLAT \citep{SPLAT}. They are listed in the Appendix~\ref{sc:splat}. 
The obtained uncertainty in the spectral type with these methods is $\pm$ 1 subtype. We added a colon to the spectral subtype to indicate a larger uncertainty. 
The objects whose spectrum, split in $J-$ and $H-$ bands, fit to different standard templates are marked as peculiar dwarfs with a ``p" added to the spectral subtype.  In order to investigate these peculiar cases, we have added standard templates for low gravity dwarfs  and  subdwarfs  from SPLAT, and  \cite{Burgasser2025}.
In consequence, we classify a spectrum as a possible subdwarf if it provides a better fit (i.e., a difference  in the value of $\chi ^2$ larger than two) with any of the added subdwarf templates compared to the standard templates.
We use the prefix of  ``d/sd" for the mild subdwarfs , and ``sd" for the subdwarfs, and ``esd" for the extreme ones  in accordance to the classification system of  \cite{Gizis1997, Lepine2007},  extended to L dwarfs by \cite{Zhang2017} and to  T dwarfs by \cite{ Burgasser2025}. The objects that we have classified as possible subdwarfs are detailed in Sect.~\ref{sc:dsd}, and marked with different symbols throughout the paper. 
For other peculiar objects, 
we also analyze the possibility of being unresolved binaries. A better fit to the composition of two templates is indicated with  a ``c" in the spectral type, as a potential  unresolved binary candidate. This could also mean a contaminated spectrum.
 This method, based on previous works \citep{Bardalez2014,Manjavacas2016,Bravo2023,BinML2026},  is described  in Sect.~\ref{sc:bin}. 
 
 The spectral types of the discovered UCDs are listed in Table~\ref{tab:find}, and for all the objects in Appendix~\ref{sc:catal}.


\section{\label{sc:template}{\it Euclid} UCD templates}

\begin{figure*}[h!]
   \centering
\includegraphics[width=1\linewidth]{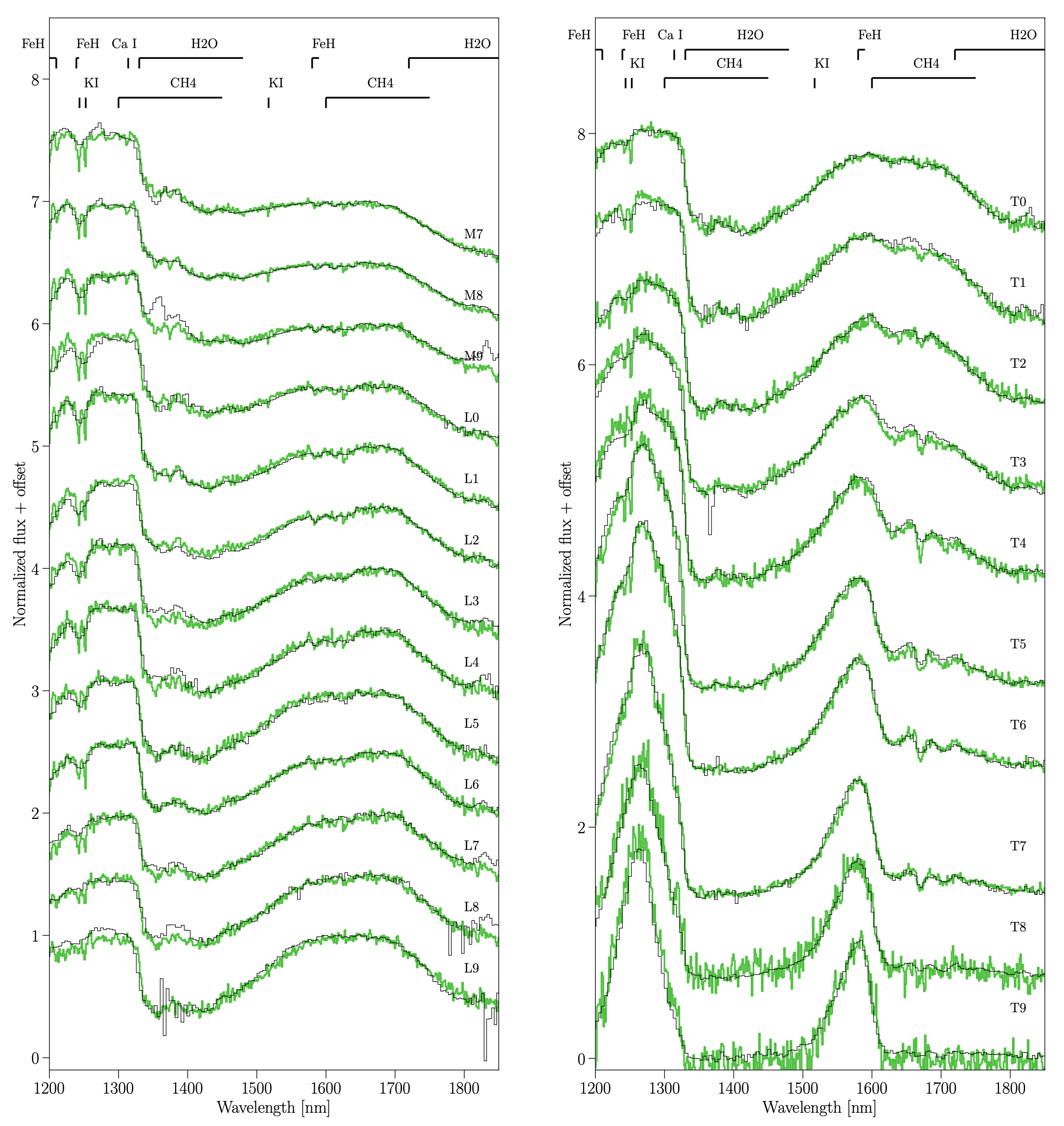} 
\caption{EUC templates (green line) for all spectral subtypes. The black lines represent the SpeX standards, which are listed in the Appendix~\ref{sc:splat}. }
\label{fig:template}
\end{figure*}

We selected the higher quality spectra from all the sample (i.e., Q1, deep observations and Wide Survey) to create telluric-free templates from \Euclid observations. Many of the spectra in the Wide Survey present artifacts as it is based on a single visit.
In order to rise the number of spectra of late-M dwarfs, we incorporated good quality spectra from the photometric search conducted by {\v Z}erjal et al. (in prep.). The templates are created by combining all the spectra of each spectral subtype by the median. The number of spectra to combine is over 10 in most of the spectral subtypes, which allowed to create high-quality \Euclid UCD templates (hereafter ``EUC templates"), shown in Fig.~\ref{fig:template}. 
L7 to L9 and T4 dwarfs have fewer numbers owing to the lack of good quality spectra, and we have only two for T8 and one for T9. We expect that these templates will improve from further discoveries and in the next data releases.
These templates are of major importance in the study of the regions affected by the telluric absorption. 
We can identify the main spectroscopic features in the wavelength range, including  molecular bands of 
water (1.35-1.55  and  1.7-1.85~$\mu$m), 
 methane (1.3-1.4 and 1.6-1.7~$\mu$m), 
 and FeH (1.20, 1.24,  and 1.6~$\mu$m). 
As well as the atomic lines K~I~(1.244, 1.253, and 1.517~$\mu$m) and Ca~I~(1.290~$\mu$m). Being the latter very weak and only present  for the M and early-L dwarfs.
The intensity of water bands increases monotonically from late M to late T dwarfs  and methane bands in the $H$- and $J$- bands appear at T0-T1 and increase monotonically down to late-T dwarfs, where they leave the flux at practically zero level. 
These telluric-free spectra clearly shows what has been widely studied \citep[see][]{Burgasser02,Burgasser06,Burgasser2025}.
It is interesting to note the fine structure of the water band around 1.48~$\mu$m  in the early-L dwarfs  (i.e., where is no longer overlapped to methane), to the limit imposed by NISP spectral resolution.
The molecular bands of FeH at 1.20 and and 1.6~$\mu$m can be well appreciated in late-M and early-L dwarfs, disappearing at late- an mid- L dwarfs, respectively. The K~I~ absorption doublet (1.244/1.253~$\mu$m)  is clearly identified from late-M to early-T dwarfs, disappearing at T

We use these templates to review the final spectral classification and, as a reference in all the subsequent spectral analysis presented in  Sect.~\ref{sc:anal}. We also did a  comparison between the EUC templates and a set of telluric-free templates from JWST/PRISM spectra, recently published  by Pritchard et al. (submitted). The comparison can be found in Appendix~\ref{sc:jwst}.

\section{\label{sc:dist} Spectrophotometric distance estimation} 

\begin{figure}[ht!]
\plottwo{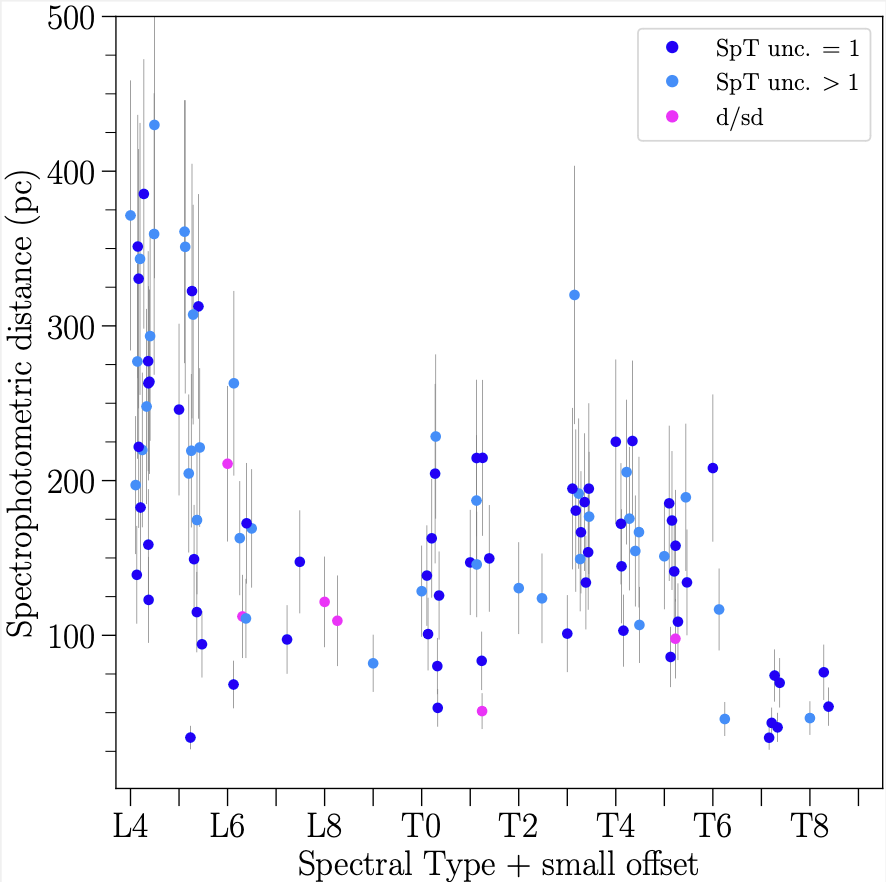}{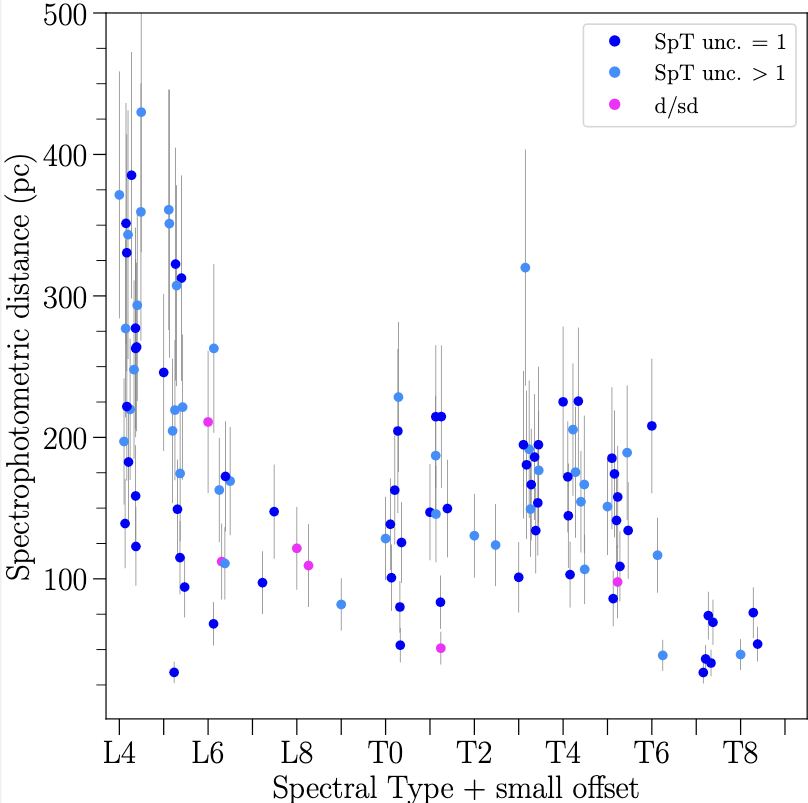}
\caption{Estimated spectrophotometric distances for the deep observations (left) and for the Wide Survey (right). The dark blue circles are for the dwarfs with an uncertainty in the spectral type of $\pm1$, the light blue circles for the dwarfs with larger uncertainties and the magenta circles for the possible mild subdwarfs (``d/sd").}
\label{fig:dist}
\end{figure}

We estimated the spectrophotometric distances for UCDs that have a precise spectral type classification (i.e. non peculiar spectra). 
The distance modulus is estimated using the semi-empirical relations of absolute magnitudes for each spectral type in field objects derived by \citet{Sanghi2024}. We use \Euclid photometry for the  \JE and \HE bands, excluding the values signaled with a flag in the \Euclid catalog. 
The distance uncertainty is computed considering the difference between the estimations from the two bands, the photometric error,  the rms dispersion reported by \citet{Sanghi2024} and the uncertainty in the spectral-type classification. The latter  is estimated by computing the distance corresponding to $\pm1$ subtypes. We use $\pm$2 for those classifications with  an added colon. We do not estimate distances for peculiar objects nor possible binaries. In the case of the possible subdwarfs, we estimate the spectrophotometric distance for a standard subtype and add a note in the reported value (see Table~\ref{tab:find}). The results, plotted in the left panel of Fig.~\ref{fig:dist} for the objects in the deep observations, show that we are finding UCDs at distances up to $440\pm100$~pc. The bulk of objects are at distances between 100 and 200~pc, except for the more distant mid-L dwarfs,  and the closer late-T dwarfs.
The closest UCD in the deep fields is E59791358, a T7 at a distance of $34\pm8$~pc. The possible mild subdwarf E27480941, is a T1  at a distance of $51\pm12$~pc. 
The right panel of Fig.~\ref{fig:dist} shows that the majority of UCDs in the Wide Survey are located at distances between 50 and 150~pc, except for the mid-L dwarfs, which are further. Most of the late-T dwarfs are closer than 75~pc.
The closest UCD in the Wide Survey is the T6 dwarf E77647802 at a distance of $17\pm4$~pc. 
Comparing the bulk of objects in left and right panels, we can see that the detections of mid-L dwarfs in the fields with deep observations are $\sim100$~pc more distant than in the Wide Survey. This could be explained by  the difference of 1 magnitude seen in Fig.\ref{fig:q1dr1magj}.

\section{\label{sc:anal}Spectral analysis}

We have divided this section in two parts, in order to analyze observations with two different depths. We start with the analysis of the UCDs in the fields with deep observations, some of which have blue-grism spectra and, thus, more indices and alkali lines available to be studied. The second part of this section will be dedicated to the Wide Survey.

\subsection{\label{sc:analdo}Analysis in the fields with deep observations}

\begin{figure*}
   \centering
 \includegraphics[width=0.95\linewidth]{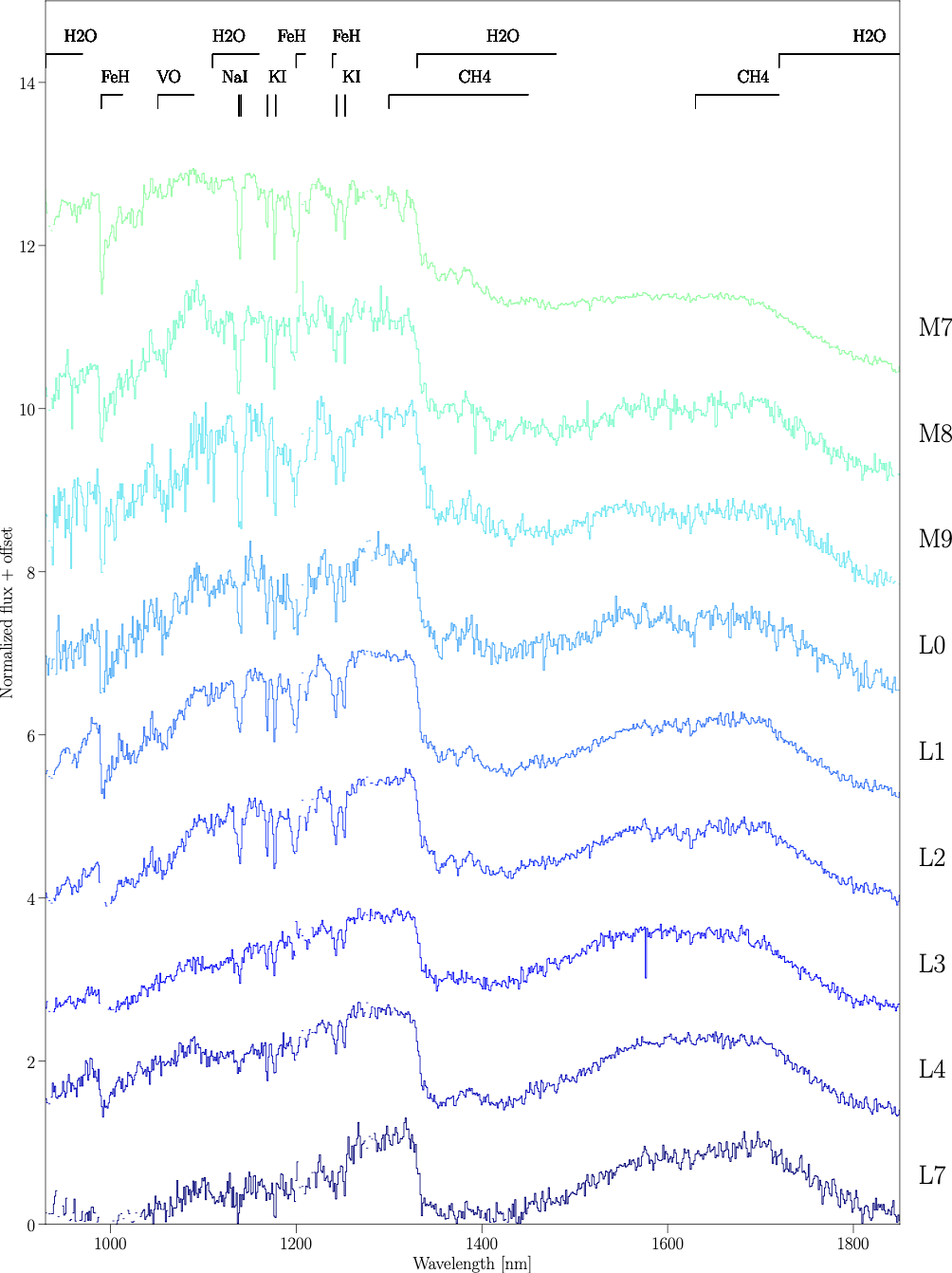} 
\caption{Spectral sequence for M and L dwarfs with the blue- and red-grism spectra. The flux is normalized at 1600~nm. The spectra are plotted without any smoothing. The missing values are due to  artifact removal.}
\label{fig:seqblueML}
\end{figure*}

\begin{figure*}
   \centering
 \includegraphics[width=0.95\linewidth]{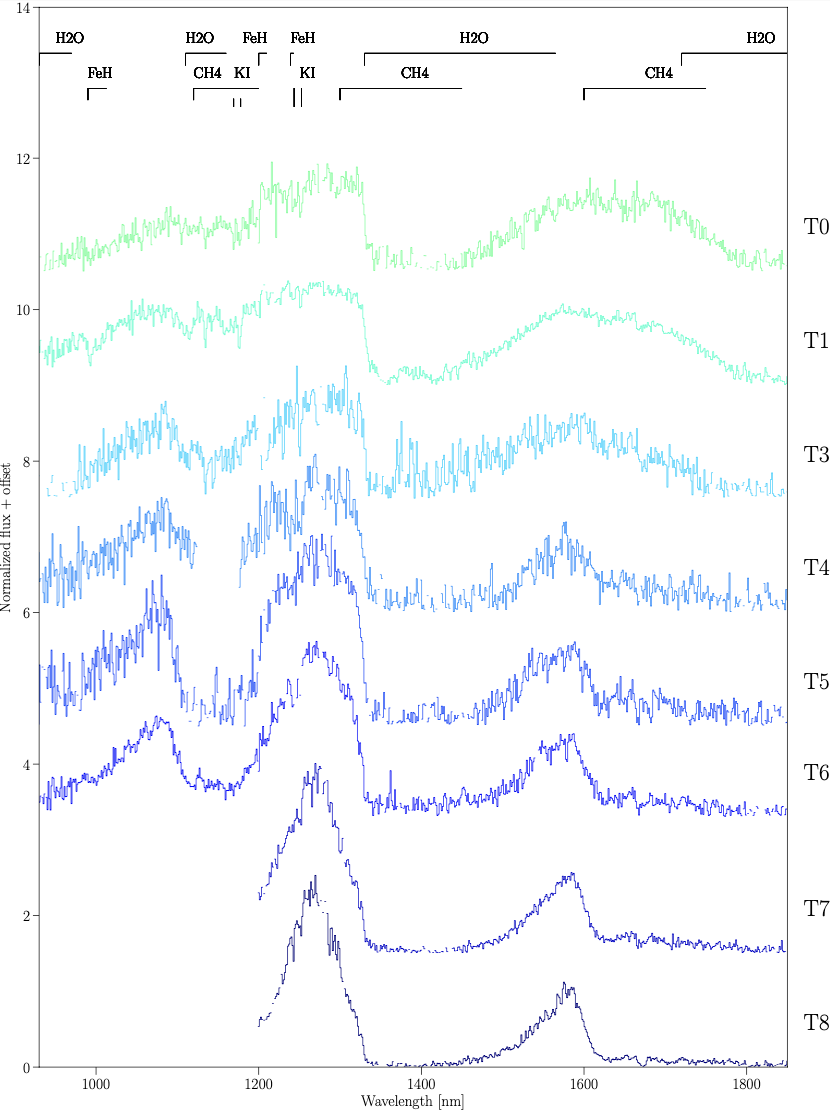} 
\caption{Spectral sequence of T dwarfs  with the blue- and red-grism spectra. T8 and T9 have no blue spectra available. The flux is normalized at 1600~nm. The spectra are plotted without any smoothing. The missing values are due to  artifact removal.}
\label{fig:seqblueT}
\end{figure*}

The good quality of the spectra from DR1 deep observations allows to compute spectral indices and to measure specific spectral features. As described in Sect.~\ref{sc:data}, EDF-N, EDF-F, and COSMOS field have blue-grism spectra.  
This gave us the opportunity to investigate the quality of the data from this grism for the analysis of UCDs. And we found that
only 19\% of the objects in these fields had good blue-grism spectra  available. We included it in the analysis to test its results.
The wavelength range of this grism goes from 926 to 1366~nm and, together with the red one, they cover a wavelength range from 926 to 1892~nm.  There is an overlap region between the two grisms from 1206 to 1366~nm. The analysis of both spectra showed some differences in their fluxes in this overlap region.  While the spectra from the red grism have been validated in the analysis of Q1 data, we focused on  the calibration of the spectra from the blue grism.
One possibility is that the differences in flux arise from a  larger background or cross contamination in blue-grism spectra owing to the reduced number of grism/angle combinations, as discussed in Sect.~\ref{sc:data}.
We reviewed the flux calibration by comparing the object integrated flux to its \YE band photometry. This band is fully covered in the wavelength range of the blue grism. For the cases where the fluxes from the spectra and  the \YE band were different, 
we shifted the level of the spectra flux to calibrate it to that from the \YE band. To verify the results, we integrated the flux again and created a look-up table with the shift values.
Figures  \ref{fig:seqblueML} and \ref{fig:seqblueT} show the best blue- and red-grism spectra of UCDs in a sequence for most spectral subtypes. There are missing subtypes as the selection was made for the highest quality spectra from the blue grism. Only the best data is plotted in the overlap wavelength region and the artifacts have been masked.

The wavelength range of the blue grism include more alkali lines: Na~I (1.138 / 1.141~$\mu$m) and K~I~absorption doublet (1.169 / 1.178~$\mu$m), plus the molecular bands of vanadium oxide (VO at 1.06~$\mu$m) and FeH  (0.99 and 1.2~$\mu$m).
There are two potassium doublets in the range of the blue grism, the one mentioned before plus K~I~(1.244 / 1.253~$\mu$m), which is also observed with the red grism. The availability of three potassium doublets  for some objects allowed us to compare the results.

\subsubsection{\label{sc:indice}Spectral indices in the blue- and red-grisms}

\begin{figure}[ht!]
\plottwo{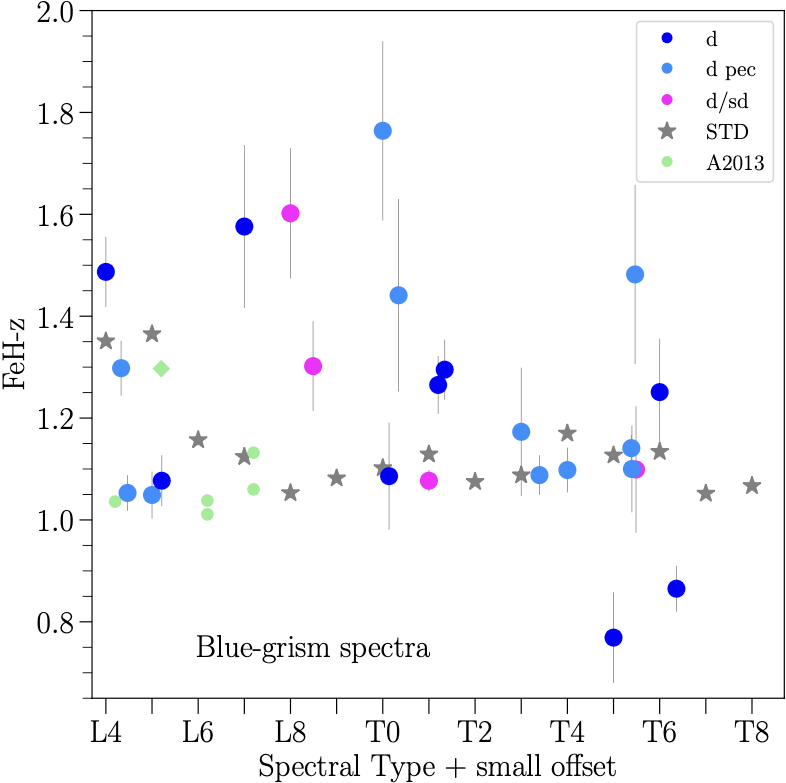}{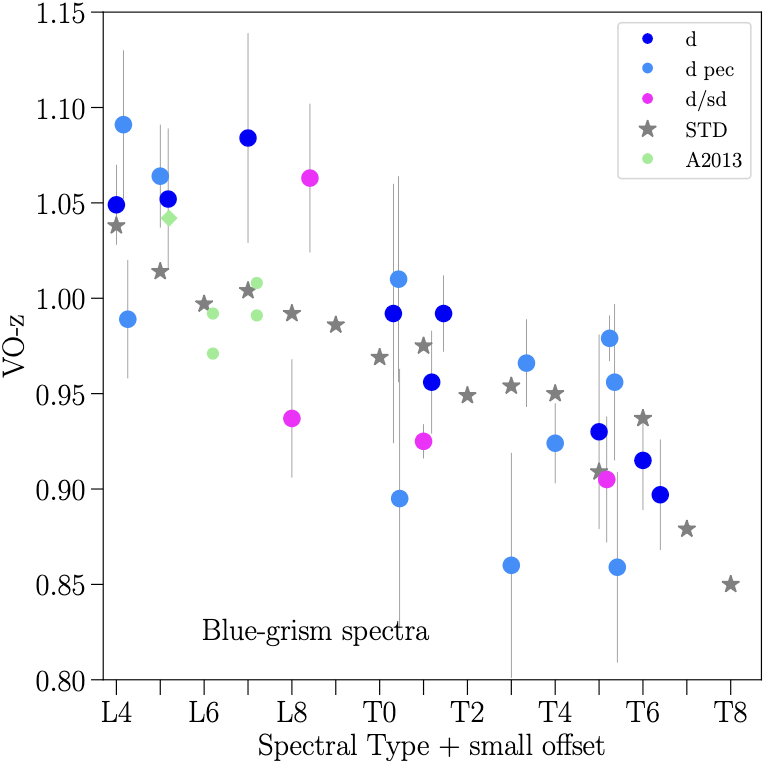}
\caption{FeH-z (left panel) and VO-z (right panel) index measurements for the UCDs in the deep observations.
 The symbols are blue circles for the dwarfs (``d"), light blue circles for the peculiar ones (``d pec"), and magenta circles for the possible mild subdwarfs (``d/sd"). The grey stars the standard (STD) templates used in this paper. 
The green symbols represent the $>10 \sigma $ measurements from \citet{Allers13} as follows: diamonds for field dwarfs,
and circles for those without classification by gravity.}
\label{fig:A13}
\end{figure}

We measured the main spectral indices that fitted the full spectral range: VO-$z$,  FeH-$z$, and FeH-$J$, defined by \citet{Allers13}; CH$_4$-$J$, CH$_4$-$H$, H$_2$O-$J$ and H$_2$O-$H$,   defined by \cite{Burgasser06},
and  NH$_3$-$H$ spectral index from \citet{Delorme08},  redefined by \cite{EGE2021amonia} for \Euclid spectral resolution. We used the SPLAT \citep{SPLAT}, as it has a full database of the indices published in the literature.
Fig.~\ref{fig:A13} shows the results for VO-$z$,  FeH-$z$, and FeH-$J$. We use different colors for the dwarfs, the possible mild subdwarfs (d/sd) and the peculiar ones. The plots include measurements to the standard templates used in this paper (``STD"; see Appendix~\ref{sc:splat}) and the estimations from \citet{Allers13} and \citet{Emily2017} as a reference, 
represented by different symbols according to 
their classification for gravity. There is no classification by gravity for the T dwarfs, since the classification is based on the  work  from \citet{Allers13},
 focused in M and L dwarfs. And there are no  intermediate-gravity dwarfs that passed the  $10 \sigma $ filter in the spectral range of the plot. The dispersion of our measurements for the FeH-$z$ index for the late-L dwarfs is too high to follow the drop seen with the measurements to the standard templates.
In contrast, we can see in the right panel of Fig.~\ref{fig:A13}, that the  measurements for the VO-$z$ index follow the decreasing trend of the standard templates.

\begin{figure}[ht!]
\plottwo{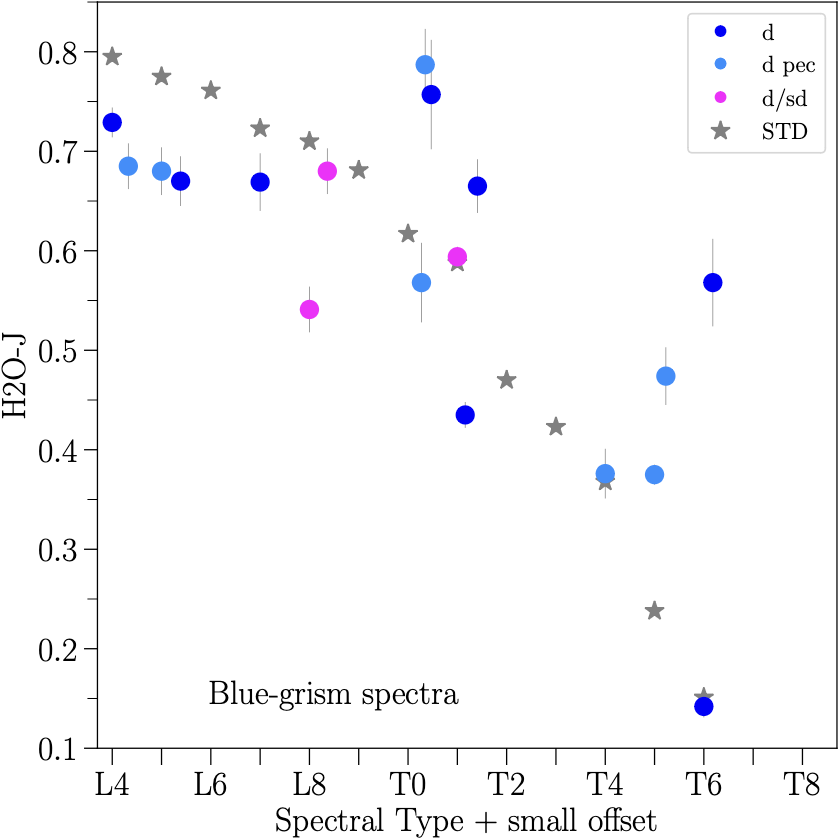}{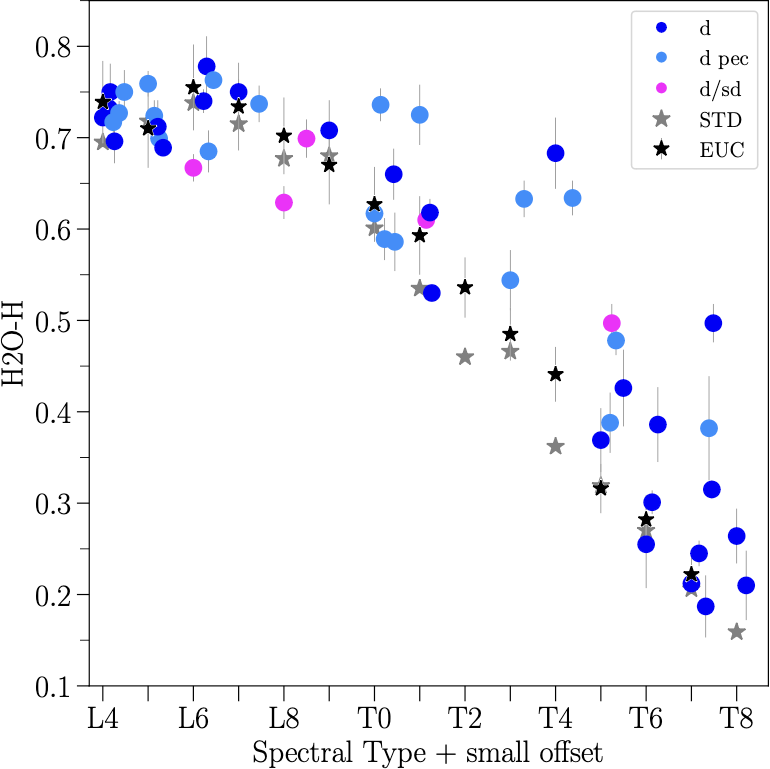}
\plottwo{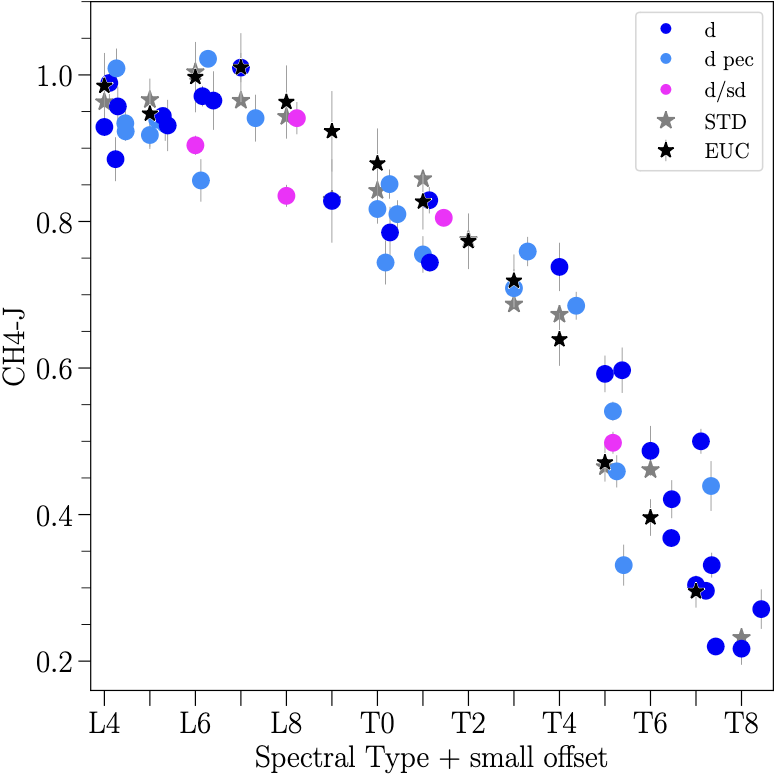}{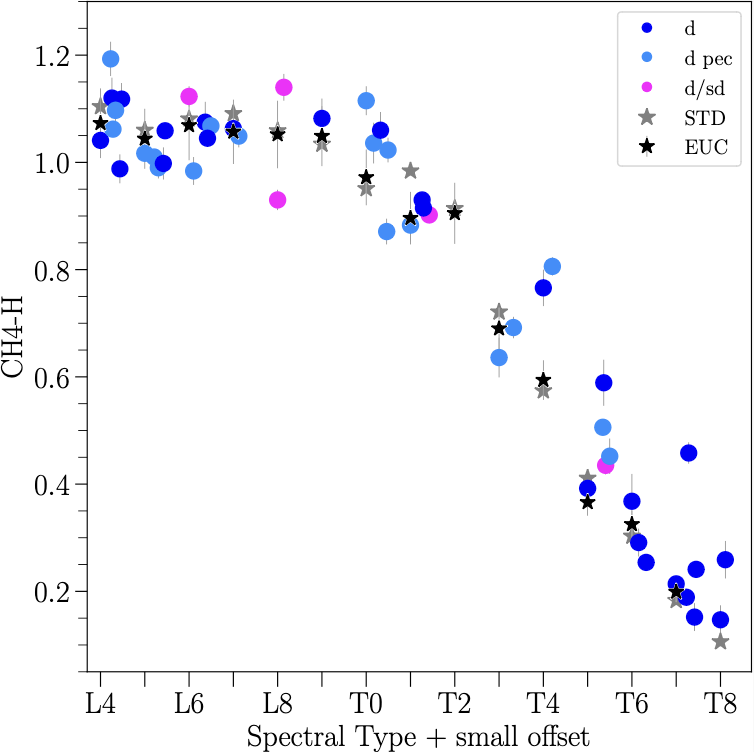}
\caption{H$_2$O (top) and CH$_4$ (bottom) index measurements for the UCDs in the deep observations.
 The symbols are blue circles for the dwarfs (``d"), light blue circles for the peculiar ones (``d pec"), and magenta circles for the possible mild subdwarfs (``d/sd"). The black stars represent the EUC templates and the grey stars the standard (STD) templates used in this paper. H$_2$O-$J$ is out of the wavelength range of the EUC templates.}
\label{fig:B24}
\end{figure}

Fig.~\ref{fig:B24} shows the values of the indices of  CH$_4$-$J$, CH$_4$-$H$, H$_2$O-$J$, and H$_2$O-$H$. The color scheme is as in the previous figure. The plots also include the measurements to the EUC templates  and to the standard templates.  We can see that the  methane absorption strength makes the indices decrease monotonically as a function of the spectral types in the T dwarfs. Such is also the case in the water index for the $H$ band, with higher dispersion on the measurements.
The water index for the $H$ band, measured in the blue-grism spectra, has fewer measurements and much higher dispersion,  owing to the lower quality of the spectra, with respect to the red one. As well, there are no values for the EUC templates for this  index as they are limited to wavelength range of the red grism, yet. Future observations with the blue grism will allow to enlarge the range of these templates. We also notice that the possible mild subdwarfs do not show any systematic difference with respect to the rest of the objects.

\begin{figure}[h!]
\centering
\includegraphics[width=0.45\hsize]{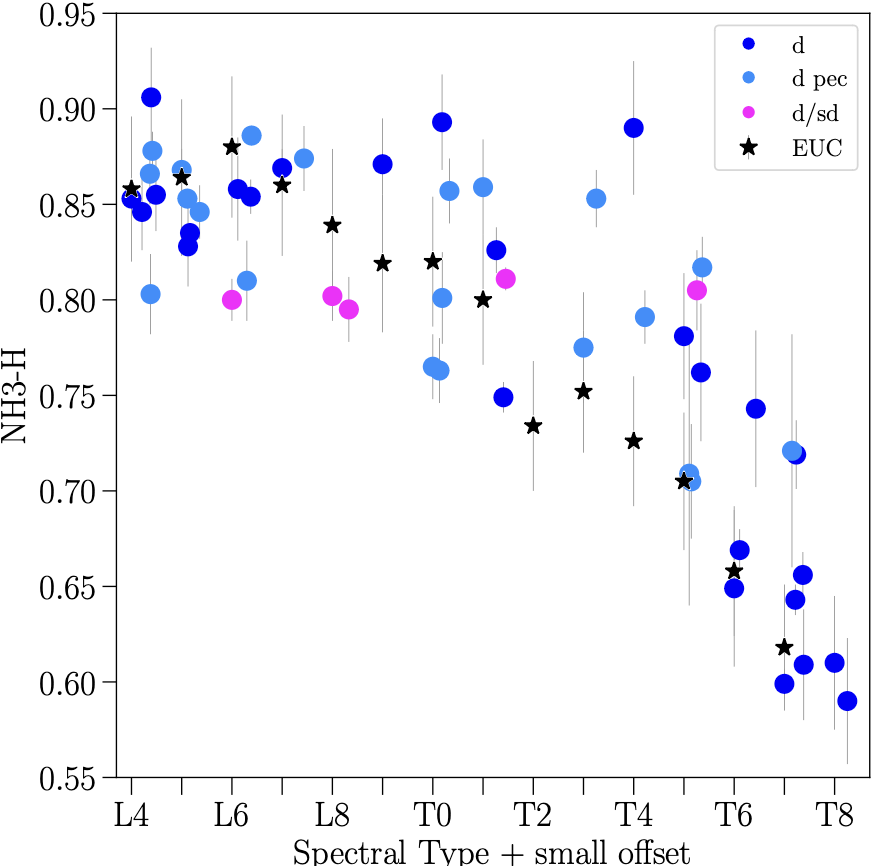}
\caption{ NH3-H index measurements for the UCDs in the deep observations.
 The symbols are blue circles for the dwarfs (``d"), light blue circles for the peculiar ones (``d pec"), and magenta circles for the possible mild subdwarfs (``d/sd"). The black stars represent the EUC templates.}
\label{fig:NH3}
\end{figure}

Fig.~\ref{fig:NH3} shows the NH$_3$-$H$ spectral index by \cite{EGE2021amonia} redefined as an indicator of the absorption of ammonia in late T and Y spectral types in \Euclid spectra. Although this index is intended for the coolest dwarfs, we can see that the values for the EUC templates follow a regular decreasing trend.


\subsubsection{\label{sc:alkali} Alkali absorption lines in the blue- and red-grisms} 

\begin{figure}[ht!]
\plottwo{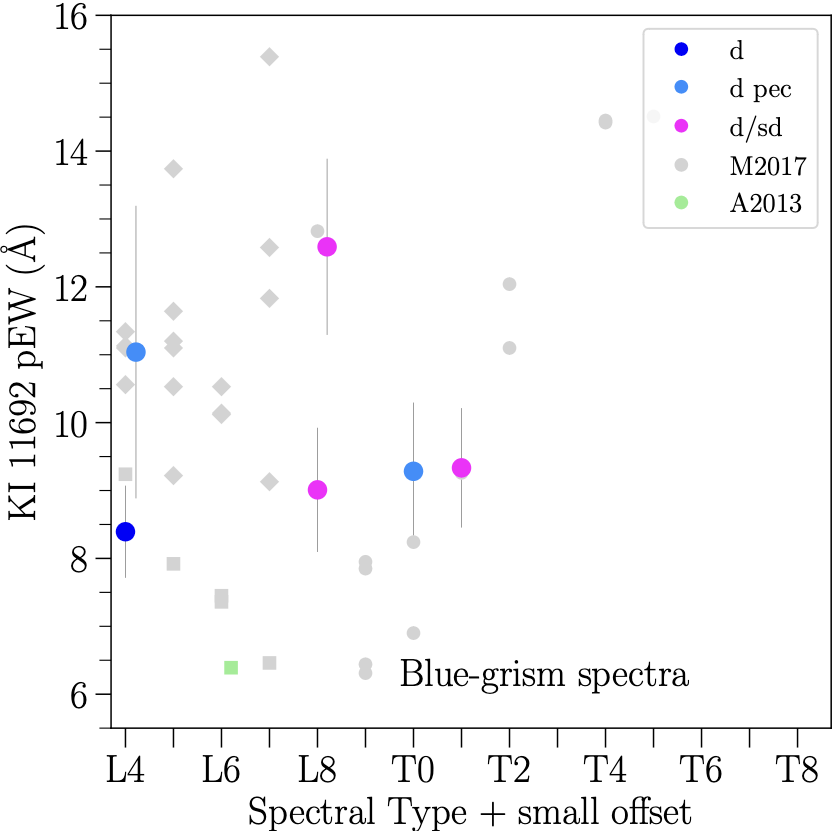}{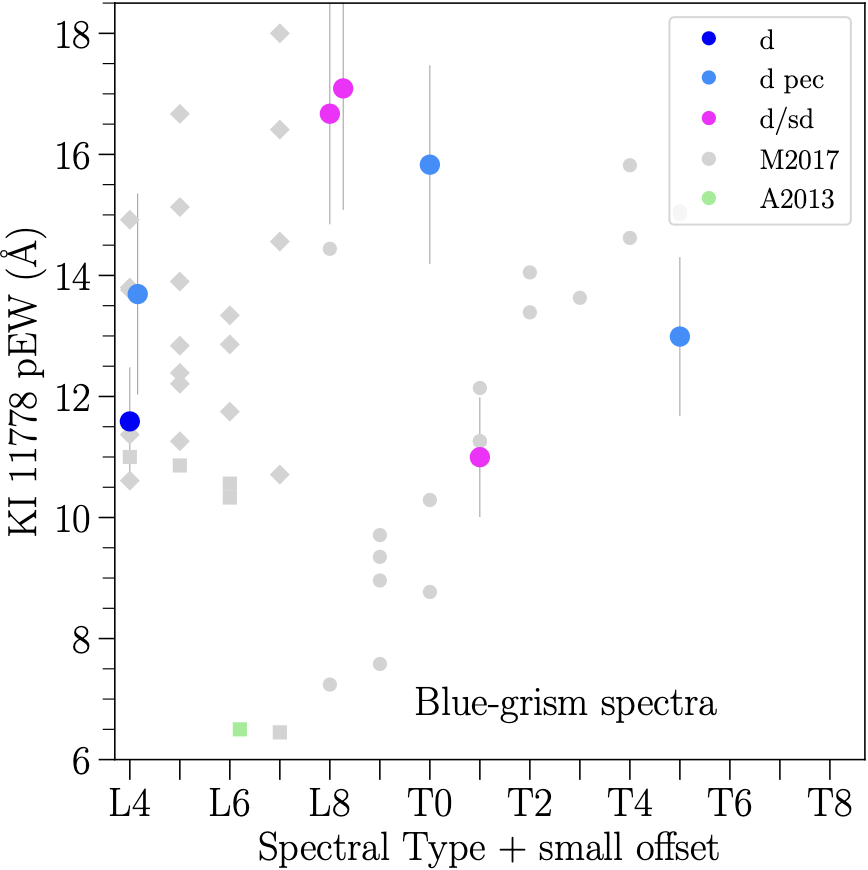}
\plottwo{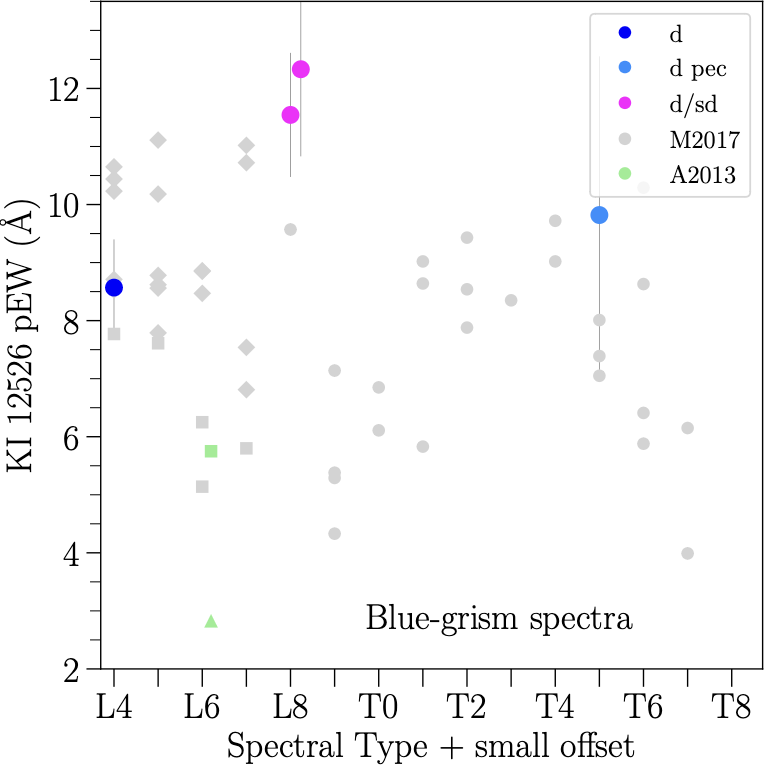}{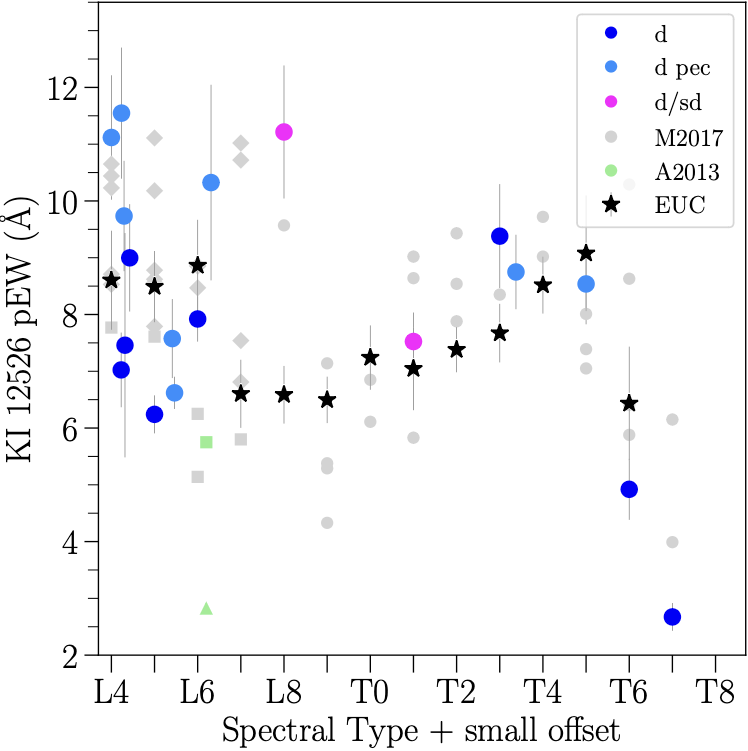}
\caption{Estimations of the K~I pEW for the deep observations of UCDs. The symbols are blue circles for the dwarfs and magenta circles for the possible mild subdwarfs. The top panels are for K~I~1.169~$\mu$m,  K~I~1.178~$\mu$m and the bottom ones for K~I~1.253~$\mu$m measured in the blue and red grisms, respectively. The black stars in the bottom right panel represent the measurements to the EUC templates. The grey symbols represent the $ >10 \sigma $  measurements from \citet{Emily2017} as follows: diamonds for field dwarfs, squares for the intermediate gravity ones, and circles for those non classified by gravity. The light green symbols are  the $ >5\sigma $  measurements from \citet{Allers13} as follows: 
squares for the intermediate gravity, and triangles for the very low ones.}
\label{fig:EW}
\end{figure}

UCDs with both blue- and red-grism spectra, allowed us to measure the gravity sensitive
potassium absorption doublets  K~I~(1.169 / 1.178~$\mu$m) and  K~I~(1.244 /  1.253~$\mu$m). The latter doublet of lines falls in the overlap wavelength region of both grisms, augmenting the possibility of reliable estimations. 
We measured the pseudo equivalent width (pEW) of these lines by fitting a Voigt profile to each of them with respect to the pseudo continuum. This continuum is estimated in a window of 20~nm out of the line central wavelength.  We do not measure the potassium line at 1.244~$\mu$m, as it is blended with FeH (1.239~$\mu$m).
Fig.~\ref{fig:EW} shows the results for the four measurements: K~I~1.169~$\mu$m,  K~I~1.178~$\mu$m, and  K~I~1.253~$\mu$m in the blue-grism spectra, plus K~I~1.253~$\mu$m in the red one. As the blue-grism spectra is not available for all the objects, or it has a poor quality, the number of measurements in the first three plots is reduced, in comparison to those in the red grism. 
The plots include the reported estimations from the works of \citet{Emily2017} and  \citet{Allers13}. There are no  field dwarfs from the latter work that passed the  $5 \sigma $ filter in the spectral range of the plot.
 A group of mid-L dwarfs have slightly low pEWs for K~I~1.253~$\mu$m: E11146411, E12348470, E12369825, E22877624, and E60042734. We inspected  their spectra and concluded that they are most probably high gravity dwarfs.
The L4 dwarf, E26553129 also has also low pEWs for K~I~1.169~$\mu$m and  K~I~1.178~$\mu$m, lines. In contrast,  E26931238, a possible d/sdL8 has high values in all the plots with the exception of K~I~1.169~$\mu$m. They are discussed in Sect.~\ref{sc:notes}.

Na~I absorption doublet (1.138 / 1.141~$\mu$m) is covered by the wavelength range of the blue-grism spectra.
The two lines cannot be resolved with NISP spectral resolution. The absorption strength of this line is strong in field objects with spectral types from late-M to early-L dwarfs, as it. was seen in figures \ref{fig:seqblueML} and \ref{fig:seqblueT}. The detection for late-L  dwarfs becomes difficult as it  broadens and
 the strength of the methane molecular  band  increases. For the  T dwarfs, the sodium doublet is even harder to be detected.

\subsection{\label{sc:analws}Analysis in the Wide Survey}

As described in Sect.~\ref{sc:data}, \Euclid visits the fields in the Wide Survey only once, and the spectroscopy is done only with the red grism. Despite these differences with respect to what we have shown in the previous sections, the large area covered by this survey  enlarges by 30 times the area of the deep fields, studied in the previous sections. The number of UCDs found by the spectral index search increased accordingly (see Sect.~\ref{sc:search}). Moreover, the very large area covered by this data allowed to find UCDs from different environments.


\subsubsection{\label{sc:indicews}Spectral indices with the red-grism spectra} 

\begin{figure}[ht!]

\plottwo{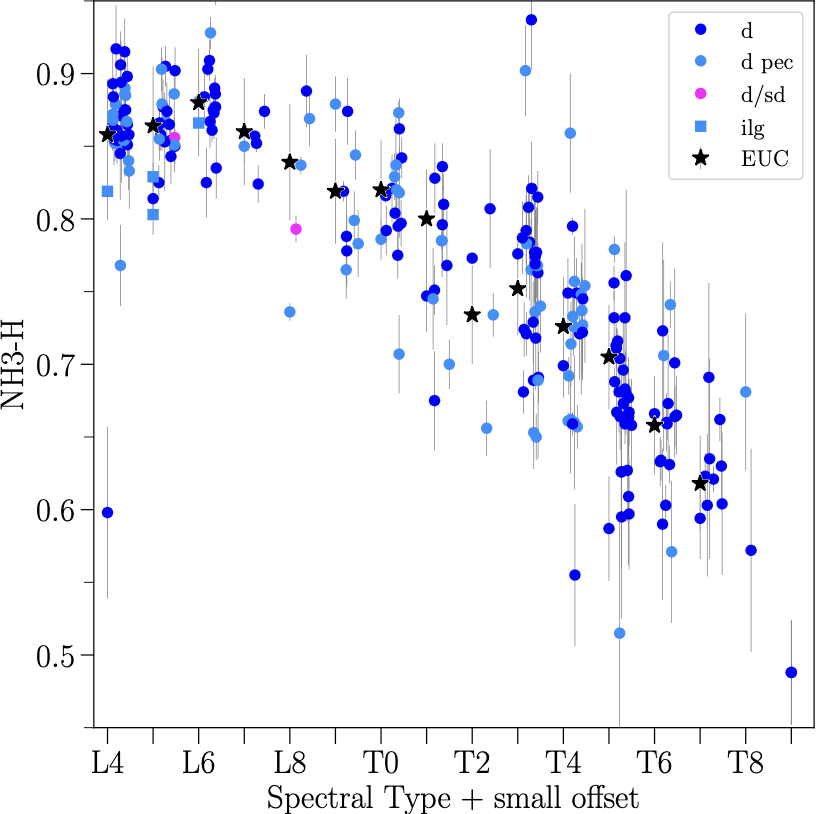}{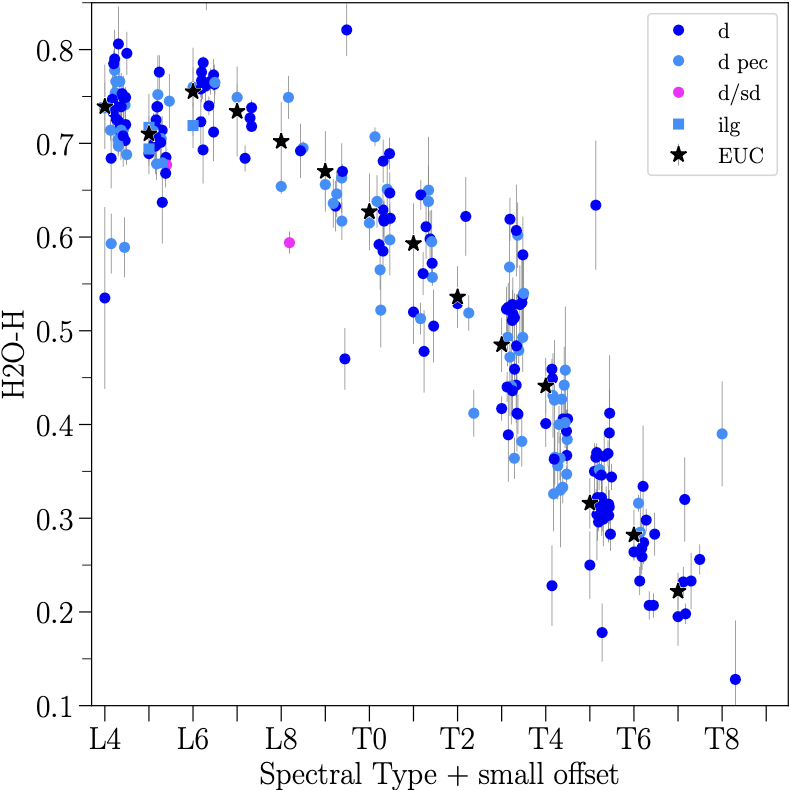}
\plottwo{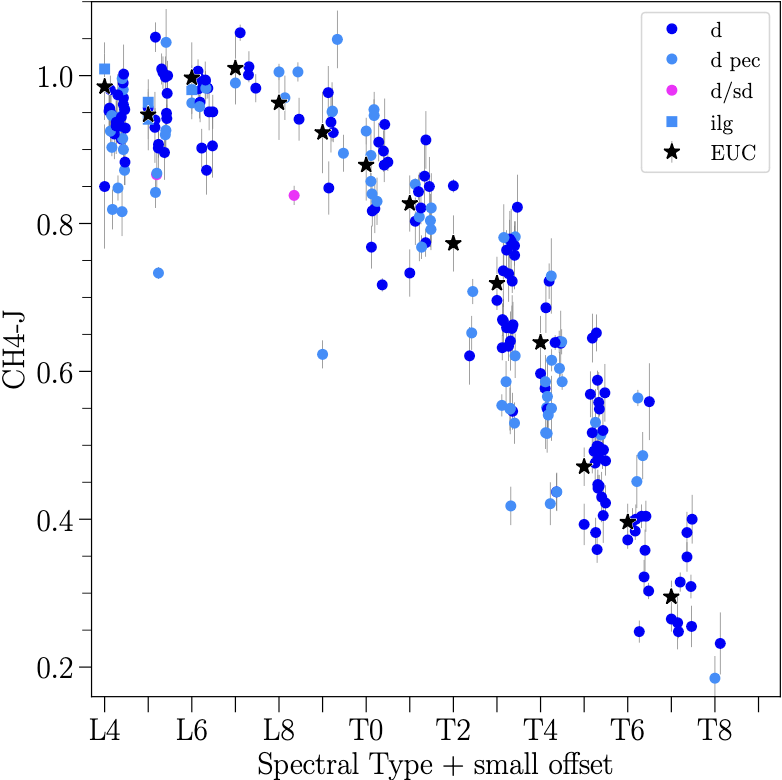}{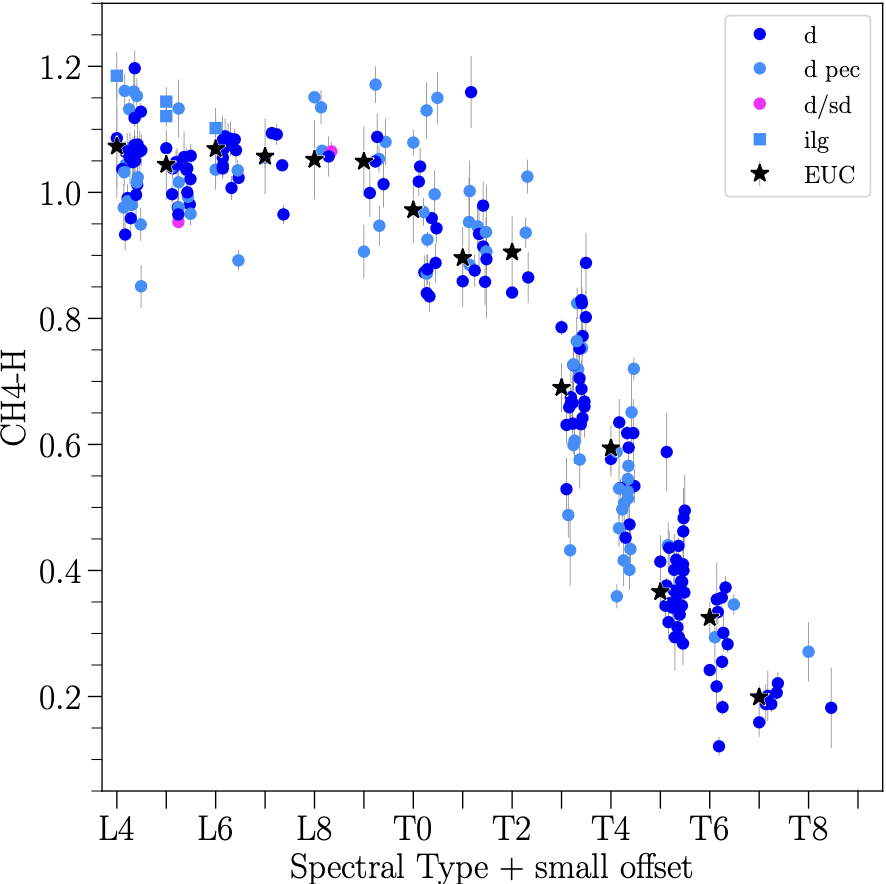}
\caption{NH$_3$-$H$ (top left), H$_2$O-H (top right),  and CH$_4$ (bottom)  index measurements for the UCDs in the Wide Survey.  
The symbols are as in previous plots: blue circles for the dwarfs (``d"), light blue circles for the peculiar ones (``d pec"), and magenta circles for the possible mild subdwarfs (``d/sd"). The black stars represent the EUC templates.}
\label{fig:B24ws}
\end{figure}

Owing to having only the red grism available, we measured only three of the spectral indices defined by \cite{Burgasser06}: CH$_4$-$J$, CH$_4$-$H$, and H$_2$O-$H$. Plus the ammonia spectral index  \citep[redefined by][]{EGE2021amonia}.  Fig.~\ref{fig:B24ws} shows the $10 \sigma$ or higher results, where the color scheme is as in the previous figures. The plots also include the measurements to the EUC templates.  
In the water and methane indices, the measurements for all the dwarfs follow the decreasing trend as a function of the spectral type, in agreement with the EUC templates. As the one-visit spectra are noisier the dispersion in the measurements  is higher than that seen for the deep observations. This is also the case of the NH$_3$-$H$ spectral index. Despite the dispersion and the larger error bars, the correlation to the spectral type in the T dwarfs is similar to what we saw in Sect.~\ref{sc:indice}.

\subsubsection{\label{sc:agews}K{\small{I}}~absorption doublet in the red grism}

\begin{figure}[ht!]
\centering
\includegraphics[width=0.55\hsize]{{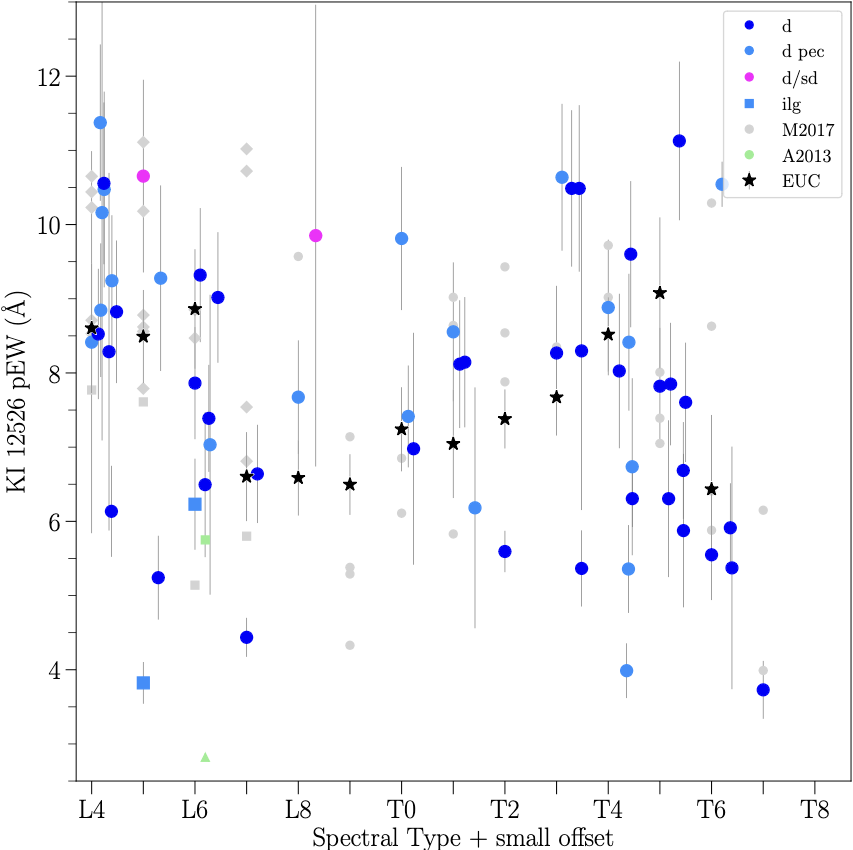}}
\caption{Estimations of the pEW for K~I~1.253~$\mu$m for the UCDs in the Wide Survey.
 The symbols are: blue circles for the dwarfs (``d"), light blue circles for the peculiar ones (``d pec"), and magenta circles for the possible mild subdwarfs (``d/sd"). The light-blue triangles represent the possible low gravity (``vlg") dwarfs. The black stars represent the measurements to the EUC templates. The grey symbols represent the $ >10 \sigma $  measurements from \citet{Emily2017} as follows: diamonds for field dwarfs, squares for the intermediate gravity ones, and circles for those non classified by gravity. The light green symbols are  the $ >5\sigma $  measurements from \citet{Allers13} as follows: diamonds for field dwarfs, squares for the intermediate gravity, and triangles for the very low ones. }
\label{fig:EWws}
\end{figure}

We measured the red line of the K~I~absorption doublet at  1.244 /  1.253~$\mu$m using the technique described in Sect~\ref{sc:alkali} for the best spectra of UCDs in the Wide Survey. Fig.~\ref{fig:EWws} shows the results with the EUC templates as a reference, and with the color scheme as in the previous figures. As these spectra from the Wide Survey are the result from only one visit, many of them are affected by artifacts, producing much larger error bars than the measurements from the deep observations (Fig.~\ref{fig:EW}). 
 We marked with light-blue-square symbol two objects the presented a red   \JE -  \HE color: the L5, E74752024 and the L6, E35269945.  
 Despite the dispersion, the pEW of the L5 indicates possible intermediate gravity. 
 These results  are discussed in Sect.~\ref{sc:young}.

\section{\label{sc:notes}Notes on some peculiar or very cool dwarfs  }

This section includes the analysis of some UCDs, including the parameters available from \Euclid data.

We are submitting observing proposals for higher resolution NIR and optical spectra follow-up for the more interesting  objects found from all these analyses.

\subsection{\label{sc:ys}The Y dwarfs}

There are two known Y dwarfs in the Wide Survey:

\begin{itemize}

   \item The Y0: dwarf WISEA J030237.52-581740.3 - This object was discovered by  \cite{Tinney2018} using WISE data. Table~\ref{tab:J0302} lists the main parameters.
The proper motion estimated by WISE is  $52.0\pm 3.5, -70.8\pm3.5 \times 10 ^{-3}$ arcsecond.   \cite{Kirkpatrick2021} estimates   a parallax of $59.9\pm 3.3  \times 10 ^{-3}$ arcsecond.
 Fig.~\ref{fig:J0302} shows the \Euclid images and spectra. We can see that it is well detected in the NIR,  at the limit of detection in the VIS image, in consequence, the \IE magnitude should be taken as an upper limit.  Despite the very low SNR of the spectra and the presence of less-reliable values (i.e., magenta lines), the zero flux in the water absorption bands makes the $J$- and $H$-band peaks to be detected. Methane in the $H$ band is  present, although not as strong as in the Y0 dwarf plotted as a reference.
      Fig.~\ref{fig:colorys} shows the  measured  \JE -  \HE  and \YE -  \HE  colors in comparison to the simulated ones from \cite{Leggett2025} for other Y dwarfs.  That work does not simulate the \IE band.  We can see in the figure  that there is a good agreement, and that the measured  \YE magnitude for this dwarf is below $10\sigma$.

   \begin{table}[h!]
   \caption{The known Y dwarfs observed in DR1}
   \centering
\begin{tabular}{lrlrrrrrrrrrrrlrrr}
\hline\hline
    WISE name & DR1 object\_id & SpT & \IE &  \YE &  \JE &  \HE &  W1 &  W2 &  \\
   &  &  & (AB) &  (AB) &  (AB) &  (AB) &   (Vega) &  (Vega)   \\\hline
  J035000.32-565830.2 & -574999749569772762 & Y1 & 25.2$\pm$0.3 & 23.2$\pm$0.1 & 23.2$\pm$0.1 & 23.4$\pm$0.1 & 18.9$\pm$0.2 & 15.9  & \\
  J030237.56-581740.2 & -456564167582948574 & Y0: & 26.9$\pm$0.6 & 23.1$\pm$0.1 & 22.8$\pm$0.1 & 23.4$\pm$0.1 & 18.5$\pm$0.1 & 14.7& \\
\hline\end{tabular}
\tablecomments{W2 lacks of error measurements.}
\label{tab:J0302}
 \end{table}

\begin{figure}[ht!]
\plottwo{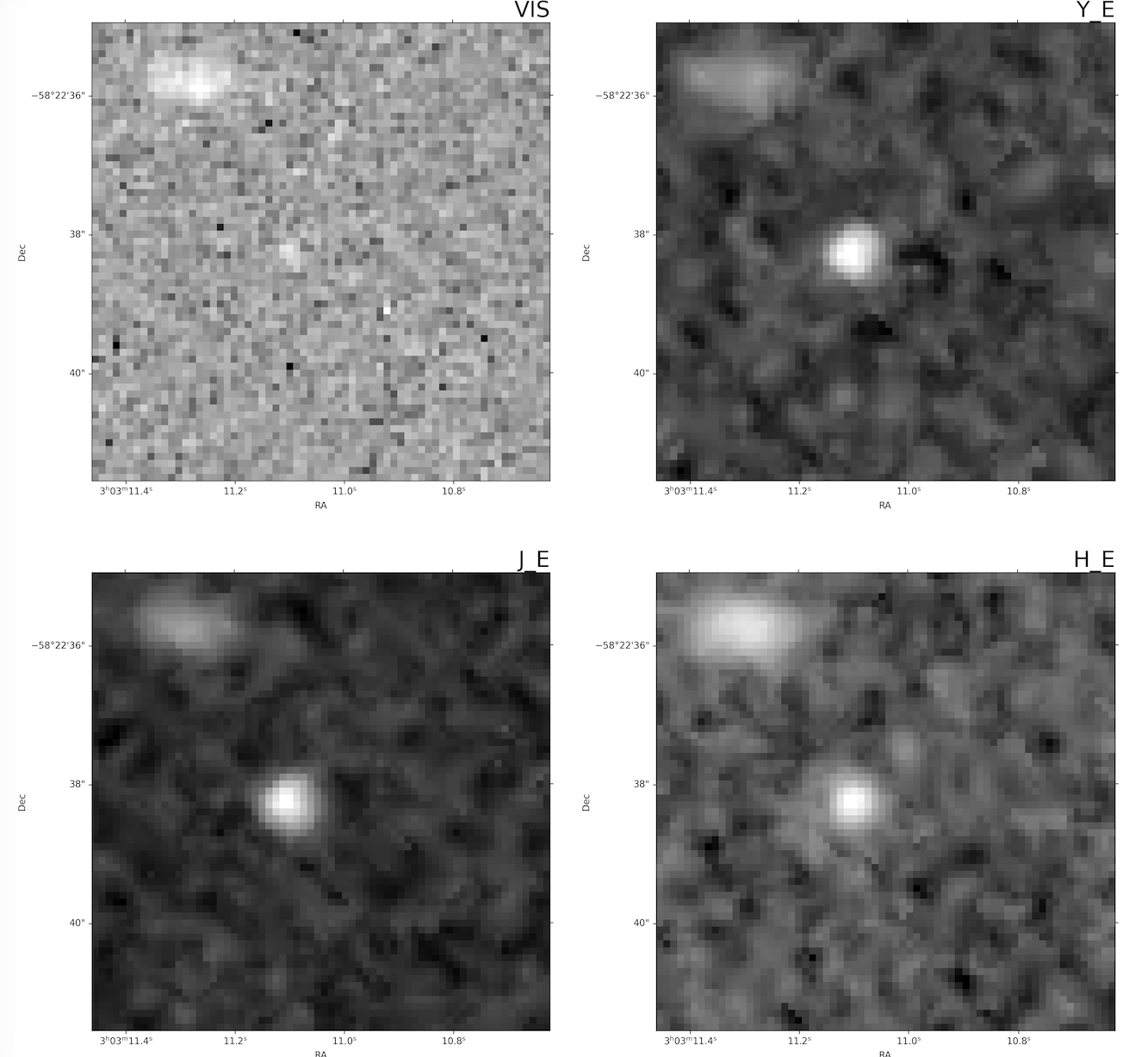}{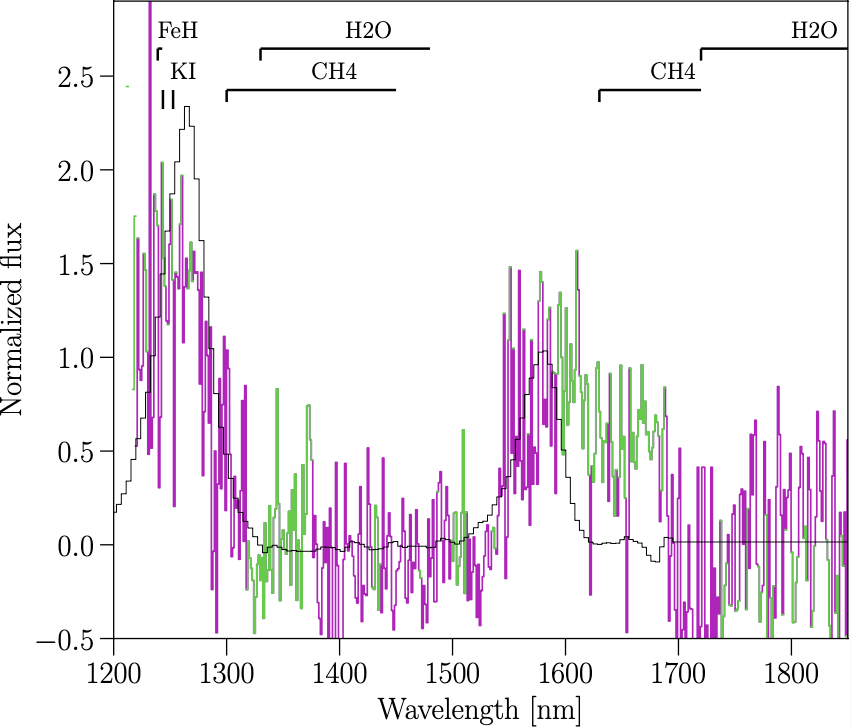}
\caption{\Euclid four-band images and spectra of the Y0: dwarf WISEA J030237.52-581740.3. The green lines represent the spectra with \texttt{NDITH}~$>$~2 and  the  magenta lines  indicate the values with  \texttt{QUALITY} $<$~0.7. The black line shows the spectra of WISE1738 observed by \cite{Schneider15} with HST/WFC3. }
\label{fig:J0302}
\end{figure}

   \item The Y1 dwarf WISE J035000.32-565830.2 - This object was discovered by  \cite{Kirkpatrick2012}  using WISE data. 
   Table~\ref{tab:J0302} lists the main parameters.
   The proper motion estimated by WISE is  $-201.7\pm 1, -575.4\pm1.1 \times 10 ^{-3}$ arcsecond. \cite{Kirkpatrick2021} estimates a parallax of $176.4\pm 2.3 \times 10 ^{-3}$ arcsecond.
   Fig.~\ref{fig:J0350} shows the spectra form \Euclid and HST/WFC3 \citep{Schneider15} of this dwarf. 
      We can see that it is well detected in the NIR,  barely detected in the VIS image. Therefore, as in the previous case, the \IE magnitude should be taken as an upper limit. The SNR of the spectra is low, at the detection limit. It shows, however a clear $J$ band peak and deep absorption in the redder water and methane absorption bands.
      Its  \JE -  \HE  and \YE -  \HE  colors are also in agreement to the simulated ones from \cite{Leggett2025} for other Y dwarfs (see Fig.~\ref{fig:colorys}). 

\begin{figure}[ht!]
\plottwo{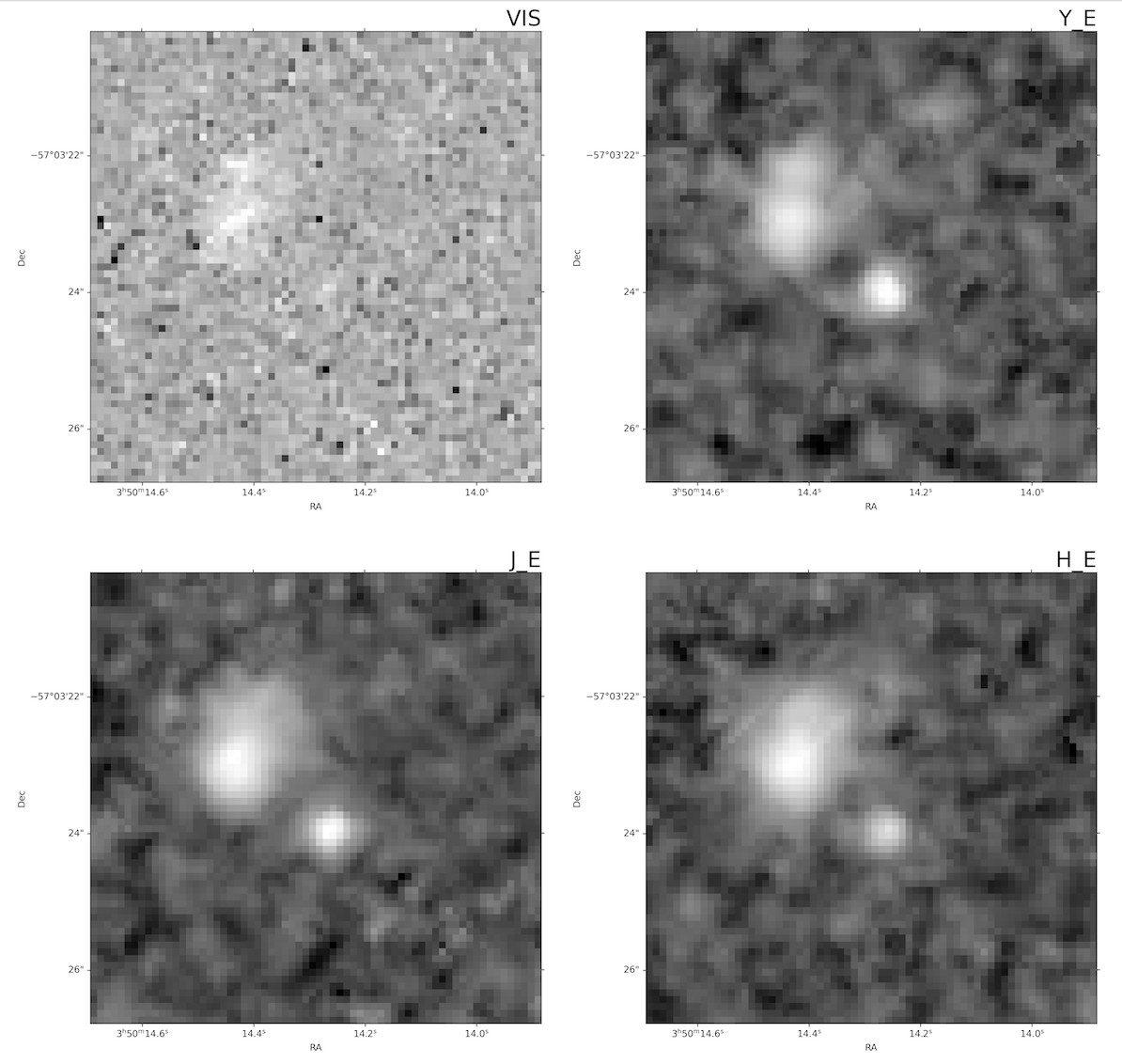}{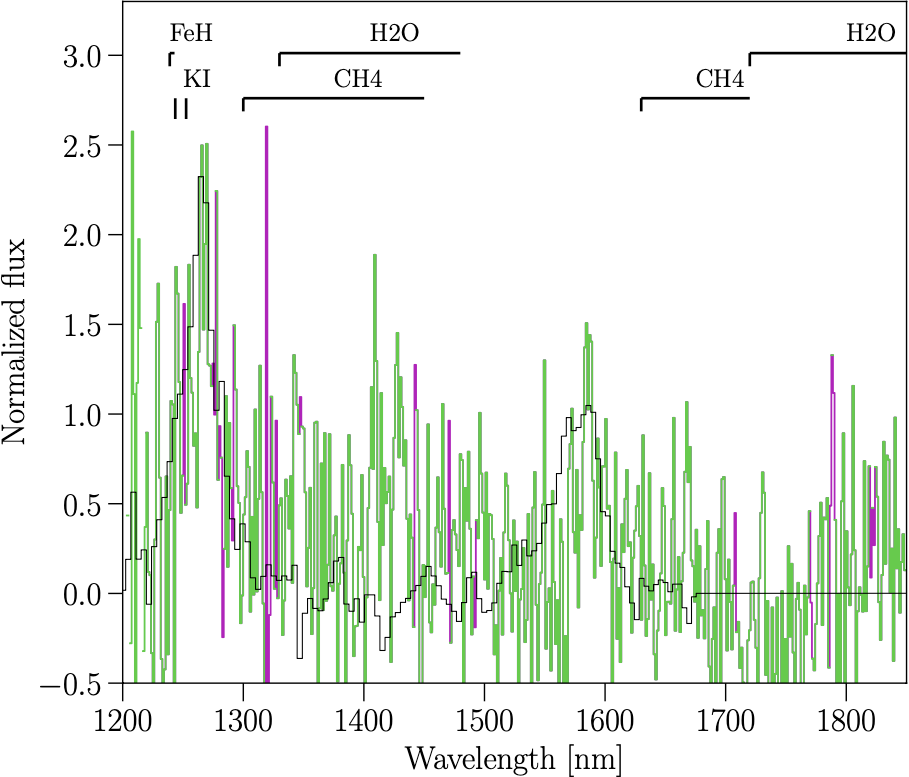}
\caption{\Euclid four-band images and spectra of the Y1 dwarf WISE J035000.32-565830.2. The green lines represent the spectra with \texttt{NDITH}~$>$~2 and  the  magenta lines  indicate the values with  \texttt{QUALITY} $<$~0.7. The black line shows the spectra of the same object observed by \cite{Schneider15} with HST/WFC3.}
\label{fig:J0350}
\end{figure}

\begin{figure}[ht!]
\plottwo{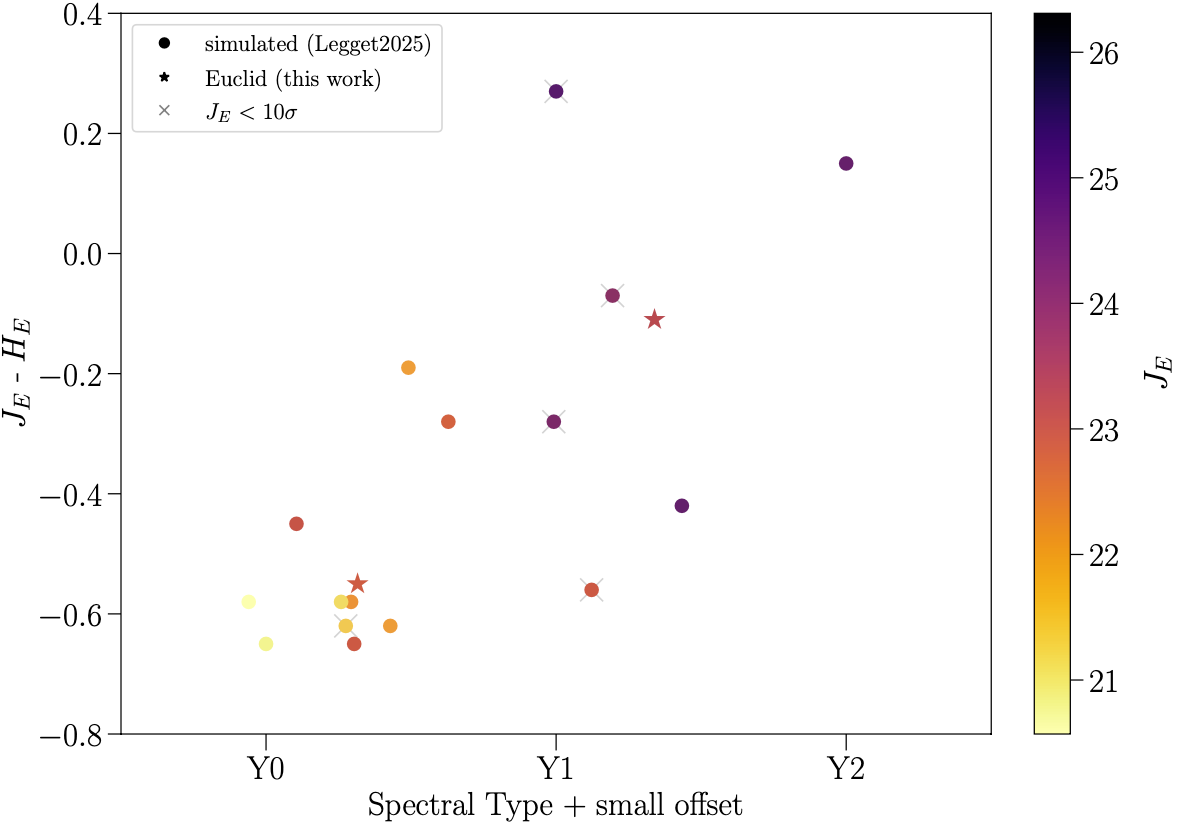}{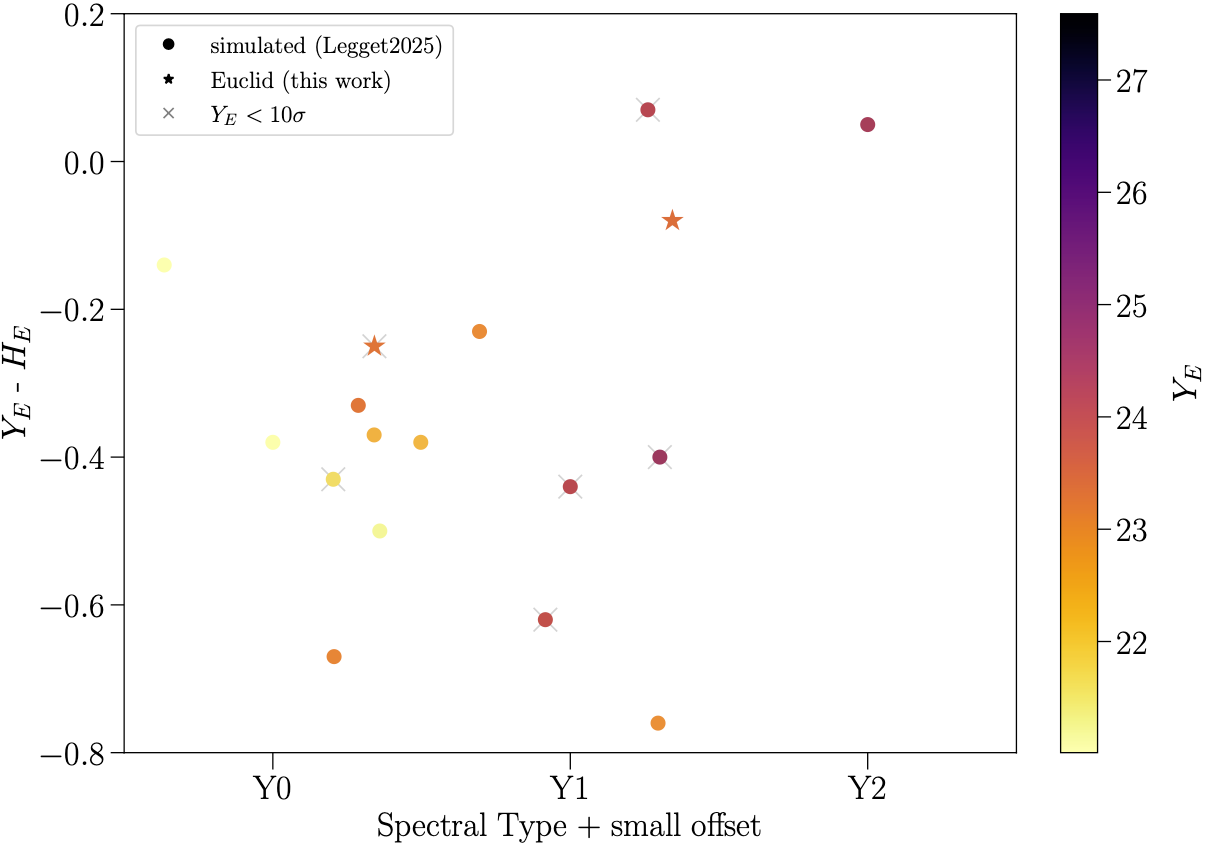}
\caption{  \JE -  \HE (left) and \YE -  \HE (right) color vs. spectral type for the Y dwarfs.
 The stars represent the real value of the two objects observed by \Euclid (this work):  WISE J03500 (Y1) and WISEA J03023 (Y0:). The circles represent the simulated colors from \citep{Leggett2025}.  The symbol color follows the measured or simulated magnitudes, respectively. A grey cross in the symbol, indicates that the SNR in the magnitude is lower than 10. }
\label{fig:colorys}
\end{figure}

\end{itemize}


\subsection{\label{sc:dsd}The possible subdwarfs }

\begin{figure}[ht!]
\plottwo{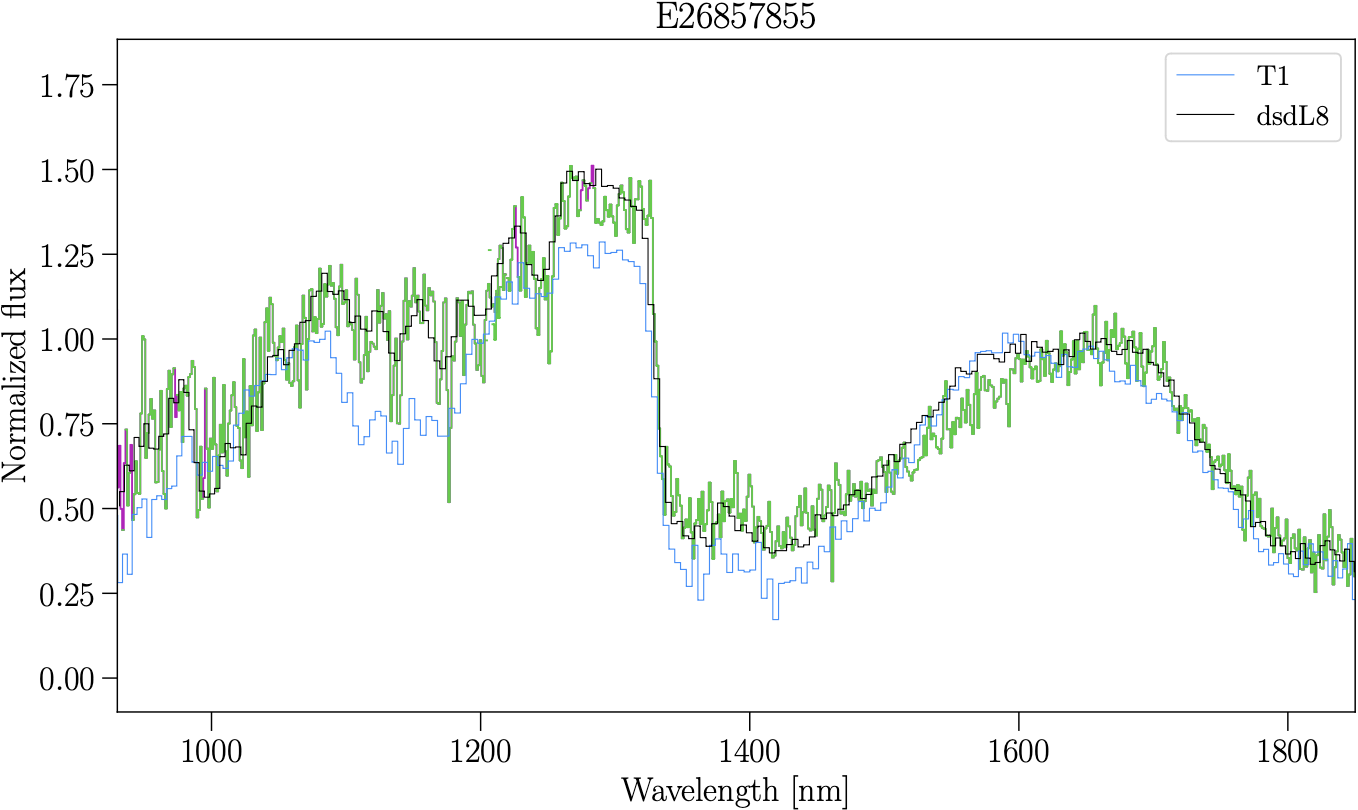}{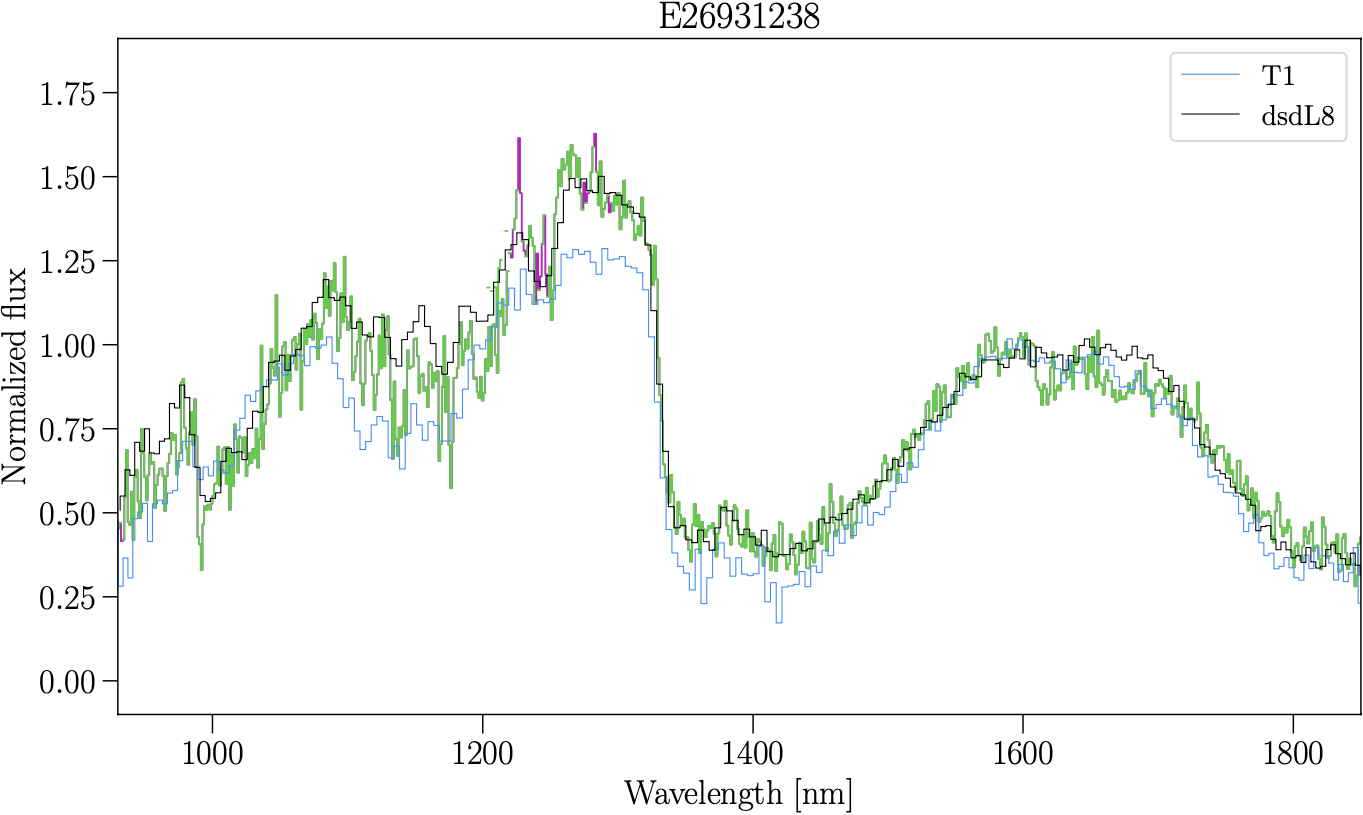}
\plottwo{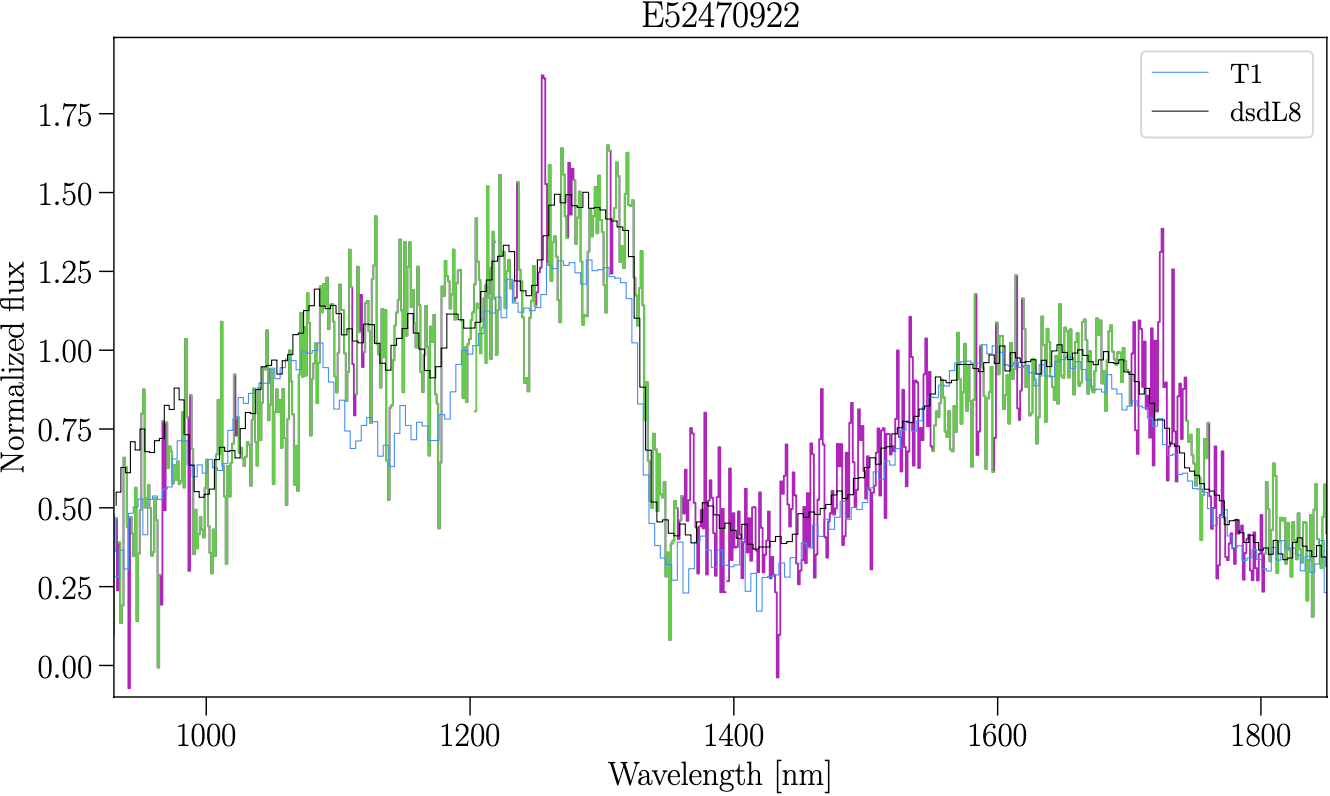}{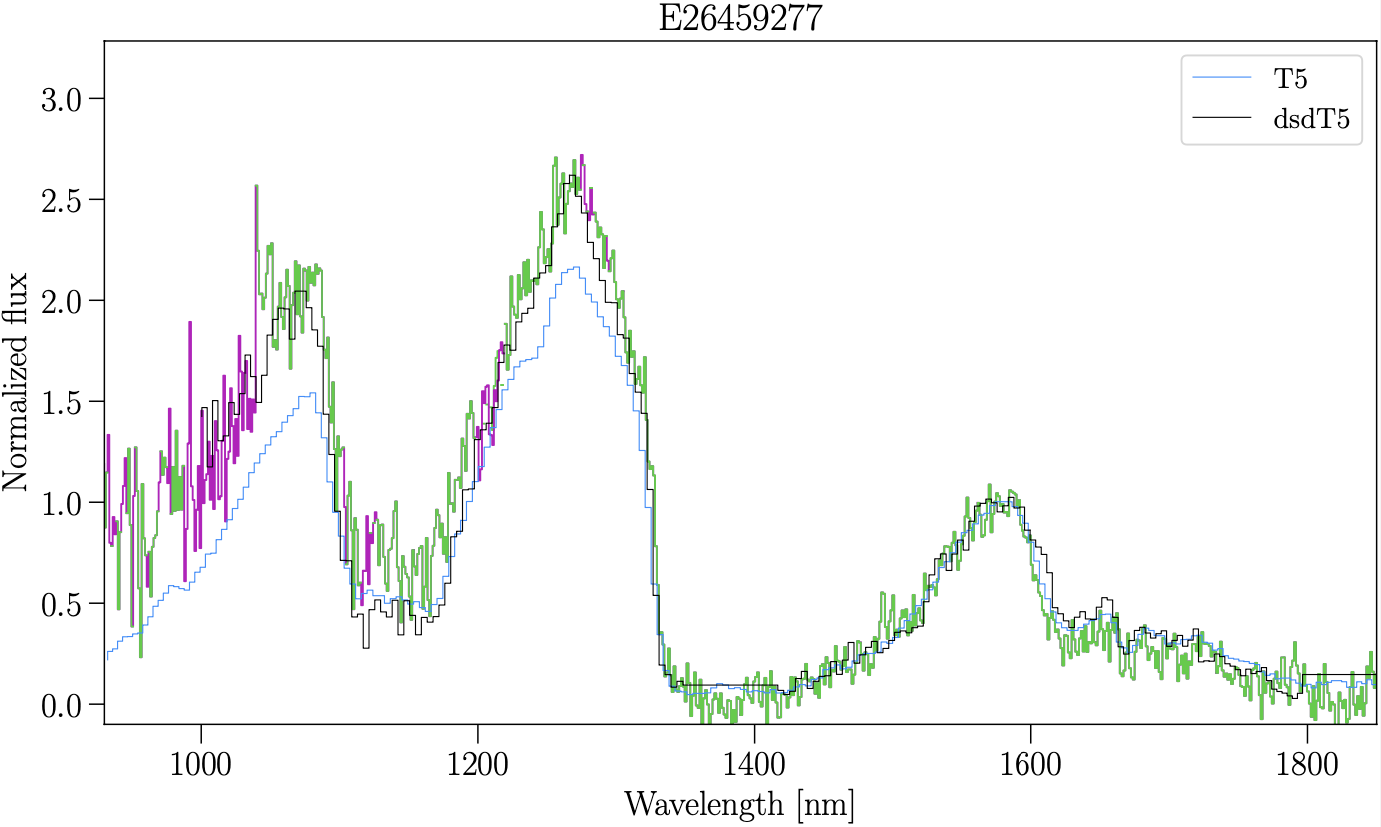}
\caption{Spectra of four possible subdwarfs listed in Table~\ref{tab:sub}. The green lines represent the spectra with \texttt{NDITH}~$>$~2 and  the  magenta lines  indicate the values with  \texttt{QUALITY} $<$~0.7. 
 The blue and black lines represent the dwarf- and subdwarf- standard templates, respectively.  }
\label{fig:dsd}
\end{figure}

\begin{figure}[ht!]
\plottwo{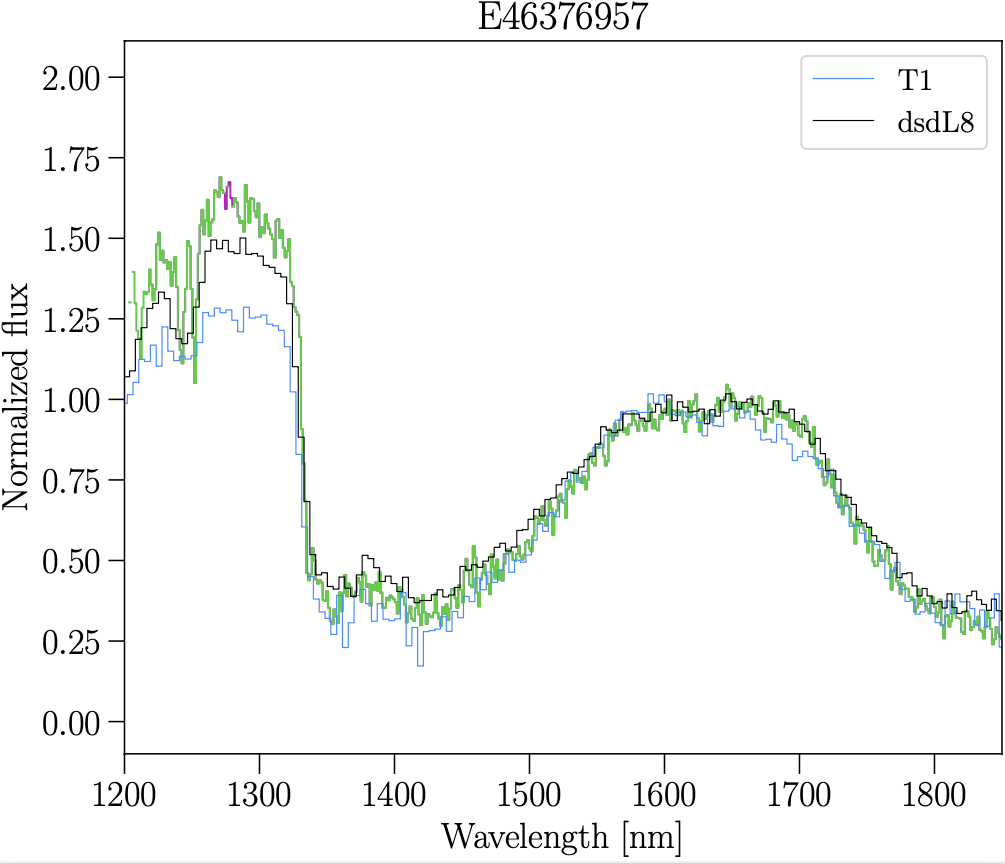}{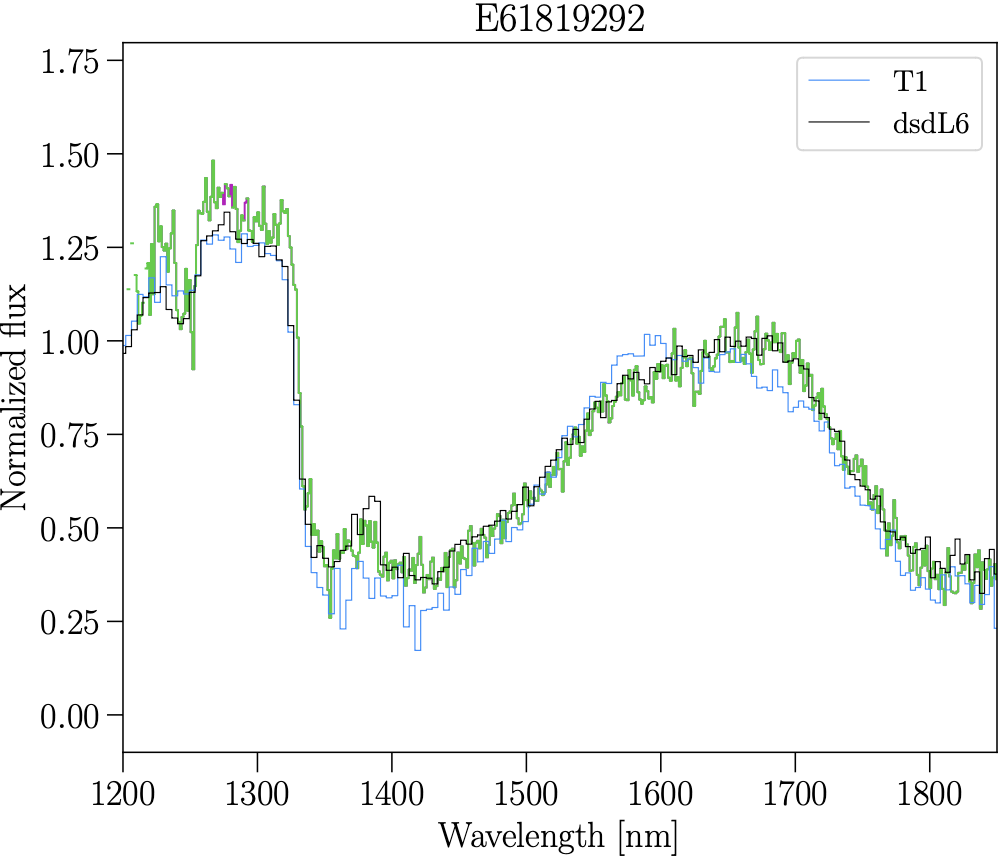}
\caption{ Spectra of the possible subdwarfs  E46376957 (left) and E61819292 (right). Only the red-grism spectra is available. The color scheme for the spectra is as in previous figures.  }
\label{fig:sd}
\end{figure}

\begin{figure}[ht!]
\centering
\includegraphics[width=0.65\hsize]{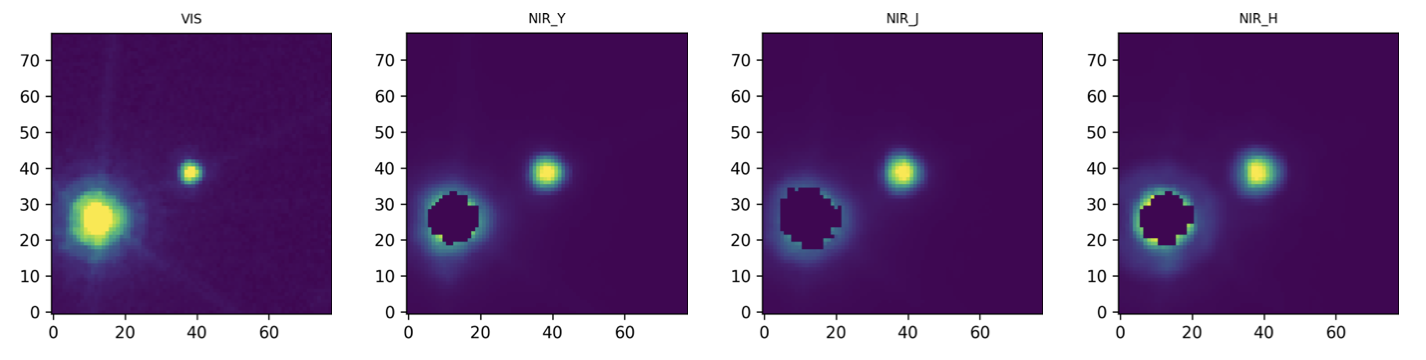}
\plottwo{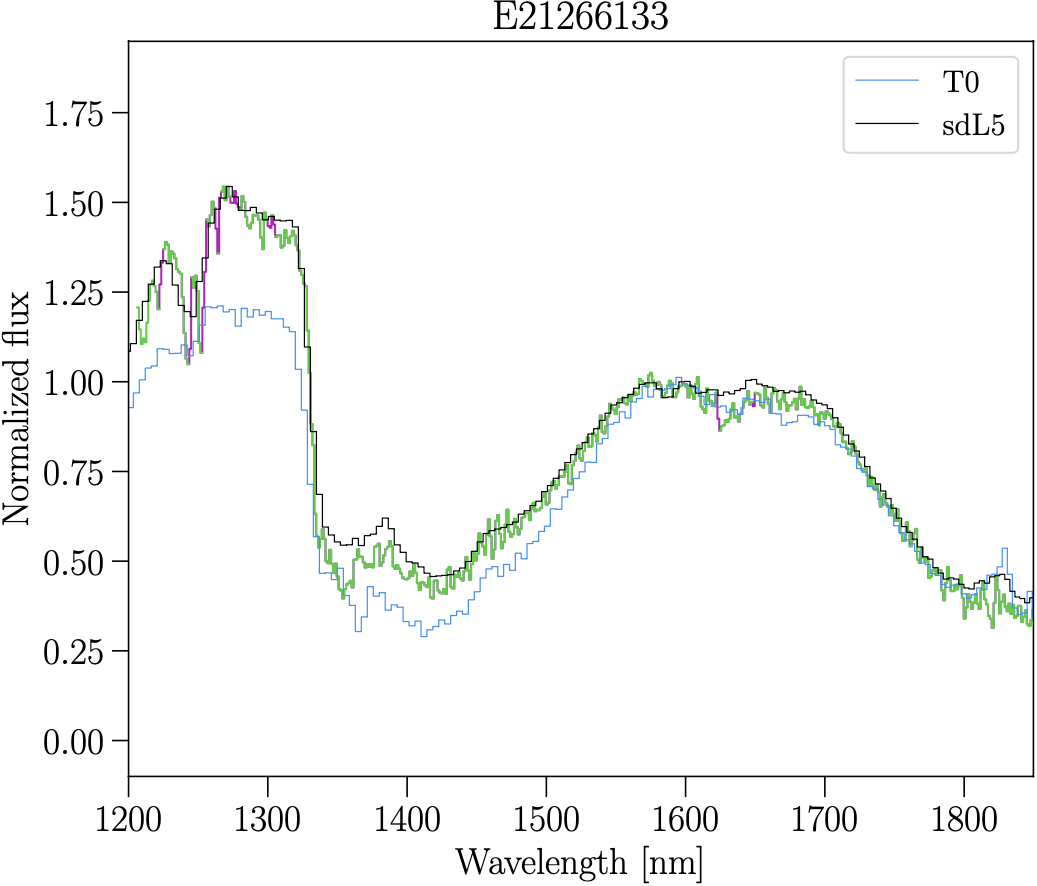}{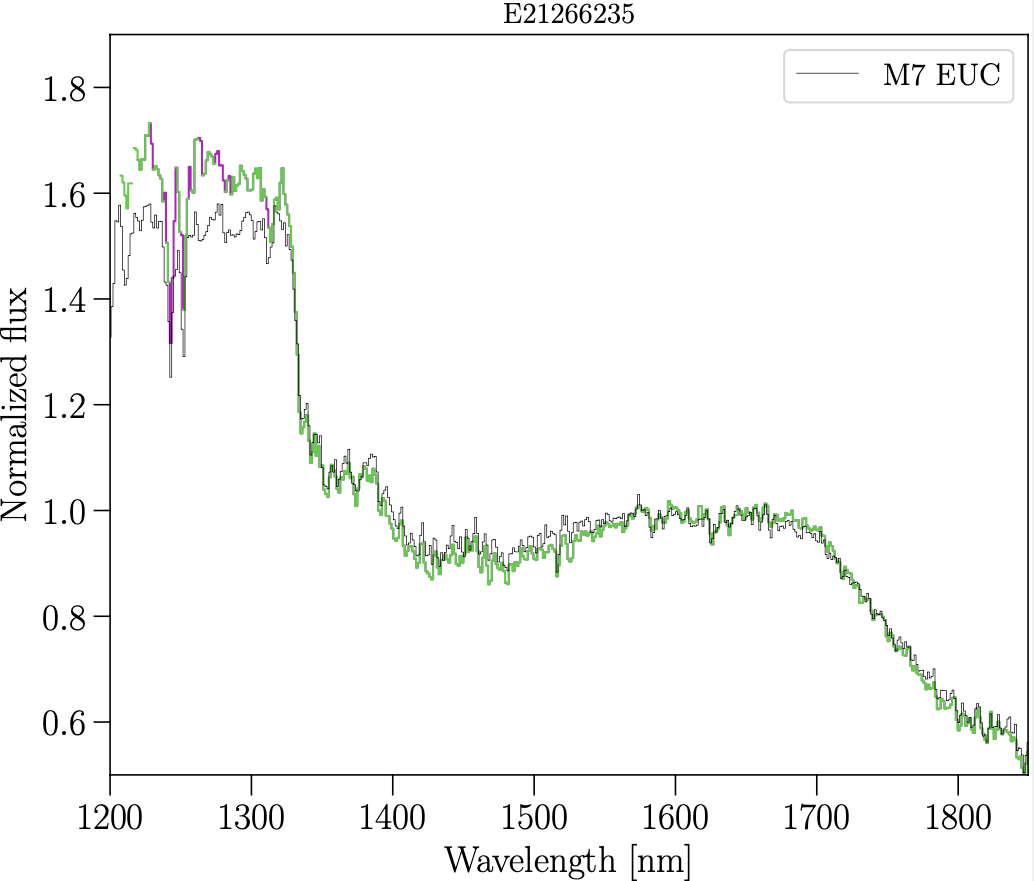}
\caption{ \Euclid four-band images and spectra of the possible subdwarf E21266133. The bright star saturates the NIR images. 
The color scheme for the spectra is as in previous figures. The  black line represents the standard and EUC templates for the left and right panels, respectively.  }
\label{fig:E21266133}
\end{figure}

In Sect.~\ref{sc:spt} we classified some objects as possible subdwarfs, following the best fit to the  subdwarf templates (i.e., a difference  in the value of $\chi ^2$ is larger than 2), as can be seen in  Table~\ref{tab:sub}. We also verified their deviating value of  \JE -  \HE color   from the rest of dwarfs. Figures \ref{fig:dsd},   \ref{fig:sd},   and  \ref{fig:E21266133}   show  the spectra of the objects classified as possible subdwarfs. Two standard templates are plotted in order to show the differences between the best fit to a normal standard and a subdwarf. The table includes the estimated spectro-photometric distances (based on a normal standard of the same spectral subtype; see Sect.~\ref{sc:dist}). And the estimated the metallicities, which are estimated by fitting the new ATMO model grid~\citep{ATMO}, which covers [M/H] from -1.5 to +0.5 dex in steps of 0.5 dex. The assigned  metallicity corresponds to the best fit to such grid, and the uncertainty is half grid spacing or 0.25 dex. As some objects have blue-grism spectrum available, the estimation  includes all the possible range for each object. The full method is discussed in detail in  Sedighi et al. (in prep), which includes all the objects in our sample from the deep observations.  In order to confirm these candidates, we are conducting  follow-up observations including spectra in the visible.

\begin{table}[h!]
\caption{Main parameters of the possible subdwarfs}
\centering
\begin{tabular}{llrrrrrrrrrr}
\hline\hline
  Alias & Designation & SpT  & $\chi ^2$  & Alt. SpT & Alt.$\chi ^2$ & \JE - \HE & SP Dist. & errDist. &Teff & [M/H]  \\
   & \texttt{Euclid DR1 +} &  &  &  &   & (AB)  & (pc) & (pc) & (K) &  (dex) \\
   \hline
E46376957 & J030530.469-400055.39 & T1p & 8.1 & d/sdL8 & 4.0 & 0.18 & 61.5 & 17.1 & 2000 & -0.5 \\
E52470922 & J032953.021-262408.40 & T1p & 10.1 & d/sdL8 & 3.7 & 0.35 & 121.6 & 29.3 & 2100 & -0.5 \\
E58663968 & J035439.352-490006.02 & T1p & 6.1 & d/sdL6 & 3.9 & 0.33 & 210.9 & 50.3 & 2000 & -0.5 \\
E61819292 & J040716.630-500134.88 & T1p & 5.8 & d/sdL6: & 2.9 & 0.32 & 112.2 & 26.9 & 2000 & -0.5 \\
E21266133 & J141038.721+435416.67 & T0p & 6.9 & sdL5 & 1.8 & ... & ... & ... & 1700 & -0.5 \\
E26459277 & J173822.265+671819.33 & T5p & 14.7 & d/sdT5 & 8.7 & -0.58 & 97.8 & 25.7 & 1200 & -1.0 \\
E26857855 & J175418.853+655123.41 & T1p & 9.8 & d/sdL8: & 3.6 & ... & ... & ... & 2100 & -0.5 \\
E26931238 & J175714.973+653316.10 & T1p & 7.9 & d/sdL8 & 3.2 & 0.22 & 109.4 & 29.3 & 2000 & -0.5 \\
E27480941 & J181914.258+670204.98 & T1p & 4.7 & d/sdT1 & 2.2 & 0.30 & 51.0 & 11.6 & 1800 & -0.5 \\
\hline
\end{tabular}
\label{tab:sub} 
\tablecomments{There are missing values for \JE - \HE and the spectrophotometric distance, owing to the flags for the \JE and \HE magnitudes in the \Euclid catalog (see the text).  }
\end{table}

\begin{itemize}

\item E26931238 - This possible s/sdL8 dwarf has a  \JE -  \HE color of 0.22, which stands out from the rest of L8 dwarfs. It is marked as such in Fig.~\ref{fig:color}. Its spectrum is shown in Fig.~\ref{fig:dsd} and shows, in the $z$- and $J$-bands a better fit to a s/sdL8 standard, than to a T1 dwarf. The estimated spectro-photometric distance of $109\pm29$~pc and the estimated metallicity is -0.5~dex for an effective temperature of 2000~K.

\item E52470922 - This possible s/sdL8 dwarf has a  \JE -  \HE color of 0.35, and also stands out from the rest of L8 dwarfs. It is marked as such in Fig.~\ref{fig:color}. Despite having a noisier spectrum (shown in Fig.~\ref{fig:dsd}),  shows a better fit to a s/sdL8 standard, than to a T1 dwarf in the $z$- and $J$-bands. The estimated spectro-photometric distance of $122\pm29$~pc and the estimated metallicity is -0.5~dex for an effective temperature of 2100~K.

\item E21266133 - The spectrum of this UCD is shown in  Fig.~\ref{fig:E21266133}. The templates for two standard, T0 and sdL5  are plotted in blue and black lines, respectively. We can see that the fit in the $J$ band is only good for the sdL5 template, while both fit well in the $H$ band. The visual inspection of the images, also shown in the figure, showed other close and bright object at 2.9 arcsec. It is saturated in the NIR images. The flag for the \JE and \HE magnitudes: "source blended with another one" 
 is set in the \Euclid catalog\footnote{\url{https://euclid.esac.esa.int/dr/q1/dpdd/}} and prevented us to compute the colors and distance. The estimated metallicity (-0.5~dex), for an effective temperature of 1700~K is in agreement with the spectral classification.
These findings opens two questions, one if E21266133 is a companion of that star (we include it, as a candidate in Sect.~\ref{sc:bin}), and the second, if the spectra is contaminated by the possible primary M7. The discussion and follow-up observations of this object are presented in Muñoz Torres et al. (in prep.).

\item E46376957- This possible s/sdL8 dwarf has a  \JE -  \HE color of 0.18, which stands out from the rest of L8 dwarfs. It is marked as such in Fig.~\ref{fig:colorws}. Its spectrum is shown in Fig.~\ref{fig:sd} and shows the better fit to the s/sdL8 standard, with respect to a T1 dwarf. The estimated spectro-photometric distance of $62\pm17$~pc and the estimated metallicity is -0.5~dex for an effective temperature of 2000~K.

\item  E27480941 - This possible s/sdT1 dwarf has an estimated spectro-photometric distance of $51\pm12$~pc. Despite having a  \JE -  \HE color that does not stands out from other T1 dwarfs, the difference between the fit to a T1 and as d/sdT1 standard, is evident in Fig.~\ref{fig:dsd}. The estimated metallicity is -0.5~dex for an effective temperature of 1800~K.
  \end{itemize}

\subsection{\label{sc:bin}Possible resolved binaries  }

The high spatial resolution of VIS (0.1 arcsecond /pixel) allows to distinguish close objects in \IE images. Such was the case of the resolved binary  E271934AB (Euclid~DR1~J180744.182+664252.80), discovered as a result of our previous work on Q1 \citep{Dominguez2025}, and which has been studied in detail from multiple follow-up observations in Muñoz Torres et al. (in prep.). 
The visual inspection of the images for the UCDs retrieved by the spectral index search have revealed numerous pairs of objects. We already
saw that E21266133 could be a possible companion to a brighter object. Here we analyze six  possible binaries. They are summarized in Table~\ref{tab:binres} and described below.

\begin{table}[h!]
\caption{Main parameters of the possible binaries }
\centering
\begin{tabular}{llrrlrc}
\hline\hline
Alias1 & Designation & SpT 1 & Ang. sep. & Alias & SpT  & References   \\
 &  Euclid DR1 + & UCD & (arcsecond) & 2nd. object  &  2nd. object  & \\
 \hline
E13075302 & J005218.072+063102.29 & L4: & 2.5 & E13075173 & M4 & (2)   \\
E50205963 & J032049.431-532659.31 & T0 & 0.2 & E50205971 & ... & (1)  \\
E52216752 & J032852.020-433655.89 & L6 & 4.4 & E52216323 & ... & (1)  \\
E15857942 & J103419.063+755358.18 & T5 & 4.4 & E15857940 & M3 & (1)   \\
E21266133 & J141038.721+435416.67 & sdL5 & 2.9 & E21266235 & M7 & (1)   \\
E26965805 & J175837.933+645333.35 & T0p & 5.9 & E26966024 & sdM6 & (1)  \\
\hline
\end{tabular}
\label{tab:binres} 
\tablecomments{ References: 1) this work, 2) \cite{Gaia2023}}
\end{table}

The  possible resolved binaries are:

\begin{itemize}

   \item  E26965805 -  This T0p dwarf could be a  binary associated with E26966024 an M5p (possible sdM6) as the primary member of the system. Their separation is 5.9 arcsecond. The spectrum of E26965805 could be contaminated by the stray-light of the much-brighter possible primary. The images and both spectra are shown in Fig.~\ref{fig:E26965805}.  Two templates are plot for the possible primary, as its spectral type could be sdM6. We are currently, conducting follow-up observations to asses the spectral type of the possible primary and the proper motion of both. The results will be presented in Muñoz Torres et al. (in prep.).

\begin{figure}[ht!]
\centering
\includegraphics[width=0.65\hsize]{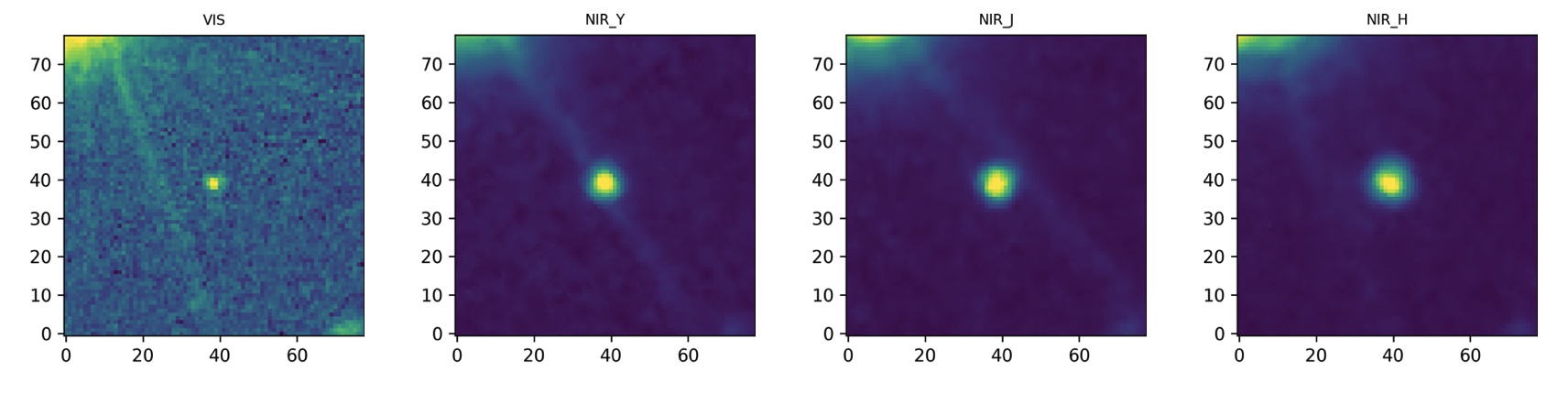}

\plottwo{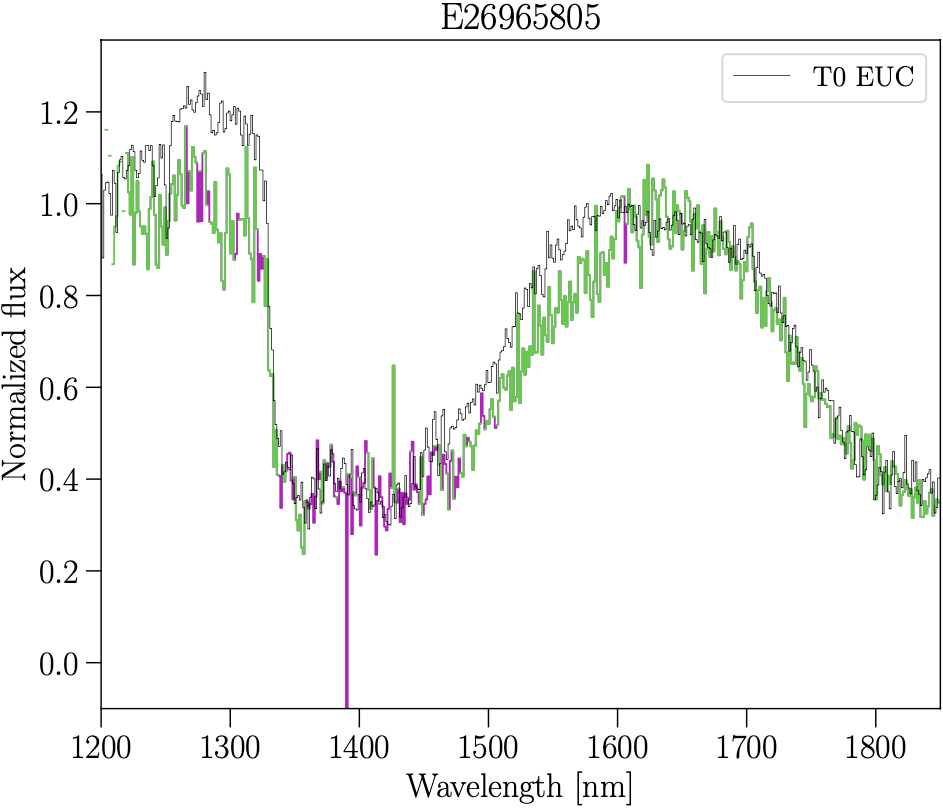}{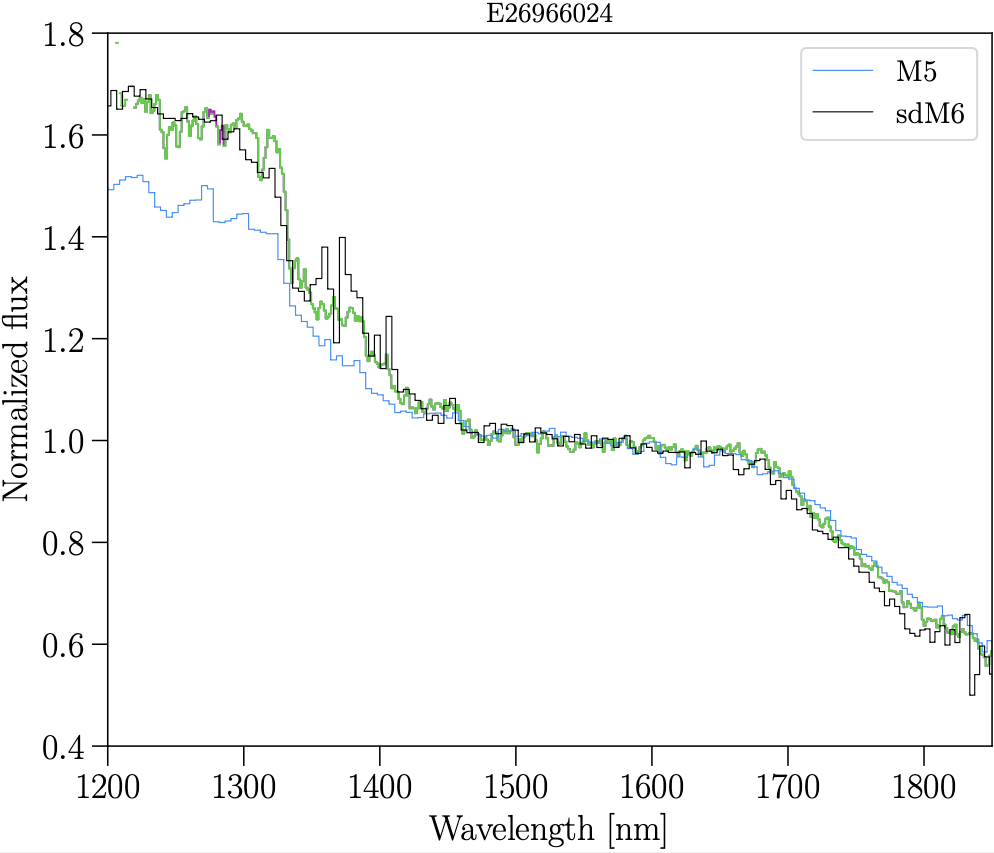}
\caption{ \Euclid four-band images (top) and spectra of E26965805 (bottom left) and the possible M5p primary (bottom right). 
The bright M5p is located in the top-left corner of the images and out of them to be able to see the faint T0. 
The color scheme for the spectra is as in previous figures.
Two templates are plot in the right panel as the spectral type could be  sdM6.  }
\label{fig:E26965805}
\end{figure}

   \item  E21266133 - This dwarf was analyzed in Sect.\ref{sc:dsd} as a possible sdL5. It was shown in  Fig.~\ref{fig:E21266133} with an M7 dwarf as the possible primary. There is however a large difference in their brightness (the primary is saturated), which points just to a possible coincidence of having two UCDs  in the same field (with an angular separation of 2.9 arcsecond). The discussion and follow-up observations of this object are presented in Muñoz Torres et al. (in prep.).
   
    \item  E13075302 - This L4 dwarf could be a  binary associated with E26966024 a possible mild subdwarf M4. It was also observed by Gaia  (DR3 2556398314497808512), which estimates the distance at 96.0$\pm$0.4~pc and a proper motion of: 37.25$\pm$0.06, -42.58$\pm$0.04~ mas/yr  \citep{Gaia2023}. Fig.~\ref{fig:E13075302} shows the images and the spectra, low quality and SNR of the L4 dwarf. As well,  the spectra of the possible d/sdM4 (a very good fit to the template). The angular separation between them is 2.5 arcsecond. 
    
    \item  E15857942 -  This T5 dwarf is vey interesting, as it is at 4.4 arcsecond of a very bright M3. Fig.~\ref{fig:E15857942} shows the images and spectra. We are currently, conducting follow-up observations to asses the spectral type of the possible primary and the proper motion of both. The results will be presented in Muñoz Torres et al. (in prep.).

    \item  E50205963 -  This  L6 dwarf is very close (0.2 arcsecond) to other object, possibly same spectral type, as the spectrum sown in Fig.~\ref{fig:E50205963} does not present any peculiarity. This  possible binary is currently being studied using the point spread function (PSF) subtraction-technique described in Muñoz Torres et al. (in prep.).

    \item  E52216752 - This T3 dwarf has some possibilities of being a binary. There is object in the images shown in Fig.~\ref{fig:E52216752}, which not completely point-like in the visible. The NIR images (0.3 arcsecond/pixel) do not resolve the two objects. The spectra and photometry are, however typical for a T3 dwarf, without any sign of contamination. It is also candidate for follow-up observations (Muñoz Torres et al. in prep). 

\end{itemize}

\begin{figure}[ht!]
\centering
\includegraphics[width=0.65\hsize]{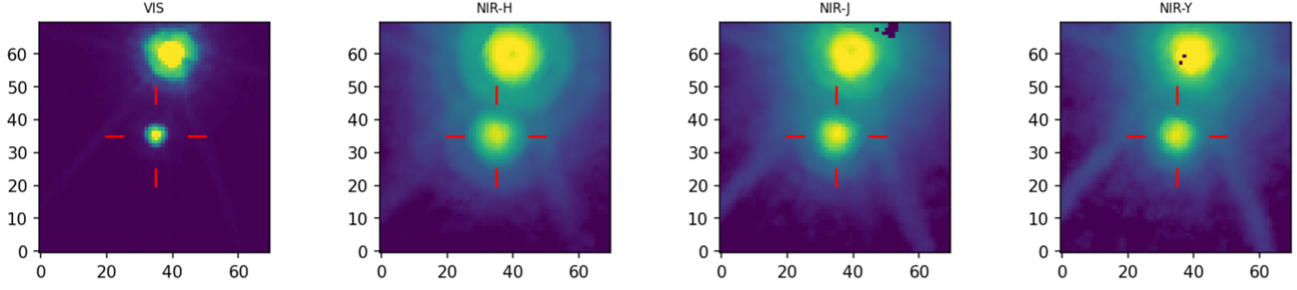}

\plottwo{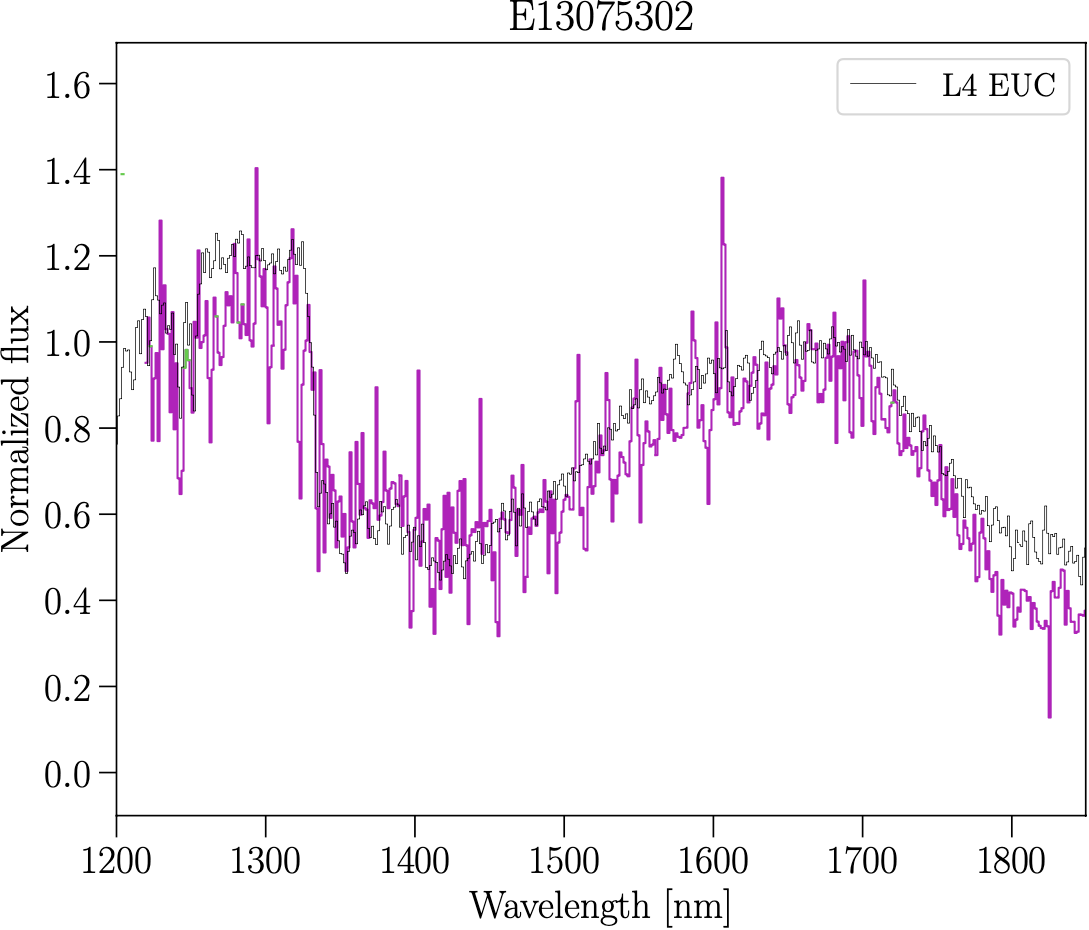}{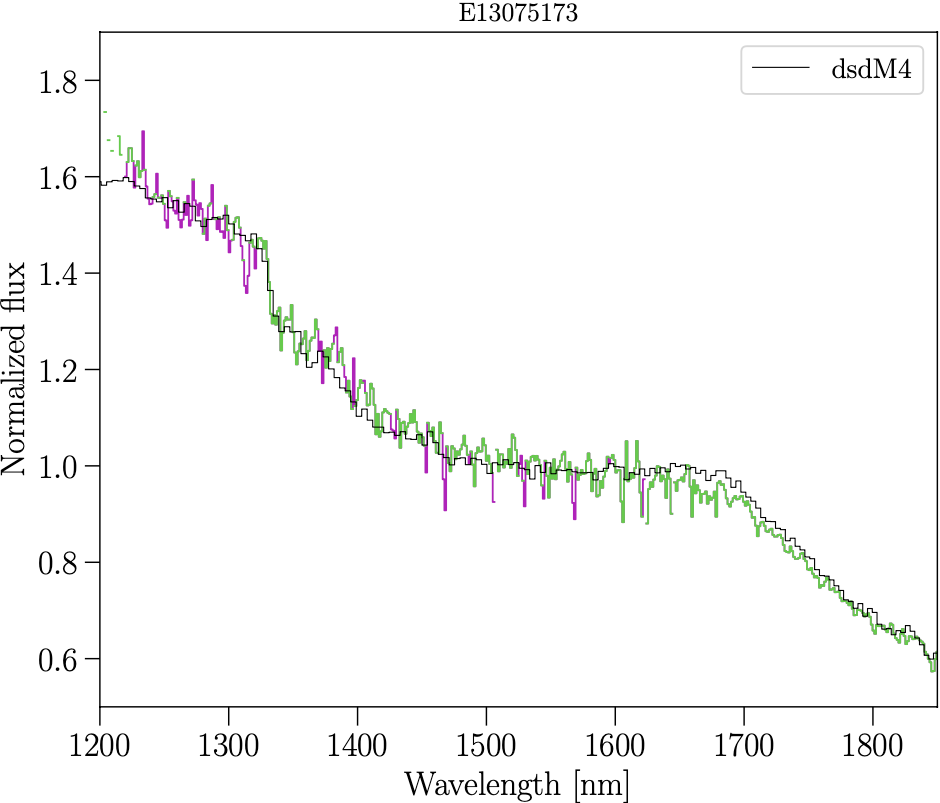}
\caption{ \Euclid four-band images (top) and spectra of E13075302 (bottom left) and the possible M4 primary (bottom right). The color scheme for the spectra is as in previous figures.    }
\label{fig:E13075302}
\end{figure}

\begin{figure}[ht!]
\centering
\includegraphics[width=0.65\hsize]{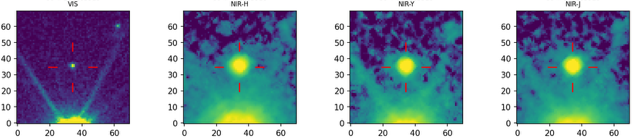}

\plottwo{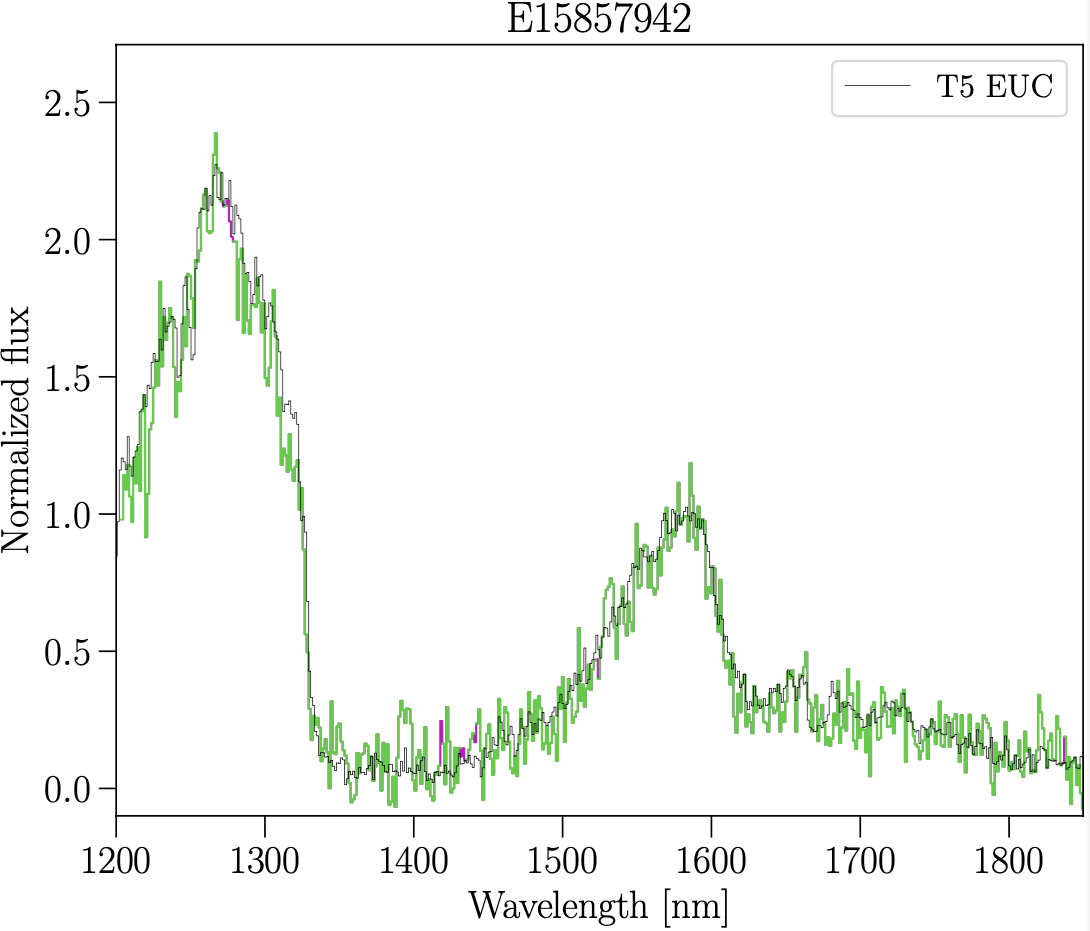}{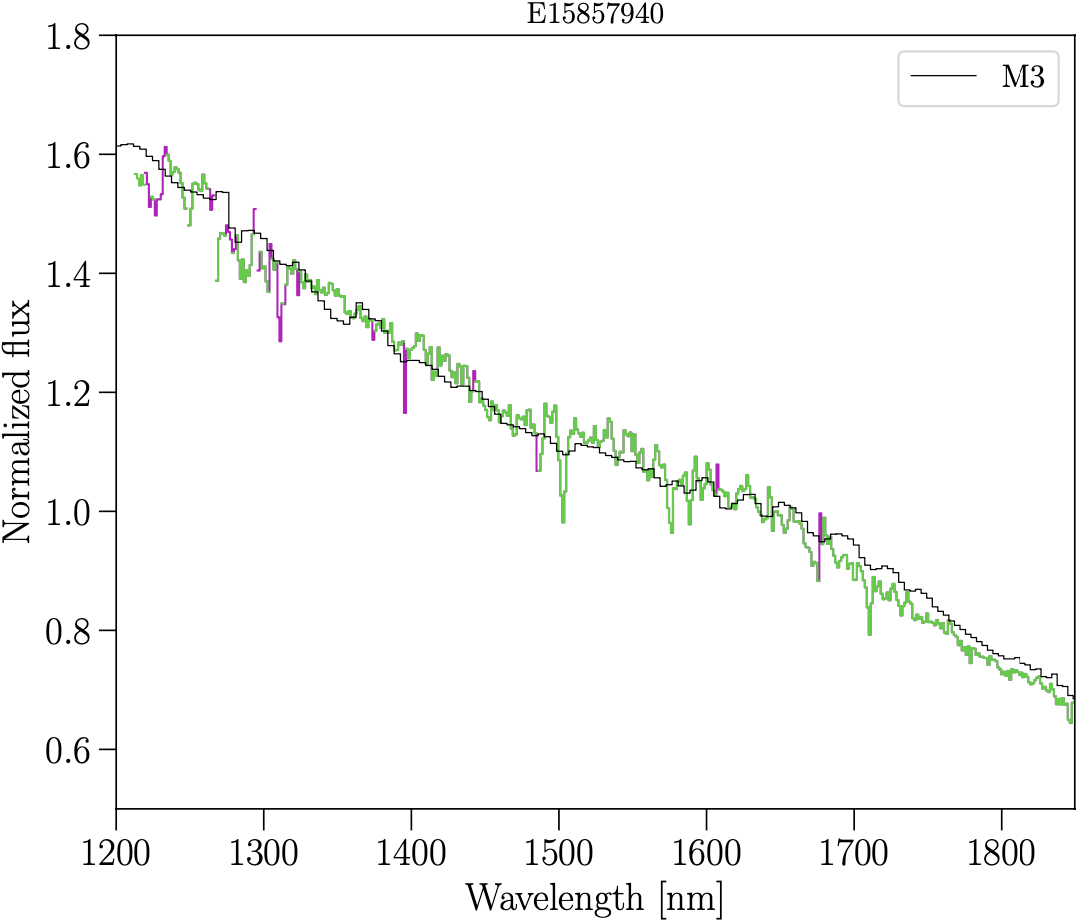}
\caption{ \Euclid four-band images (top) and spectra of the T5 dwarf E15857942 (bottom left) and the possible M3 primary (bottom right). The color scheme for the spectra is as in previous figures.  }
\label{fig:E15857942}
\end{figure}

\begin{figure}[ht!]
\plottwo{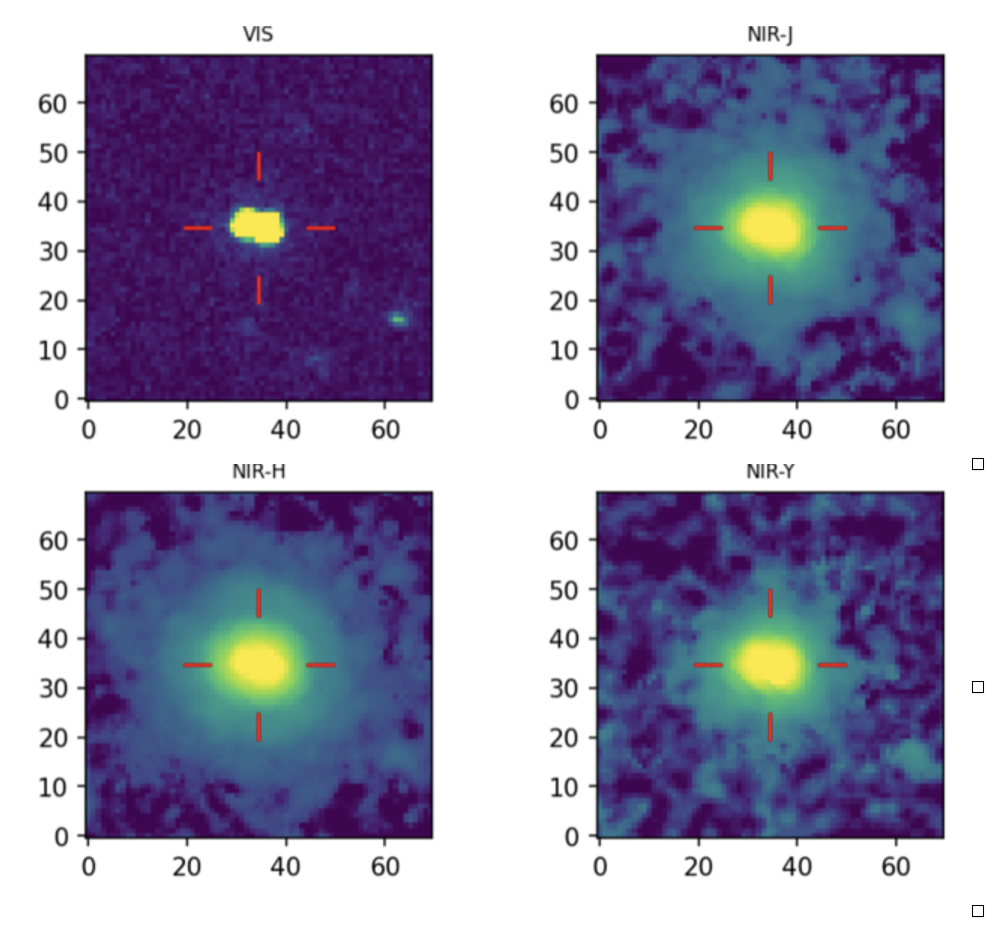}{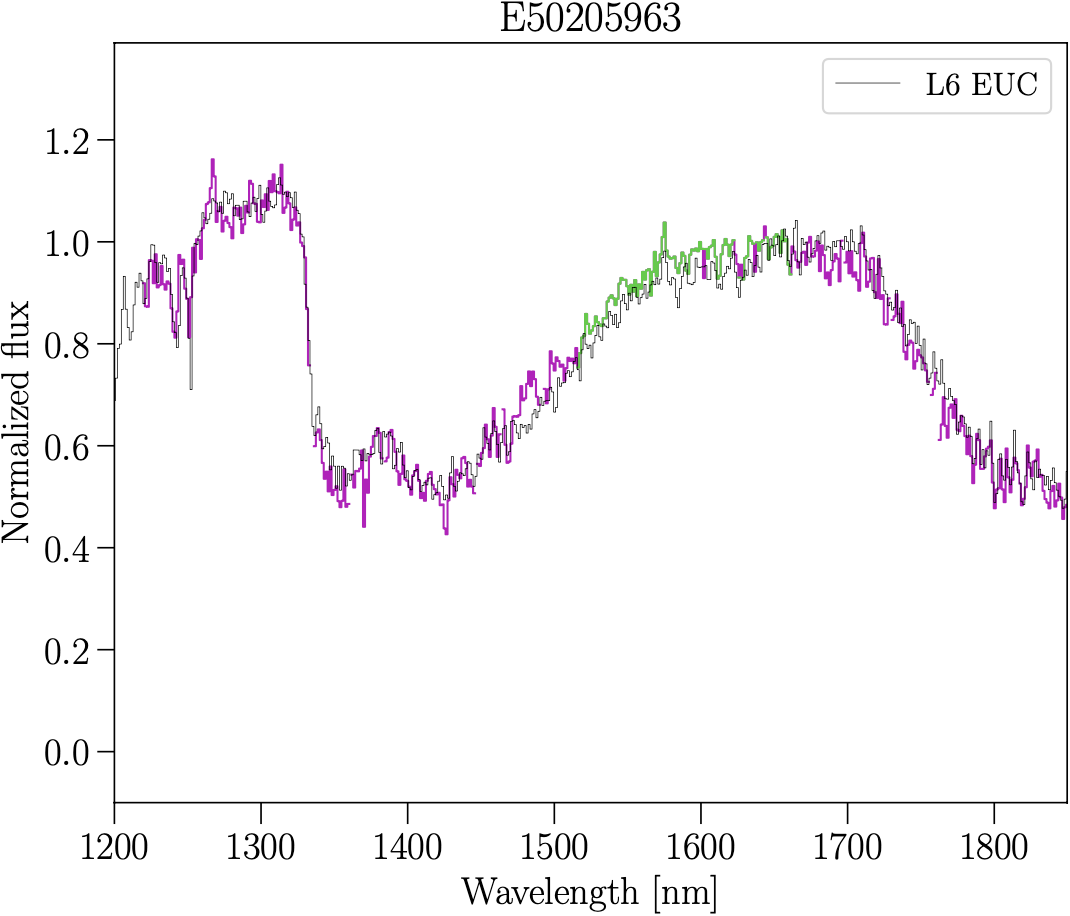}
\caption{ \Euclid four-band images and spectra of the possible binary E50205963. The color scheme for the spectra is as in previous figures.  }
\label{fig:E50205963}
\end{figure}

\begin{figure}[ht!]
\plottwo{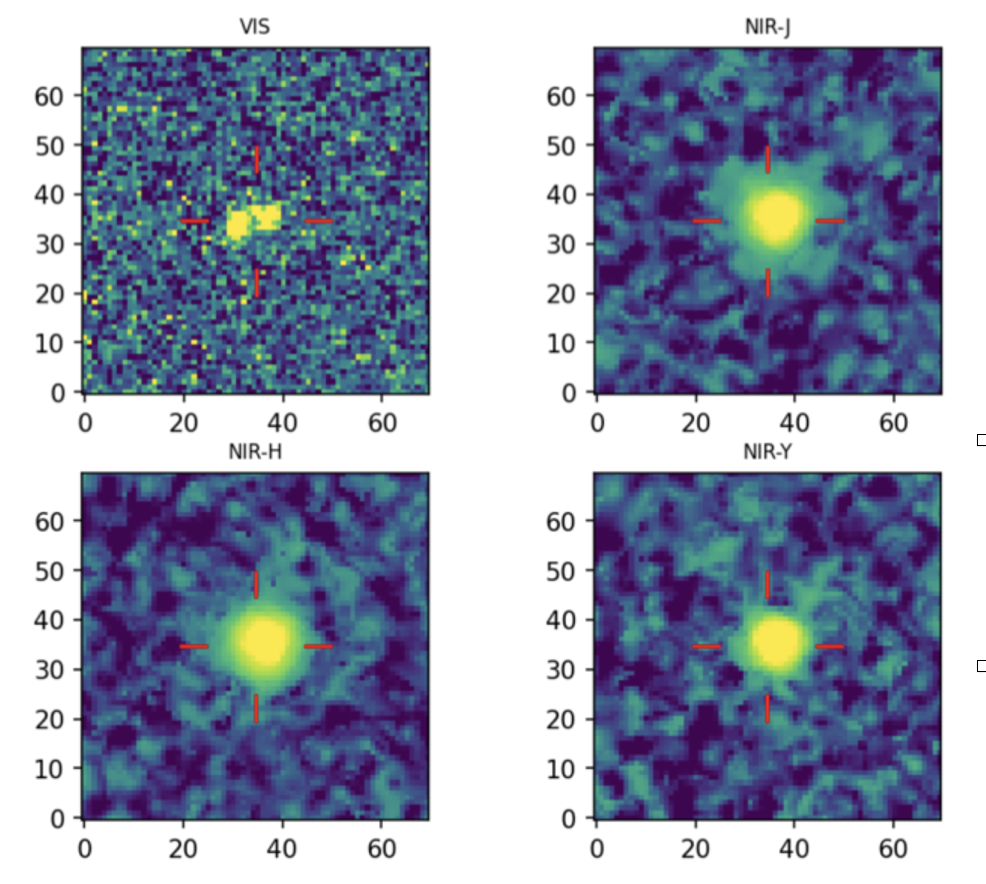}{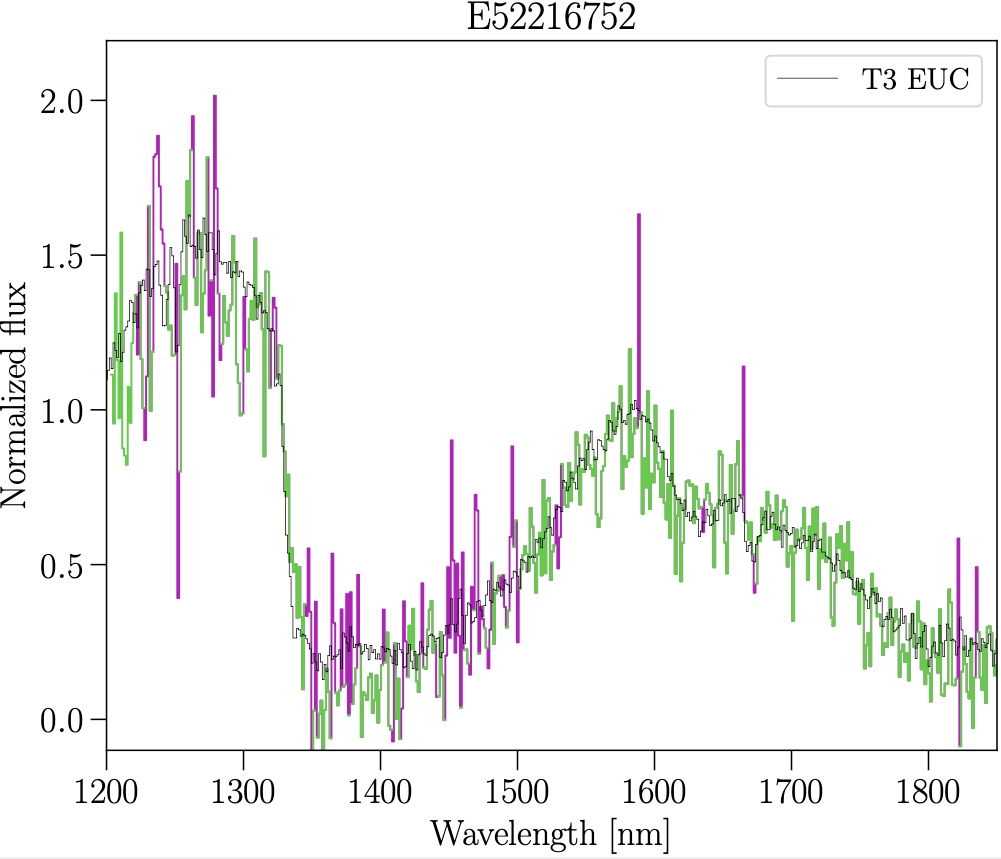}
\caption{ \Euclid four-band images and spectra of the possible binary E52216752. The color scheme for the spectra is as in previous figures.  }
\label{fig:E52216752}
\end{figure}

\begin{figure}[ht!]
\plottwo{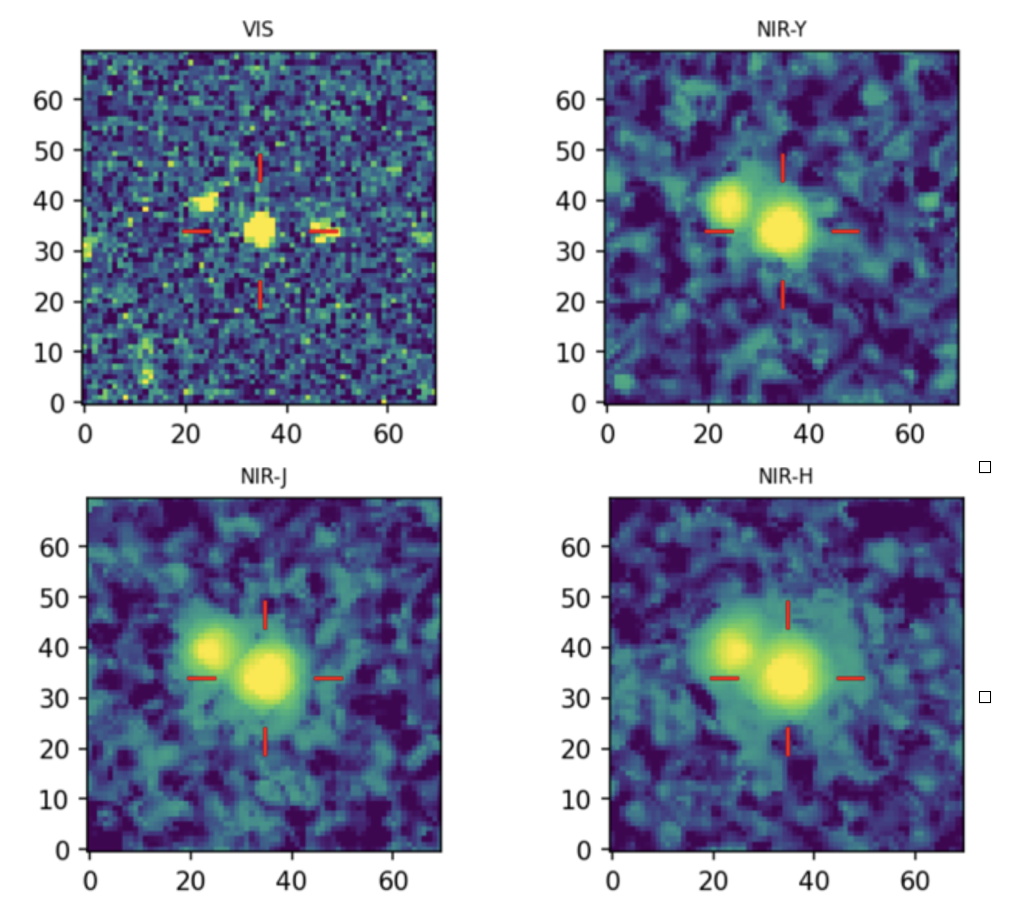}{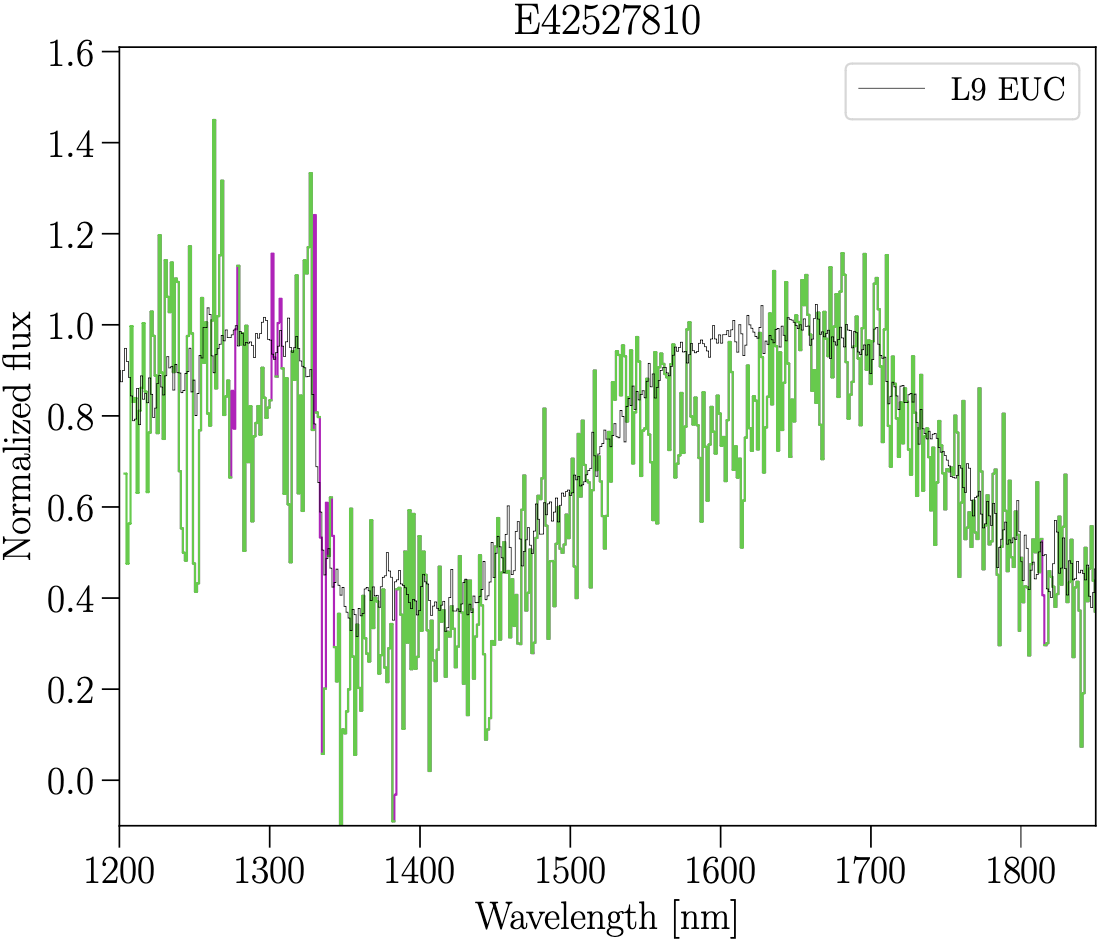}
\caption{ \Euclid four-band images and spectra of the possible binary E42527810. The color scheme for the spectra is as in previous figures. }
\label{fig:E42527810}
\end{figure}

%
%

\subsection{\label{sc:bin}Possible unresolved binaries }

In Sect.~\ref{sc:spt}, we did the classification by spectral type and marked with a ``c" the peculiar objects that had a better fit to a weighted blend of two templates. The composite templates are created combing  M,L,T standard templates for the primary with later types (L and T) from the same templates. Each template is weighted by the relation of absolute magnitudes, between primary and secondary. The absolute magnitudes for each spectral type are taken from the semi-empirical relations for field objects by \citet{Sanghi2024}. The blend is done by averaging the weighted values.

\begin{figure}[ht!]
\centering
\includegraphics[width=0.45\hsize]{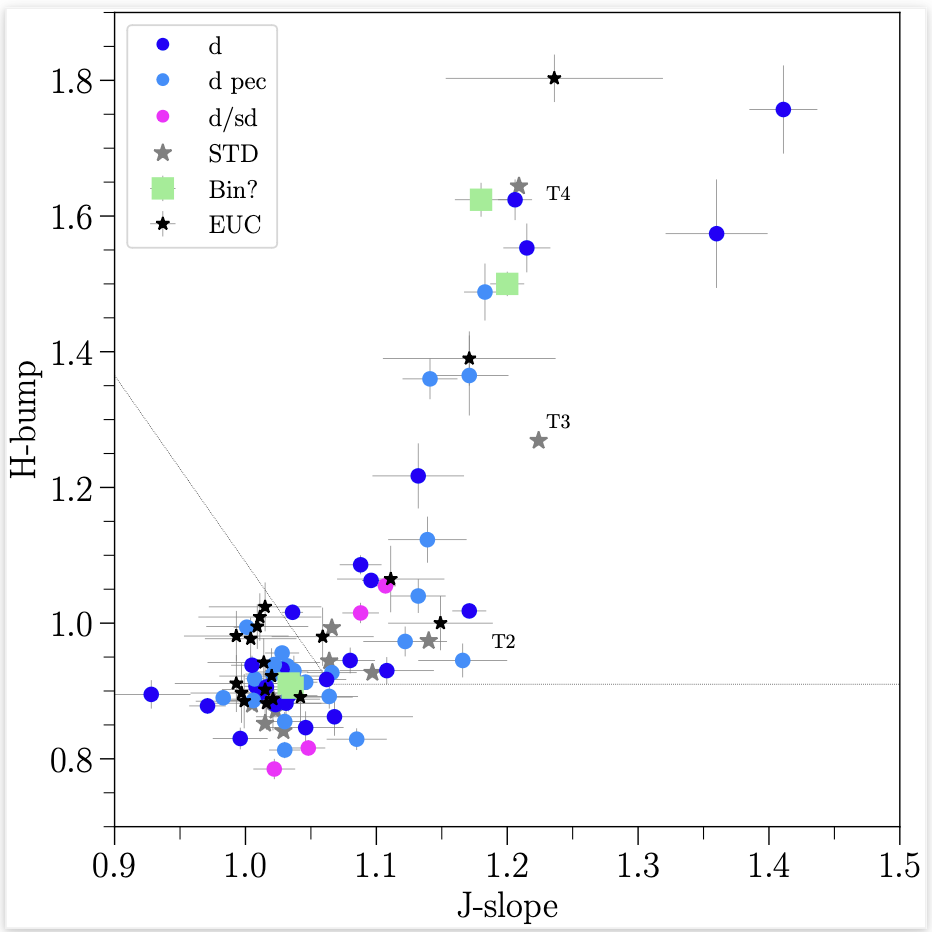}
\caption{ $H$-bump vs.  $J$-slope index ratio for the sample of UCDs in the deep observations, plus the possible  binaries (``Bin?"; green squares). The other symbols  are as in previous plots: blue circles for the dwarfs, light-blue circles for the peculiar ones, and magenta circles for the possible mild subdwarfs, The grey stars represent the standard (STD) templates used in this paper, T2 to T4 have a label close to the symbol. The black stars represent the EUC templates. Two lines are plotted following the criteria of \cite{Bardalez2014}. }
\label{fig:hbump}
\end{figure}

In order to  investigate the possibility of  being an unresolved binary we selected the objects where the fit to any of these composite templates is very good (i.e. with a difference larger than two in the value of $\chi ^2$, with respect to the fit with the standard templates). Then, we assign an alternative spectral type formed by the two components of the blended template. 
As the number of blended templates is considerably higher than that for the standard templates, the comparison is made with a set of templates
 for all subtypes, including 20 intermediate types (i.e., by interpolation),
 using the EUC templates described in Sect.~\ref{sc:template}. We created a list of candidates as possible spectral binaries and examined other parameters such as their ellipticity, the VIS image (which has a pixel size of 0.1~arcsecond), and the H-bump index defined by \cite{Bardalez2014}. The values for these parameters are listed in Table~\ref{tab:bin}. It includes the candidates from the deep observations and the Wide Survey.
Fig.~\ref{fig:hbump} shows the  $H$-bump vs.  $J$-slope plot  and the region in the plot towards the top of the two lines, indicates the objects with the higher probability of being binaries \citep{Bardalez2014}.  Two of the candidates fall in this region  E57105624  and  E61249736 , while three fall below E23449751, E61864817 and E64627617.  
The standard templates are also plotted and show how mid-T dwarfs fall inside the higher probability region. Further studies are required to confirm the candidates. From the literature, only E23449751 is known (2MASS J01334773-5439445).

\begin{figure}[ht!]
\plottwo{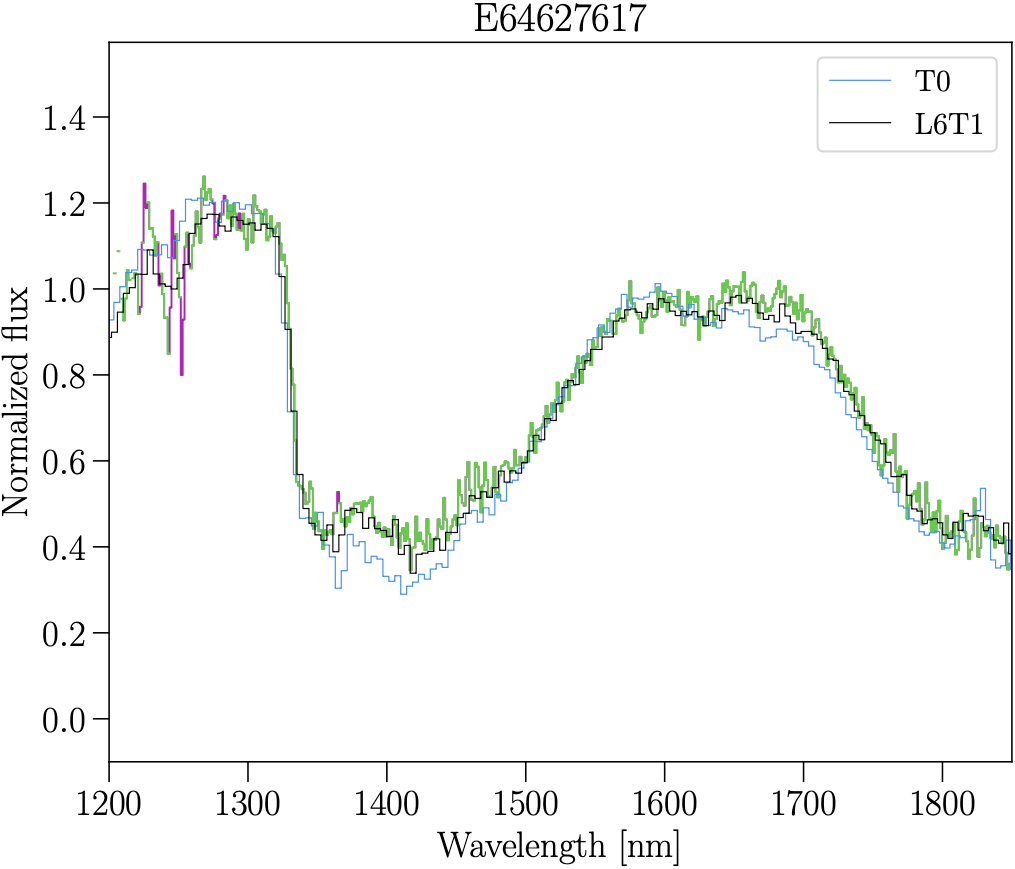}{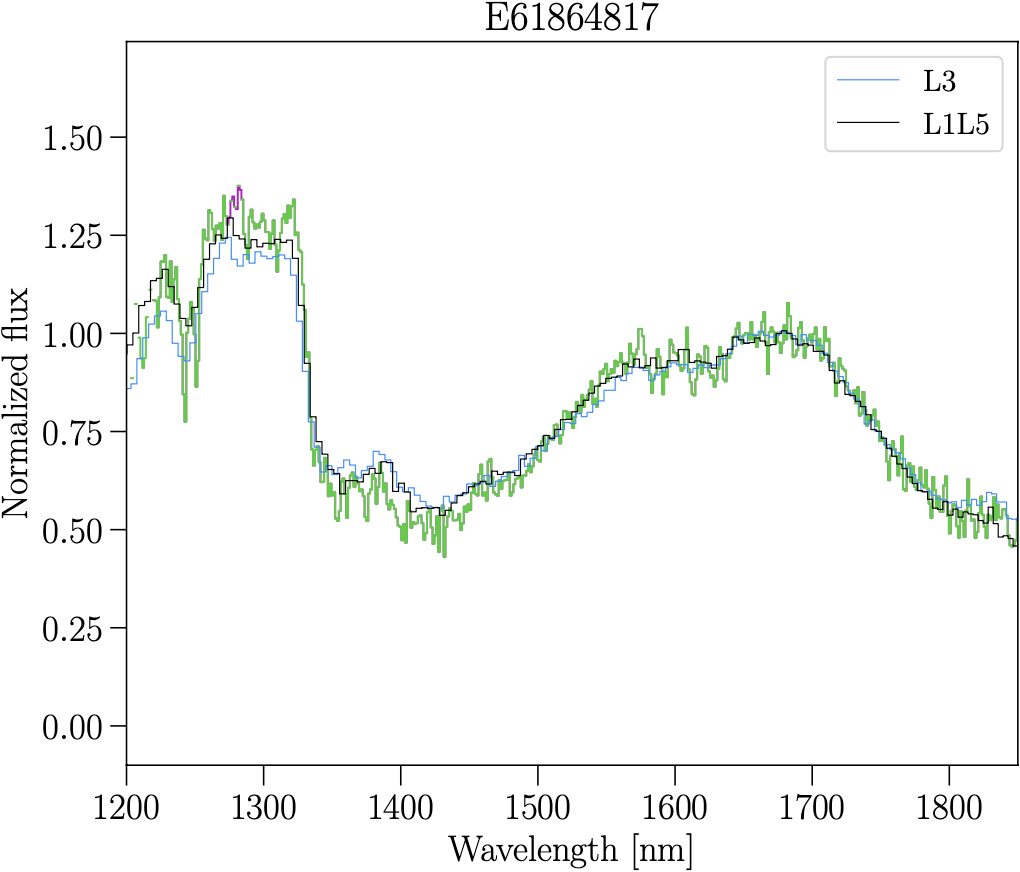}
\plottwo{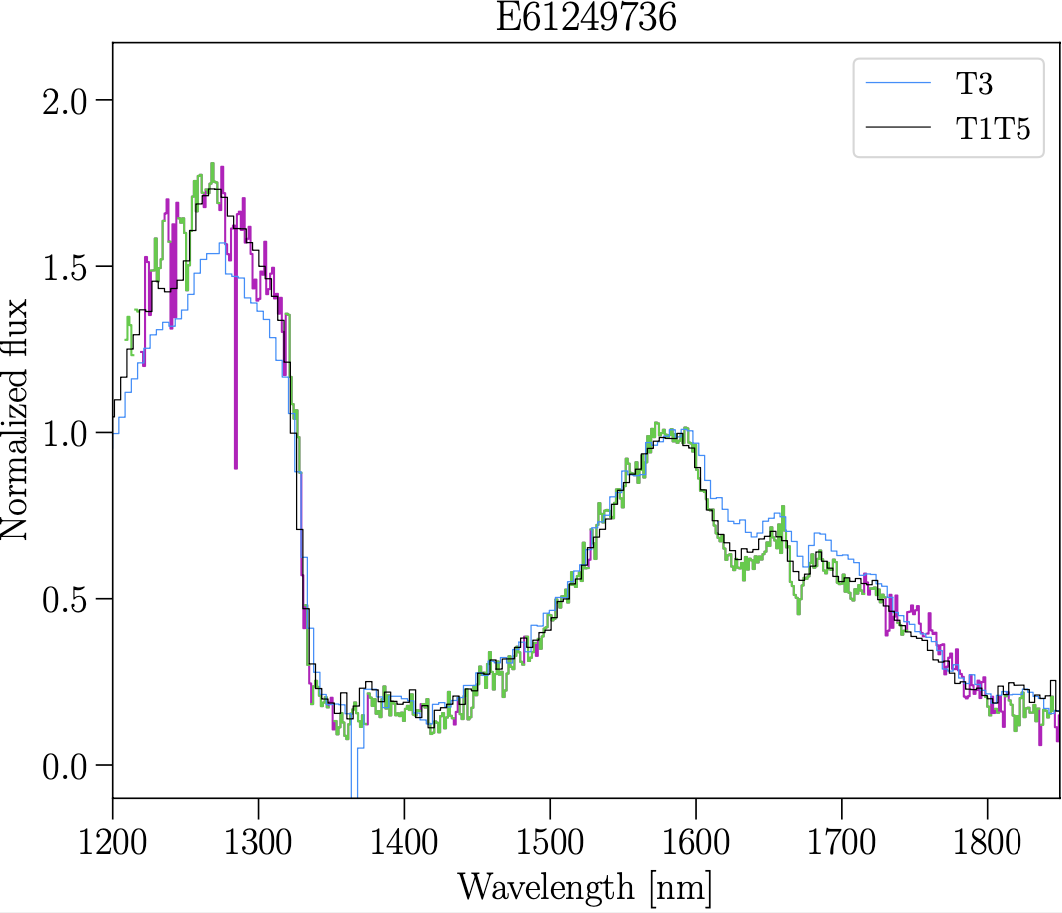}{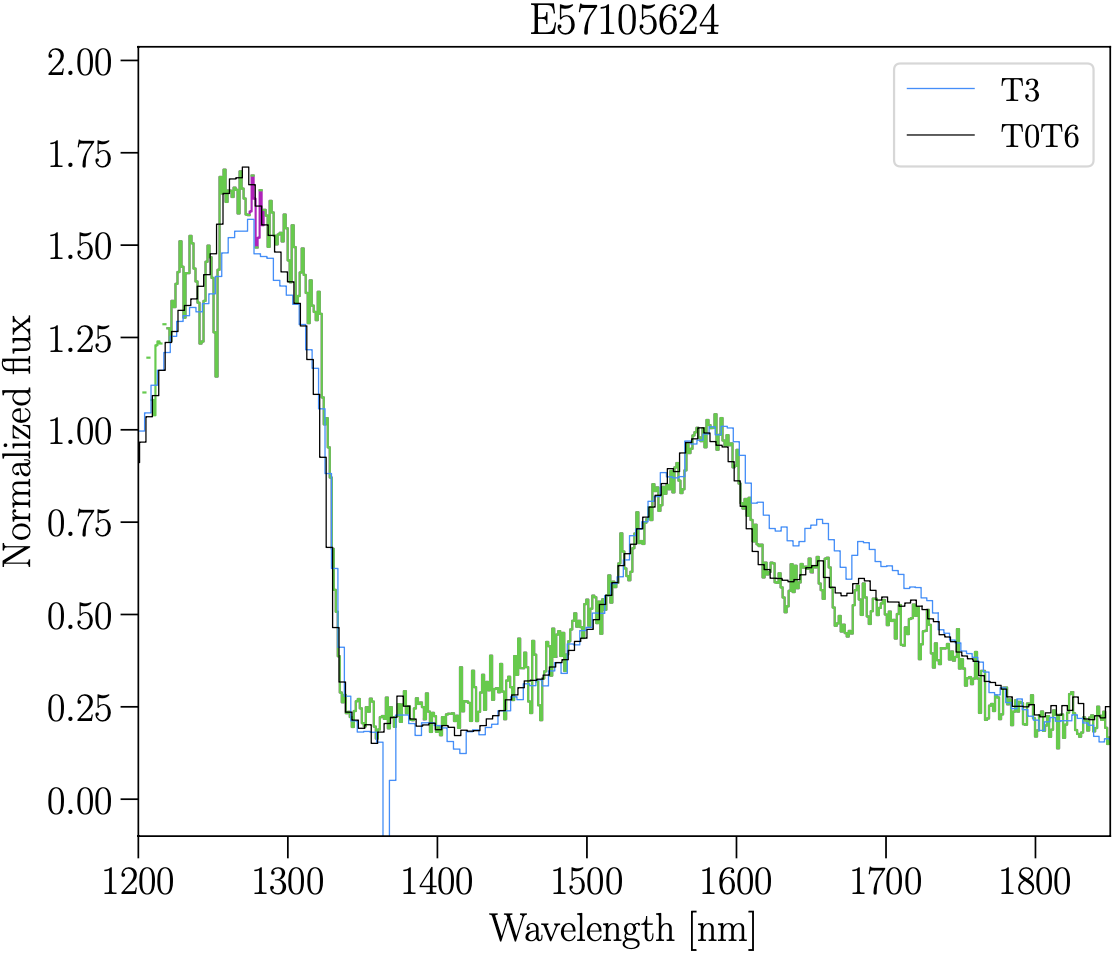}
\caption{ Spectra of four possible spectral unresolved binaries. The color scheme for the spectra is as in previous figures.    }
\label{fig:unresbin}
\end{figure}

\begin{table}[h!]
\caption{Main parameters of the possible spectral unresolved binaries }
\centering
\begin{tabular}{llcrcrrlrr}
\hline\hline
Alias   &  Designation  &    SpT  &  $\chi ^2$ &  Alt. SpT &  Alt. $\chi ^2$  & Ellipticity &   VIS image   &   $H$-bump &  $J$-slope    \\
  &    \texttt{Euclid DR1 +}  &     &     &     &     &     &   (contour)  &  &  \\
\hline
E23449751 & J013347.940-543944.17 & T0c & 3.7 & L6T3 & 1.2 & 0.03 & circular ? & 1.1 & 1.0   \\
E57105624 & J034825.349-501644.67 & T3c & 5.7 & T0T6 & 3.3 & 0.06 & circular & 1.6 & 1.2   \\
E61249736 & J040459.936-643249.84 & T3c & 4.0 & T1T5 & 2.0 & 0.10 & circular & 1.5 & 1.2   \\
E61864817 & J040727.556-483657.70 & L3c & 4.7 & L1L5 & 3.1 & 0.02 & not circular & 0.9 & 1.0   \\
E64627617 & J041830.628-492504.72 & T0c & 3.5 & L6T1 & 1.6 & 0.06 & circular ? & 0.9 & 1.0   \\
\hline
\end{tabular}
\label{tab:bin} 
\end{table}

\begin{figure}[ht!]
\plottwo{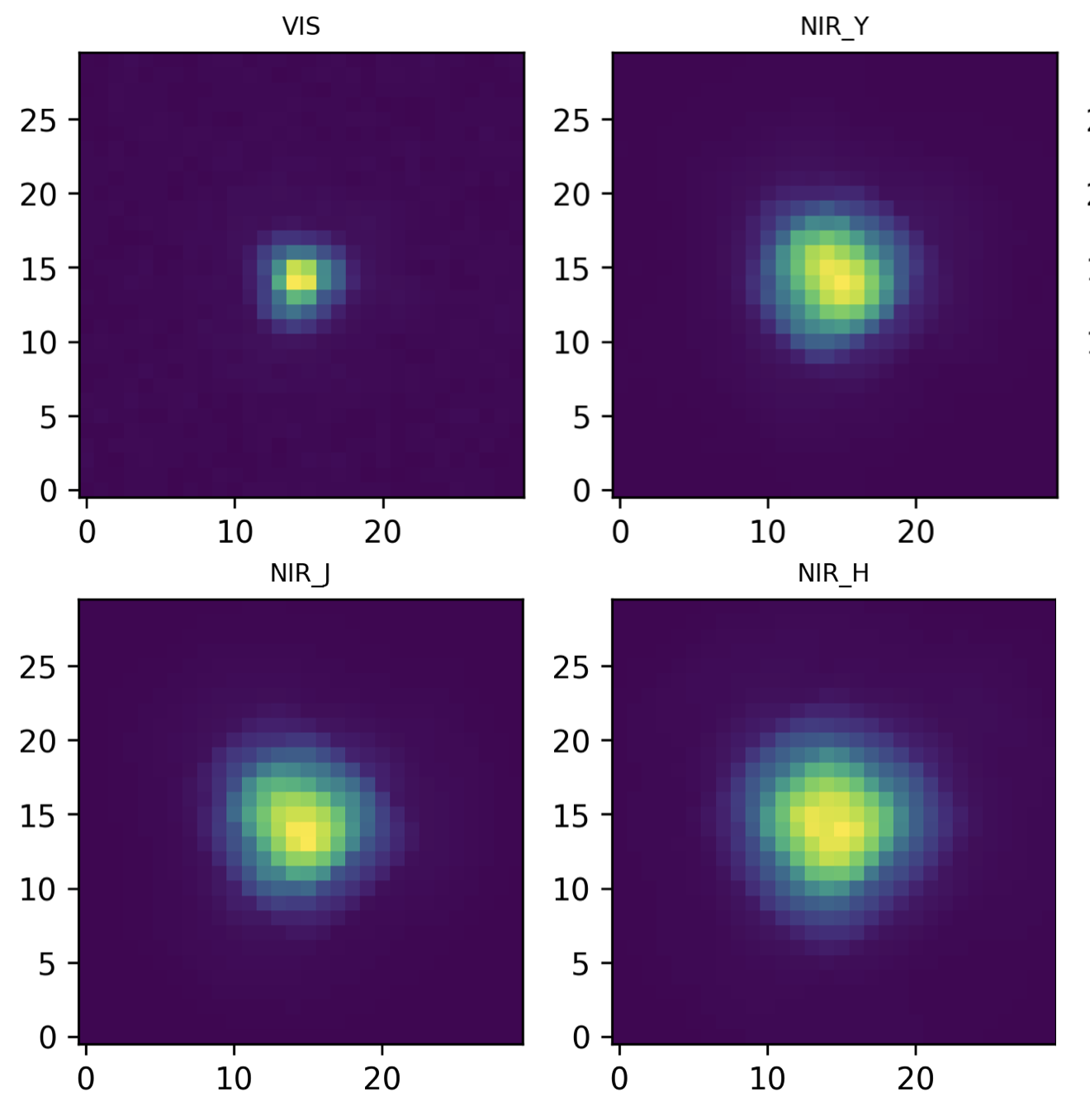}{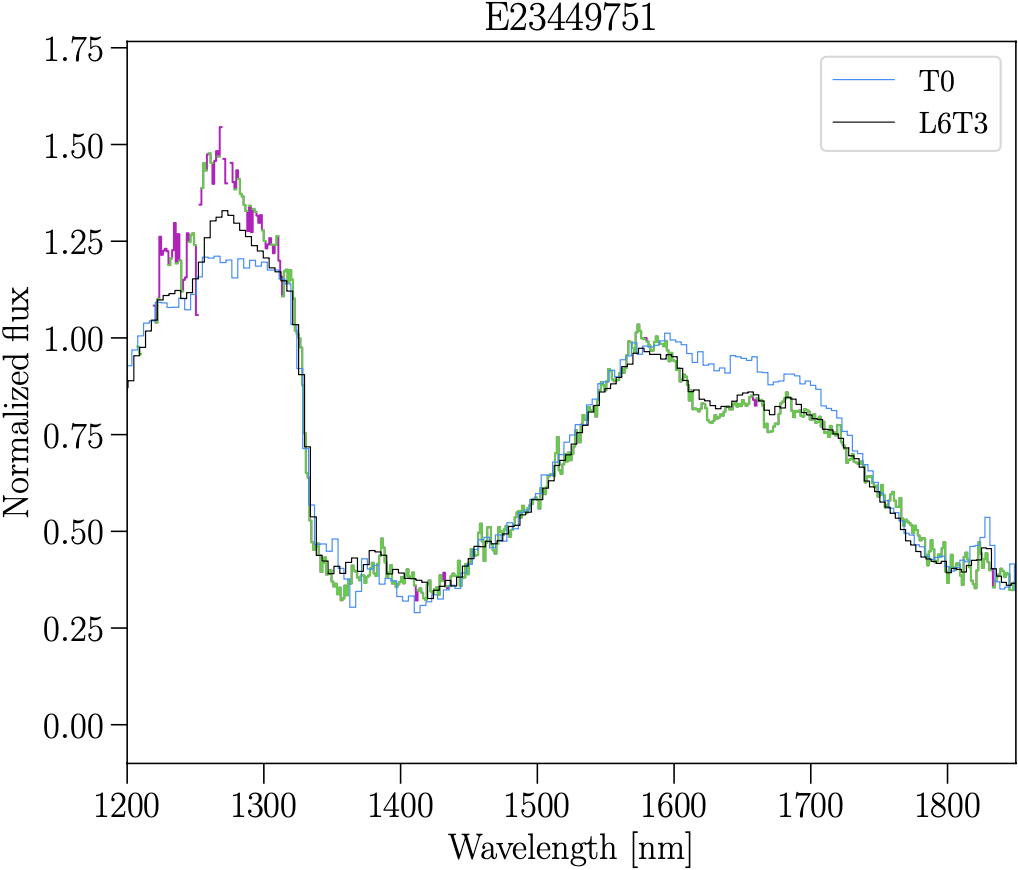}
\caption{ \Euclid four-band images and spectra of the possible unresolved binary E23449751. The color scheme for the spectra is as in previous figures.  }
\label{fig:E23449751}
\end{figure}

 
\subsection{\label{sc:young}Possible young dwarfs}

\begin{figure}[ht!]
\plottwo{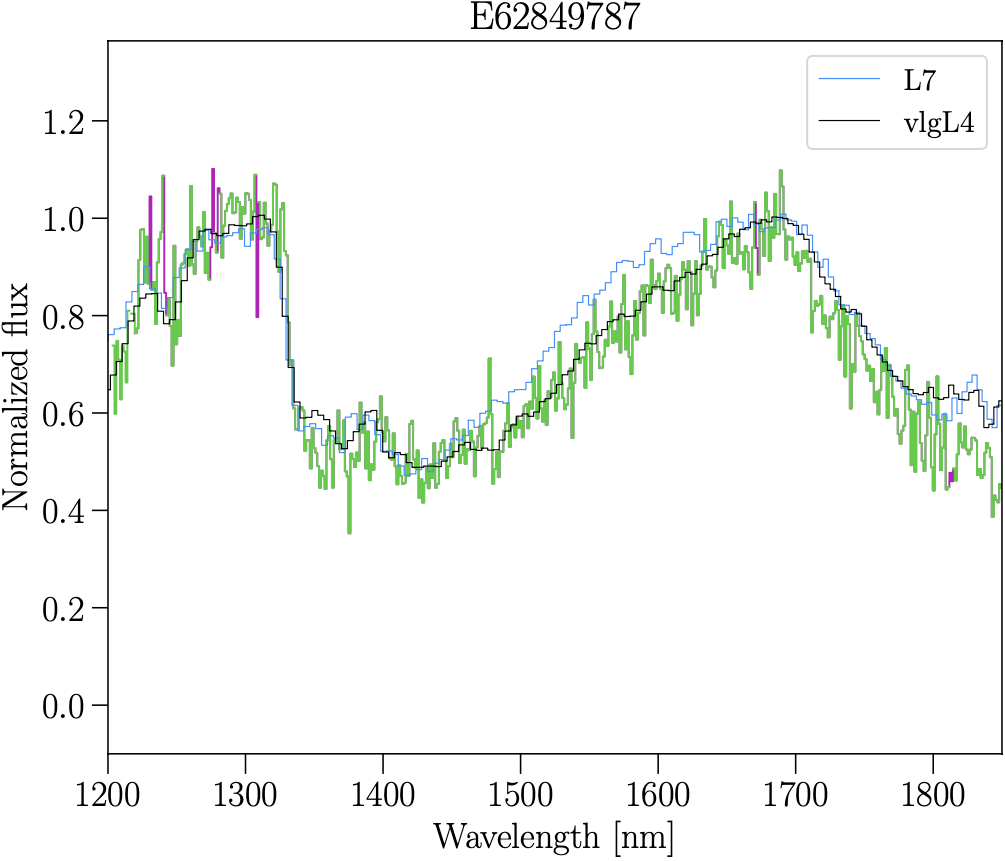}{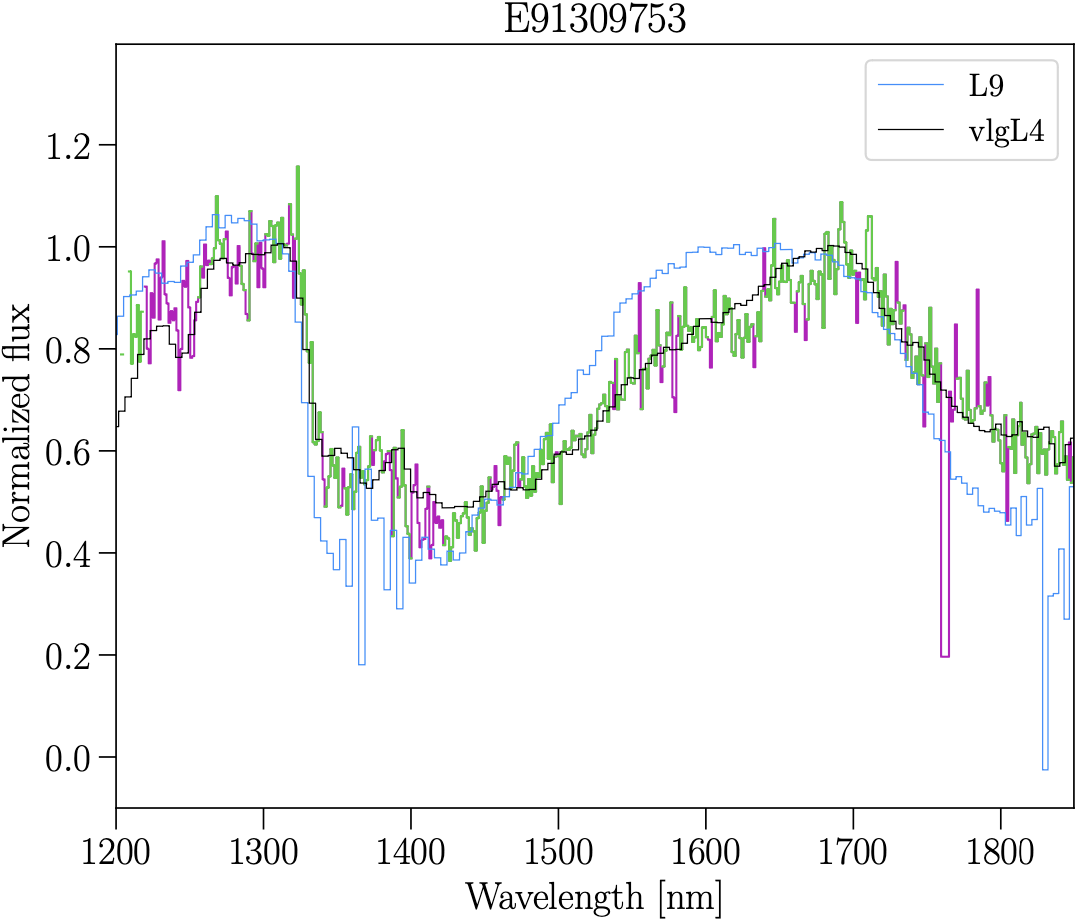}
\caption{ Spectra of two L4$\gamma$ dwarfs E62849787 and E91309753 (right). The color scheme for the spectra is as in previous figures. }
\label{fig:young}
\end{figure}

We have identified some UCDs with spectral features indicating the possibility of having low gravity: a low pEW, a red \JE~-~\HE color or a triangular-shaped $H$-band peak. Among the candidates, we found some objects already known in the literature. We selected those with the most clear features to classify them using the available low-gravity standards. Fig.~\ref{fig:young} shows an example of two possible L4$\gamma$ dwarfs
The result is shown in Table~\ref{tab:young} summarizes the results as these objects are  analyzed in detail in Lodieu et al. (submitted).

\begin{table}[h!]
\caption{Possible young dwarfs }
\centering
\begin{tabular}{llrlcll}
\hline\hline
Alias & Designation & SpT &  J-H & Alt. SpT & Lit. SpT & References  \\
  &    \texttt{Euclid DR1 +}  &        &     &     &     &    \\
\hline
E58125670 &   J035230.160-492546.91 & L9p & 0.77 & young L & ... & 1, 4 \\
E62849787 &   J041123.948-301023.47 & L7p & 0.58 & L4$\gamma$      & ... &  1,5  \\
E70446447 &   J044147.147-212716.63 & L3p & 0.59 & L3$\gamma$        & ... & 1, 4 \\
E74752024 &   J045900.485-285337.24 & L5p & 0.84 & L4$\gamma$      & L7:red,  L4$\gamma$ & 1, 2,   3,  4 \\
E75327173 &   J050118.521-451214.10 & T2p & 0.48 & young T & ... & 1, 4 \\
E89824328 &   J055917.838-493056.86 & L8p & 0.97 & L5$\gamma$/$\beta$     & ... & 1, 4 \\
E91309753 &   J060514.340-341242.80 & L9p & 0.50 & L4$\gamma$       & ... & 1, 5  \\
E13574484 &   J090258.762+670829.96 & L8p & 0.73 & young L & L7red & 1, 2,  4 \\
E14209729 &   J092823.350+682404.21 & L5p & 0.79 & L5$\gamma$/$\beta$    & ... & 1, 4 \\
E35269945 &   J233047.869-330351.11 & L6p & 0.78 &  L6$\gamma$, T0 &  L7$\gamma$ & 1, 3,  4 \\
E35942591 &   J235742.219-641513.97 & T0p & 0.85 & young T & ... & 1, 4 \\
\hline
\end{tabular}
\label{tab:young} 
\tablecomments{ The references are 1) this work, 2) \cite{Schneider2017}, 3) \cite{Gagne2026}, 4) Lodieu et al. (submitted).  }
\end{table}

%


\section{\label{sc:concl}Conclusions  }

We have analyzed DR1 spectra of UCDs, starting with the cross-match of the dwarfs from our previous works. We analyzed the differences between the Q1 and the new deeper DR1 data falling in the same fields. And we extended the analysis to other deep fields (auxiliary and calibration) and to the one-visit very large area (2100 deg$^2$) of the Wide Survey.
We searched for new UCDs with spectral types from  mid-L to T directly in the 250 million spectra from DR1 and discovered more than 600  dwarfs.
We also searched for Y dwarfs, analyzed the absence of new discoveries and concluded that the possibility of finding new Y dwarfs in future data releases, with more visits in the deep fields, remains open.

We did the spectral-type classification and gained detail in the classification of peculiar objects by including templates for subdwarfs and low-gravity dwarfs. We found 9 possible subdwarfs. We created a full set  of \Euclid templates from M7 to T8 with the red grism. We used them as a reference for the spectral classification of candidates and their analysis.  
 
 We did the spectral analysis using specific spectral indices for L and T dwarfs, we measured the pEW of the gravity-sensitive absorption doublet K~I~(1.244 /  1.253~$\mu$m). We found 6 new dwarfs with possible low surface gravity.
 We investigated the new data from the blue-grism spectra, compared it with the red one, and extended the spectral analysis for the  EDF-N, EDF-F, and COSMOS fields. The analysis included K~I~absorption doublet (1.169 / 1.178~$\mu$m),  the vanadium oxide (VO at 1060~$\mu$m) and FeH absorption bands (990 and 1200~$\mu$m). 
 We found 5 possible resolved binaries, including one possible metal poor.  We created a catalog with all the objects and all the measurements presented in this paper.
 
 As in our previous work  \citep{Dominguez2025}, we demonstrated that NISP's sensitivity and spectral resolution are sufficient for the study  of UCDs. We analyzed dwarfs down to a magnitude \JE$ \sim 23 $  as a borderline case, and \JE$ \sim 22 $ for the majority of spectral types in our study.  It makes us to foresee the discovery of even fainter UCDs in future data releases, which will continue to gain depth throughout the \Euclid nominal mission. And future observations with the blue grism will allow us  to enlarge the range of the \Euclid  templates to 1206-1892~nm.

This paper is the first of a series in the study of UCDs using \Euclid DR1 spectra. We will continue to analyze the data, and extend it to peculiar cases that require higher resolution NIR spectroscopy by follow-up observations.



\begin{acknowledgments}
Funding for CDT, SMT, NS,  M{\v Z},  TS, ELM, KB was provided by the European Union (ERC Advanced Grant, SUBSTELLAR, project number 101054354).
VB, NL,  PMB acknowledge financial support from the Agencia Estatal de Investigaci\'on (AEI/10.13039/501100011033) of the Ministerio de Ciencia e Innovaci\'on  through the following projects:
 PID2022-137241NB-C41 (for VB), 
  PID2025-176415NB-I00  for NL,
 PMB has made use of the Spanish Virtual Observatory (https://svo.cab.inta-csic.es) project funded by MCIN/AEI/10.13039/501100011033 through grant PID2023-146210NB-I00.
The authors wish to acknowledge the contribution of the IAC High-Performance Computing support team and hardware facilities to the results of this research.
This work has made use of the \Euclid DR1 and  \Euclid Quick Release Q1 data from the \Euclid\/ mission of the European Space Agency (ESA),  \cite{https://doi.org/10.57780/esa-2853f3b}. 
The authors acknowledge the Euclid Consortium,
  the European Space Agency, and a number of agencies and
  institutes that have supported the development of \Euclid, in
  particular
  the Agenzia Spaziale Italiana,
  the Austrian Forschungsf\"orderungsgesellschaft funded through BMK,
  the Belgian Science Policy,
  the Canadian Euclid Consortium,
  the Deutsches Zentrum f\"ur Luft- und Raumfahrt,
  the DTU Space and the Niels Bohr Institute in Denmark,
  the French Centre National d'Etudes Spatiales,
  the Funda\c{c}\~{a}o para a Ci\^{e}ncia e a Tecnologia,
  the Hungarian Academy of Sciences,
  the Ministerio de Ciencia, Innovaci\'{o}n y Universidades,
  the National Aeronautics and Space Administration,
  the National Astronomical Observatory of Japan,
  the Netherlandse Onderzoekschool Voor Astronomie,
  the Norwegian Space Agency,
  the Research Council of Finland,
  the Romanian Space Agency,
  the State Secretariat for Education, Research, and Innovation (SERI) at the Swiss
  Space Office (SSO),
  and the United Kingdom Space Agency.
  A complete and detailed list is available on the \Euclid\ web site
  (\url{www.euclid-ec.org}).
We thank the anonymous referee for their insightful comments which improved the paper.

\end{acknowledgments}





%
\facilities{\Euclid space mission}

\software{SPLAT \citep{SPLAT} (\url{https://splat.physics.ucsd.edu/splat/index.html}, astropy \citep{2022ApJ...935..167A}, TOPCAT \citep{2005ASPC..347...29T},
splinecoeff (\url{github.com/nikolavitas/idl/}), TexShop (\url{https://pages.uoregon.edu/koch/texshop/texshop.html}.
          }



\appendix

\section{\label{sc:catal} Full catalog of UCDs in this work and their measurements }

\begin{table}[h!]
\caption{Description of the columns for the catalog to be upload to CDS.}
\centering
\begin{tabular}{lllll}
\hline\hline
Parameter & Format & Unit & Comment & Example  \\ 
\hline
ALIAS & a14 &  ... & Alias name & E51657931  \\ 
DESIGNATION & a27 &  ... & IAU name  & Euclid DR1 J032637.903-264051.71  \\ 
OBJECT\_ID  & i20 &  ... & Euclid DR1 identifier  & -516579318266810311  \\ 
RIGHT ASCENSION  & f11.7 &  deg & Right Ascension & 51.6579318404208  \\ 
DECLINATION  & f11.7 &  deg & Declination & -26.6810311488915  \\ 
SPT  & a10 &  ... & SpT from this work  & M7  \\ 
\IE & f6.3 & AB & \IE magnitude (2 FWHM aperture) & 22.93  \\ 
\IE ERR & f6.3 & AB & \IE magnitude error & 0.02  \\ 
\YE & f6.3 & AB & \YE magnitude (2 FWHM aperture) & 20.17  \\ 
\YE ERR & f6.3 & AB & \YE magnitude error & 0.02  \\ 
\JE & f6.3 & AB & \JE magnitude (2 FWHM aperture) & 19.87  \\ 
\JE ERR & f6.3 & AB & \JE magnitude error & 0.02  \\ 
\HE & f6.3 & AB & \HE magnitude (2 FWHM aperture) & 19.74  \\ 
\HE ERR & f6.3 & AB & \HE magnitude error & 0.02  \\ 
SP DIST  & f7.1 &  pc  & Spectro-photometric distance & 452.5  \\ 
SP DIST ERR  & f7.1 &  pc  & Spectro-photometric distance error & 15.3  \\ 
\hline
\end{tabular}
\end{table}

\section{\label{sc:Q1600} UCDs from M7 to L3   from previous works updated}

Table \ref{tab:Q1600} shows the list of Q1 UCDs with updated spectral types from M7 to L3.

\begin{table}[h!]
\digitalasset
\caption{UCDs  from previous works with updated spectral types from M7 to L3.}
\centering
\begin{tabular}{llrrlrrrrr}
\hline\hline
Alias   &  Designation  &     R.A. (J2000)   &    Dec. (J2000)   &   SpT       &   \IE    &  \YE  & \JE  & \HE &  SP Dist. \\
    &    \texttt{Euclid DR1 +}        &      &       &      &     (AB)   &  (AB)    &  (AB)    &  (AB) &  (pc)   \\ 
\hline
  E50775779 & J032306.187-275210.51 & 50.7757798 & -27.8695851 & L1 & 22.89 & 20.12 & 19.82 & 19.72 & 252.3\\
  E50847588 & J032323.421-270238.97 & 50.8475888 & -27.0441590 & M8 & 21.5 & 19.02 & 18.77 & 18.69 & 244.9\\
  E50942246 & J032346.139-281144.76 & 50.9422466 & -28.1957656 & M8 & 23.74 & 21.16 & 20.89 & 20.77 & 643.7\\
  E50991044 & J032357.850-275820.27 & 50.9910445 & -27.9722964 & M8 & 22.55 & 20.14 & 19.86 & 19.72 & 399.5\\
  E51145027 & J032434.806-283316.84 & 51.1450275 & -28.5546775 & L1 & 23.57 & 21.08 & 20.74 & 20.56 & 378.5\\
  E51152091 & J032436.501-272653.63 & 51.1520912 & -27.4482301 & T0 & 22.83 & 19.61 & 19.11 & 18.75 & 53.1\\
  E51164169 & J032439.400-283249.09 & 51.1641697 & -28.5469690 & M9 & 22.34 & 19.89 & 19.65 & 19.55 & 320.0\\
  E51271385 & J032505.132-270759.69 & 51.2713856 & -27.1332467 & M7 & 23.50 & 20.97 & 20.63 & 20.46 & 637.1\\
  E51301844 & J032512.442-285908.17 & 51.3018448 & -28.9856035 & L1 & 22.25 & 19.46 & 19.16 & 19.04 & 185.6\\
  E51360463 & J032526.511-270213.81 & 51.3604631 & -27.0371701 & M9 & 21.79 & 19.07 & 18.77 & 18.63 & 211.1\\\
     &       &      &       &      &       &     &    &   &   \\ 
\hline
\end{tabular}
\label{tab:Q1600} 
\tablecomments{ This table is published in its entirety in the electronic edition of the {\it Astrophysical Journal}.  A portion is shown here 
for guidance regarding its form and content.  }
\end{table}

\section{\label{sc:splat} Standard templates}       

The list of  standard templates can be found in Table \ref{tab:std}.

\begin{table}[h!]
\caption{Standard templates used in this paper}
\centering
\begin{tabular}{llll}
\hline\hline
SpT  &  Name  &    SpT reference  &  Data reference \\
\hline
M6.0  &  Wolf 359  &    \cite{Kirkpatrick1991}  &  \cite{burgasser2008} \\
M7.0  &  VB 8  &    \cite{Kirkpatrick1991}  &  \cite{burgasser2008} \\
M8.0  &  VB 10  &    \cite{Kirkpatrick1991}  &  \cite{2004AJ....127.2856B} \\
M9.0  &  LHS 2924  &    \cite{Kirkpatrick1991}  &  \cite{2006AJ....131.1007B} \\
L0.0  &  2MASP J0345432+254023  &    \cite{Kirkpatrick99}  &  \cite{2006AJ....131.1007B} \\
L1.0  &  2MASSW J1439284+192915  &   \cite{Kirkpatrick99}  &  \cite{2004AJ....127.2856B} \\
L2.0  &  Kelu-1  &    \cite{Kirkpatrick99}  &  \cite{2007ApJ...658..557B} \\
L3.0  &  2MASSW J1506544+132106  &    \cite{2000AJ....120.1085G}  &  \cite{2007ApJ...659..655B} \\
L4.0  &  2MASS J21580457-1550098  &  \cite{2008ApJ...689.1295K}  &  \cite{2014ApJ...794..143B} \\
L5.0  &  SDSS J083506.16+195304.4  &    \cite{2006AJ....131.2722C}  &  \cite{2006AJ....131.2722C} \\
L6.0  &  2MASSI J1010148-040649  &   \cite{2003AJ....126.2421C}  &  \cite{2006AJ....132..891R} \\
L7.0  &  2MASSI J0103320+193536  &    \cite{2022AandA...664A.111B}  &  \cite{2003AJ....126.2421C} \\
L8.0  &  2MASSW J1632291+190441  &    \cite{Kirkpatrick99}  &  \cite{2004AJ....127.2856B} \\
L9.0  &  DENIS-P J0255-4700  &    \cite{2008ApJ...689.1295K}  &  \cite{Burgasser06} \\
T0.0  &  SDSS J120747.17+024424.8  &    \cite{2002AJ....123.3409H}  &  \cite{2007AJ....134.1162L} \\
T1.0  &  SDSSp J083717.22-000018.3  &    \cite{Burgasser06}  &  \cite{Burgasser06} \\
T2.0  &  SDSSp J125453.90-012247.4  &   \cite{Burgasser06}  &  \cite{2004AJ....127.2856B} \\
T3.0  &  2MASS J12095613-1004008  &    \cite{Burgasser06}  &  \cite{2004AJ....127.2856B} \\
T4.0  &  2MASSI J2254188+312349  &   \cite{Burgasser06}  &  \cite{2004AJ....127.2856B} \\
T5.0  &  2MASS J15031961+2525196  &   \cite{Burgasser06}  &  \cite{2004AJ....127.2856B} \\
T6.0  &  SDSSp J162414.37+002915.6  &   \cite{Burgasser06}  &  \cite{2004AJ....127.2856B} \\
T7.0  &  2MASSI J0727182+171001  &    \cite{Burgasser06}  &  \cite{2004AJ....127.2856B} \\
T8.0  &  2MASSI J0415195-093506  &    \cite{Burgasser06}  &  \cite{2004AJ....127.2856B} \\
T9.0  &  UGPS J072227.51-054031.2  &   \cite{Burgasser06}  & \cite{SPLAT} \\
\hline
\end{tabular}
\label{tab:std} 
\end{table}

\section{\label{sc:jwst} Comparison to JWST templates}       

\begin{figure*}
   \centering
 \includegraphics[width=0.5\linewidth]{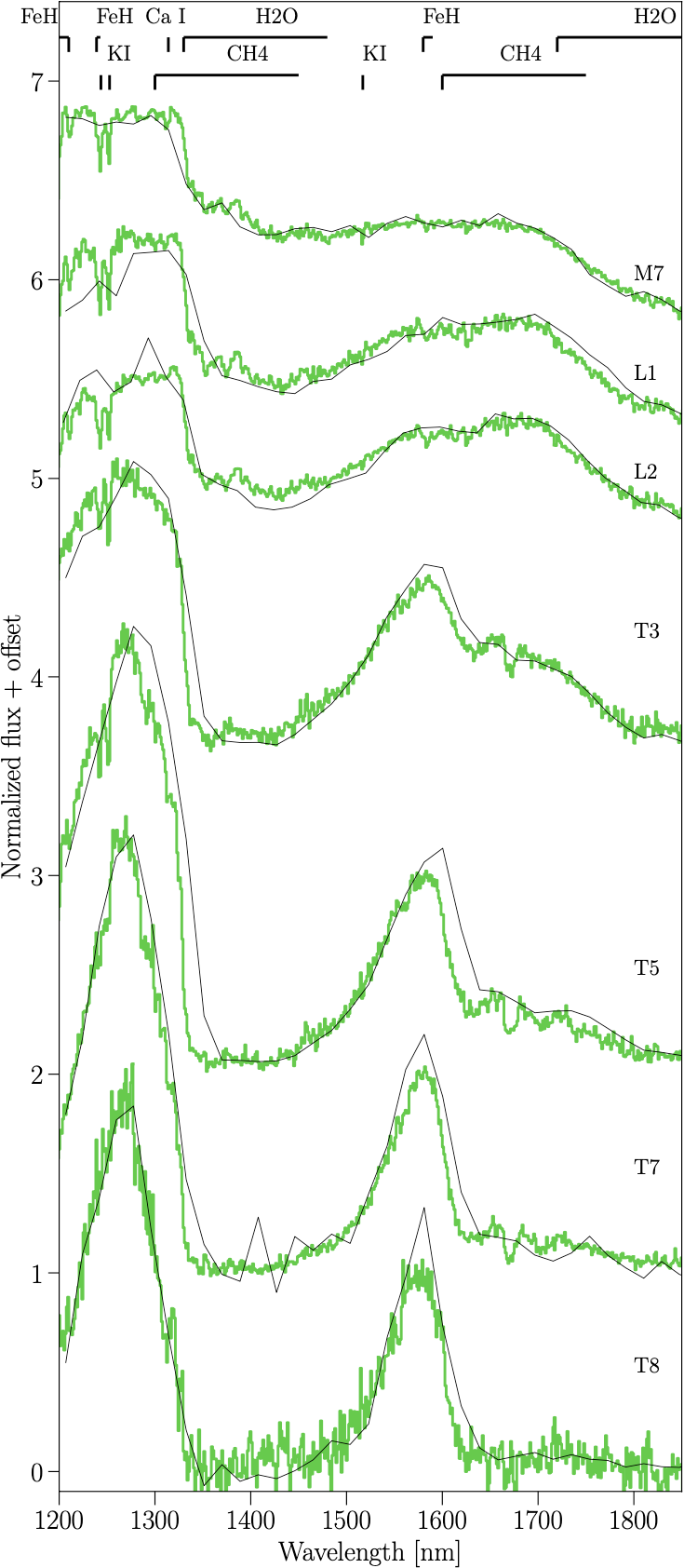}
\caption{EUC templates (green line) for selected spectral subtypes to be compared with the JWST templates by Pritchard et al. (submitted).}
\label{fig:JWSTtemplates}
\end{figure*}

Very recently, Pritchard et al. (submitted) published a set of very low resolution telluric-free templates from JWST/PRISM spectra. Fig.~\ref{fig:JWSTtemplates} shows a comparison with the  EUC templates described in Sect.~\ref{sc:template}. Being the sequence in spectral subtypes from JWST still limited, we have plotted only the spectral types with available data. We can see in the figure that there are some differences in the width of the H-band peak, in flux of the J-band,  or that the JWST template for subtype T5 is asymmetric with respect to the \Euclid one, which is in deed also noted in the Figure 1a from Pritchard et al.

\bibliography{all}{}

\begin{thebibliography}{}
\expandafter\ifx\csname natexlab\endcsname\relax\def\natexlab#1{#1}\fi
\providecommand{\url}[1]{\href{#1}{#1}}
\providecommand{\dodoi}[1]{doi:~\href{http://doi.org/#1}{\nolinkurl{#1}}}
\providecommand{\doeprint}[1]{\href{http://ascl.net/#1}{\nolinkurl{http://ascl.net/#1}}}
\providecommand{\doarXiv}[1]{\href{https://arxiv.org/abs/#1}{\nolinkurl{https://arxiv.org/abs/#1}}}

\bibitem[{K.~N. {Allers} {\&} M.~C. {Liu}(2013){Allers} \& {Liu}}]{Allers13}
{Allers}, K.~N., \& {Liu}, M.~C. 2013, \bibinfo{title}{{A Near-infrared
  Spectroscopic Study of Young Field Ultracool Dwarfs},} \apj, 772, 79,
  \dodoi{10.1088/0004-637X/772/2/79}

\bibitem[{{}{Astropy Collaboration} {et~al.}(2022){Astropy Collaboration},
  {Price-Whelan}, {Lim}, {Earl}, {Starkman}, {Bradley}, {Shupe}, {Patil},
  {Corrales}, {Brasseur}, {N{\"o}the}, {Donath}, {Tollerud}, {Morris},
  {Ginsburg}, {Vaher}, {Weaver}, {Tocknell}, {Jamieson}, {van Kerkwijk},
  {Robitaille}, {Merry}, {Bachetti}, {G{\"u}nther}, {Aldcroft},
  {Alvarado-Montes}, {Archibald}, {B{\'o}di}, {Bapat}, {Barentsen},
  {Baz{\'a}n}, {Biswas}, {Boquien}, {Burke}, {Cara}, {Cara}, {Conroy},
  {Conseil}, {Craig}, {Cross}, {Cruz}, {D'Eugenio}, {Dencheva}, {Devillepoix},
  {Dietrich}, {Eigenbrot}, {Erben}, {Ferreira}, {Foreman-Mackey}, {Fox},
  {Freij}, {Garg}, {Geda}, {Glattly}, {Gondhalekar}, {Gordon}, {Grant},
  {Greenfield}, {Groener}, {Guest}, {Gurovich}, {Handberg}, {Hart},
  {Hatfield-Dodds}, {Homeier}, {Hosseinzadeh}, {Jenness}, {Jones}, {Joseph},
  {Kalmbach}, {Karamehmetoglu}, {Ka{\l}uszy{\'n}ski}, {Kelley}, {Kern},
  {Kerzendorf}, {Koch}, {Kulumani}, {Lee}, {Ly}, {Ma}, {MacBride}, {Maljaars},
  {Muna}, {Murphy}, {Norman}, {O'Steen}, {Oman}, {Pacifici}, {Pascual},
  {Pascual-Granado}, {Patil}, {Perren}, {Pickering}, {Rastogi}, {Roulston},
  {Ryan}, {Rykoff}, {Sabater}, {Sakurikar}, {Salgado}, {Sanghi}, {Saunders},
  {Savchenko}, {Schwardt}, {Seifert-Eckert}, {Shih}, {Jain}, {Shukla}, {Sick},
  {Simpson}, {Singanamalla}, {Singer}, {Singhal}, {Sinha}, {Sip{\H{o}}cz},
  {Spitler}, {Stansby}, {Streicher}, {{\v{S}}umak}, {Swinbank}, {Taranu},
  {Tewary}, {Tremblay}, {de Val-Borro}, {Van Kooten}, {Vasovi{\'c}}, {Verma},
  {de Miranda Cardoso}, {Williams}, {Wilson}, {Winkel}, {Wood-Vasey}, {Xue},
  {Yoachim}, {Zhang}, {Zonca}, \& {Astropy Project
  Contributors}}]{2022ApJ...935..167A}
{Astropy Collaboration}, {Price-Whelan}, A.~M., {Lim}, P.~L., {et~al.} 2022,
  \bibinfo{title}{{The Astropy Project: Sustaining and Growing a
  Community-oriented Open-source Project and the Latest Major Release (v5.0) of
  the Core Package},} \apj, 935, 167, \dodoi{10.3847/1538-4357/ac7c74}

\bibitem[{D.~C. {Bardalez Gagliuffi} {et~al.}(2014{\natexlab{a}}){Bardalez
  Gagliuffi}, {Burgasser}, {Gelino}, {Looper}, {Nicholls}, {Schmidt}, {Cruz},
  {West}, {Gizis}, \& {Metchev}}]{Bardalez2014}
{Bardalez Gagliuffi}, D.~C., {Burgasser}, A.~J., {Gelino}, C.~R., {et~al.}
  2014{\natexlab{a}}, \bibinfo{title}{{SpeX Spectroscopy of Unresolved Very Low
  Mass Binaries. II. Identification of 14 Candidate Binaries with
  Late-M/Early-L and T Dwarf Components},} \apj, 794, 143,
  \dodoi{10.1088/0004-637X/794/2/143}

\bibitem[{D.~C. {Bardalez Gagliuffi} {et~al.}(2014{\natexlab{b}}){Bardalez
  Gagliuffi}, {Burgasser}, {Gelino}, {Looper}, {Nicholls}, {Schmidt}, {Cruz},
  {West}, {Gizis}, \& {Metchev}}]{2014ApJ...794..143B}
{Bardalez Gagliuffi}, D.~C., {Burgasser}, A.~J., {Gelino}, C.~R., {et~al.}
  2014{\natexlab{b}}, \bibinfo{title}{{SpeX Spectroscopy of Unresolved Very Low
  Mass Binaries. II. Identification of 14 Candidate Binaries with
  Late-M/Early-L and T Dwarf Components},} \apj, 794, 143,
  \dodoi{10.1088/0004-637X/794/2/143}

\bibitem[{H. {Bouy} {et~al.}(2022){Bouy}, {Tamura}, {Barrado}, {Motohara},
  {Castro Rodr{\'\i}guez}, {Miret-Roig}, {Konishi}, {Koyama}, {Takahashi},
  {Hu{\'e}lamo}, {Bertin}, {Olivares}, {Sarro}, {Berihuete}, {Cuillandre},
  {Galli}, {Yoshii}, \& {Miyata}}]{2022AandA...664A.111B}
{Bouy}, H., {Tamura}, M., {Barrado}, D., {et~al.} 2022,
  \bibinfo{title}{{Infrared spectroscopy of free-floating planet candidates in
  Upper Scorpius and Ophiuchus},} \aap, 664, A111,
  \dodoi{10.1051/0004-6361/202243850}

\bibitem[{A. {Bravo} {et~al.}(2023){Bravo}, {Schneider}, {Bardalez Gagliuffi},
  {Burgasser}, {Meisner}, {Kirkpatrick}, {Faherty}, {Kuchner}, {Caselden},
  {Sainio}, {Hamlet}, \& {Backyard Worlds: Planet 9 Collaboration}}]{Bravo2023}
{Bravo}, A., {Schneider}, A.~C., {Bardalez Gagliuffi}, D., {et~al.} 2023,
  \bibinfo{title}{{An Investigation of New Brown Dwarf Spectral Binary
  Candidates From the Backyard Worlds: Planet 9 Citizen Science Initiative},}
  \aj, 166, 226, \dodoi{10.3847/1538-3881/acffc1}

\bibitem[{A.~J. {Burgasser}(2007){Burgasser}}]{2007ApJ...659..655B}
{Burgasser}, A.~J. 2007, \bibinfo{title}{{Binaries and the L Dwarf/T Dwarf
  Transition},} \apj, 659, 655, \dodoi{10.1086/511027}

\bibitem[{A.~J. {Burgasser} {et~al.}(2006){Burgasser}, {Geballe}, {Leggett},
  {Kirkpatrick}, \& {Golimowski}}]{Burgasser06}
{Burgasser}, A.~J., {Geballe}, T.~R., {Leggett}, S.~K., {Kirkpatrick}, J.~D.,
  \& {Golimowski}, D.~A. 2006, \bibinfo{title}{{A Unified Near-Infrared
  Spectral Classification Scheme for T Dwarfs},} \apj, 637, 1067,
  \dodoi{10.1086/498563}

\bibitem[{A.~J. {Burgasser} {et~al.}(2008){Burgasser}, {Liu}, {Ireland},
  {Cruz}, \& {Dupuy}}]{burgasser2008}
{Burgasser}, A.~J., {Liu}, M.~C., {Ireland}, M.~J., {Cruz}, K.~L., \& {Dupuy},
  T.~J. 2008, \bibinfo{title}{{Subtle Signatures of Multiplicity in Late-type
  Dwarf Spectra: The Unresolved M8.5 + T5 Binary 2MASS J03202839-0446358},}
  \apj, 681, 579, \dodoi{10.1086/588379}

\bibitem[{A.~J. {Burgasser} {et~al.}(2007){Burgasser}, {Looper}, {Kirkpatrick},
  \& {Liu}}]{2007ApJ...658..557B}
{Burgasser}, A.~J., {Looper}, D.~L., {Kirkpatrick}, J.~D., \& {Liu}, M.~C.
  2007, \bibinfo{title}{{Discovery of a High Proper Motion L Dwarf Binary:
  2MASS J15200224-4422419AB},} \apj, 658, 557, \dodoi{10.1086/511518}

\bibitem[{A.~J. {Burgasser} {\&} M.~W. {McElwain}(2006){Burgasser} \&
  {McElwain}}]{2006AJ....131.1007B}
{Burgasser}, A.~J., \& {McElwain}, M.~W. 2006, \bibinfo{title}{{Resolved
  Spectroscopy of M Dwarf/L Dwarf Binaries. I. DENIS J220002.05-303832.9AB},}
  \aj, 131, 1007, \dodoi{10.1086/499042}

\bibitem[{A.~J. {Burgasser} {et~al.}(2004){Burgasser}, {McElwain},
  {Kirkpatrick}, {Cruz}, {Tinney}, \& {Reid}}]{2004AJ....127.2856B}
{Burgasser}, A.~J., {McElwain}, M.~W., {Kirkpatrick}, J.~D., {et~al.} 2004,
  \bibinfo{title}{{The 2MASS Wide-Field T Dwarf Search. III. Seven New T Dwarfs
  and Other Cool Dwarf Discoveries},} \aj, 127, 2856, \dodoi{10.1086/383549}

\bibitem[{A.~J. {Burgasser} {\&} {}{Splat Development Team}(2017){Burgasser} \&
  {Splat Development Team}}]{SPLAT}
{Burgasser}, A.~J., \& {Splat Development Team}. 2017, \bibinfo{title}{{The
  SpeX Prism Library Analysis Toolkit (SPLAT): A Data Curation Model},} in
  Astronomical Society of India Conference Series, Vol.~14, Astronomical
  Society of India Conference Series, 7--12, \dodoi{10.48550/arXiv.1707.00062}

\bibitem[{A.~J. {Burgasser} {et~al.}(2002){Burgasser}, {Kirkpatrick}, {Brown},
  {Reid}, {Burrows}, {Liebert}, {Matthews}, {Gizis}, {Dahn}, {Monet}, {Cutri},
  \& {Skrutskie}}]{Burgasser02}
{Burgasser}, A.~J., {Kirkpatrick}, J.~D., {Brown}, M.~E., {et~al.} 2002,
  \bibinfo{title}{{The Spectra of T Dwarfs. I. Near-Infrared Data and Spectral
  Classification},} \apj, 564, 421, \dodoi{10.1086/324033}

\bibitem[{A.~J. {Burgasser} {et~al.}(2025){Burgasser}, {Schneider}, {Meisner},
  {Caselden}, {Hsu}, {Gerasimov}, {Aganze}, {Softich}, {Karpoor}, {Theissen},
  {Brooks}, {Bickle}, {Gagn{\'e}}, {Artigau}, {Marsset}, {Rothermich},
  {Faherty}, {Kirkpatrick}, {Kuchner}, {Andersen}, {Beaulieu}, {Colin},
  {Gantier}, {Gramaize}, {Hamlet}, {Hinckley}, {Kabatnik}, {Kiwy}, {Martin},
  {Massat}, {Pendrill}, {Sainio}, {Sch{\"u}mann}, {Th{\'e}venot}, {Walla},
  {W{\k{e}}dracki}, \& {Backyard Worlds: Planet 9
  Collaboration}}]{Burgasser2025}
{Burgasser}, A.~J., {Schneider}, A.~C., {Meisner}, A.~M., {et~al.} 2025,
  \bibinfo{title}{{New Cold Subdwarf Discoveries from Backyard Worlds and a
  Metallicity Classification System for T Subdwarfs},} \apj, 982, 79,
  \dodoi{10.3847/1538-4357/adb39f}

\bibitem[{K. {Chiu} {et~al.}(2006){Chiu}, {Fan}, {Leggett}, {Golimowski},
  {Zheng}, {Geballe}, {Schneider}, \& {Brinkmann}}]{2006AJ....131.2722C}
{Chiu}, K., {Fan}, X., {Leggett}, S.~K., {et~al.} 2006,
  \bibinfo{title}{{Seventy-One New L and T Dwarfs from the Sloan Digital Sky
  Survey},} \aj, 131, 2722, \dodoi{10.1086/501431}

\bibitem[{K.~L. {Cruz} {et~al.}(2003){Cruz}, {Reid}, {Liebert}, {Kirkpatrick},
  \& {Lowrance}}]{2003AJ....126.2421C}
{Cruz}, K.~L., {Reid}, I.~N., {Liebert}, J., {Kirkpatrick}, J.~D., \&
  {Lowrance}, P.~J. 2003, \bibinfo{title}{{Meeting the Cool Neighbors. V. A
  2MASS-Selected Sample of Ultracool Dwarfs},} \aj, 126, 2421,
  \dodoi{10.1086/378607}

\bibitem[{M.~C. {Cushing} {et~al.}(2011){Cushing}, {Kirkpatrick}, {Gelino},
  {Griffith}, {Skrutskie}, {Mainzer}, {Marsh}, {Beichman}, {Burgasser},
  {Prato}, {Simcoe}, {Marley}, {Saumon}, {Freedman}, {Eisenhardt}, \&
  {Wright}}]{Cushing11}
{Cushing}, M.~C., {Kirkpatrick}, J.~D., {Gelino}, C.~R., {et~al.} 2011,
  \bibinfo{title}{{The Discovery of Y Dwarfs using Data from the Wide-field
  Infrared Survey Explorer (WISE)},} \apj, 743, 50,
  \dodoi{10.1088/0004-637X/743/1/50}

\bibitem[{P. {Delorme} {et~al.}(2008){Delorme}, {Delfosse}, {Albert},
  {Artigau}, {Forveille}, {Reyl{\'e}}, {Allard}, {Homeier}, {Robin}, {Willott},
  {Liu}, \& {Dupuy}}]{Delorme08}
{Delorme}, P., {Delfosse}, X., {Albert}, L., {et~al.} 2008,
  \bibinfo{title}{{CFBDS J005910.90-011401.3: reaching the T-Y brown dwarf
  transition?},} \aap, 482, 961, \dodoi{10.1051/0004-6361:20079317}

\bibitem[{C. {Dominguez-Tagle} {et~al.}(2025){Dominguez-Tagle}, {{\v{Z}}erjal},
  {Sedighi}, {Mas-Buitrago}, {Martin}, {Zhang}, {Vitas}, {B{\'e}jar}, {Tsilia},
  {Mu{\~n}oz Torres}, {Lodieu}, {Barrado}, {Solano}, {Cruz}, {Tata},
  {Phan-Bao}, \& {Burgasser}}]{Dominguez2025}
{Dominguez-Tagle}, C., {{\v{Z}}erjal}, M., {Sedighi}, N., {et~al.} 2025,
  \bibinfo{title}{{Euclid Quick Data Release (Q1){\textemdash}Spectroscopic
  Search, Classification, and Analysis of Ultracool Dwarfs in the Deep
  Fields},} \apj, 991, 84, \dodoi{10.3847/1538-4357/adf72d}

\bibitem[{J.~D. {Draxl Giannoni} {et~al.}(2026){Draxl Giannoni}, {Desai},
  {Burgasser}, {Dunning}, {Aganze}, {McDermott}, {Theissen}, \& {Bardalez
  Gagliuffi}}]{BinML2026}
{Draxl Giannoni}, J.~D., {Desai}, M., {Burgasser}, A.~J., {et~al.} 2026,
  \bibinfo{title}{{Identifying and Characterizing Very Low-mass Spectral Blend
  Binaries with Machine Learning Methods},} \aj, 171, 118,
  \dodoi{10.3847/1538-3881/ae3086}

\bibitem[{P.~R. {Eisenhardt} {et~al.}(2004){Eisenhardt}, {Wright}, \& {WISE
  Team}}]{WISE}
{Eisenhardt}, P.~R., {Wright}, E.~L., \& {WISE Team}. 2004,
  \bibinfo{title}{{The Wide-field Infrared Survey Explorer},} in American
  Astronomical Society Meeting Abstracts, Vol. 205, American Astronomical
  Society Meeting Abstracts, 05.08

\bibitem[{{}{Euclid Collaboration} {et~al.}(2026){Euclid Collaboration},
  {Aussel}, {Tereno}, {Schirmer}, {Alguero}, {Altieri}, {Balbinot}, {de Boer},
  {Casenove}, {Corcho-Caballero}, {Furusawa}, {Furusawa}, {Hudson}, {Jahnke},
  {Libet}, {Macias-Perez}, {Masoumzadeh}, {Mohr}, {Odier}, {Scott}, {Vassallo},
  {Verdoes Kleijn}, {Zacchei}, {Aghanim}, {Amara}, {Andreon}, {Auricchio},
  {Awan}, {Azzollini}, {Baccigalupi}, {Baldi}, {Balestra}, {Bardelli},
  {Basset}, {Battaglia}, {Belikov}, {Bender}, {Biviano}, {Bonchi}, {Bonino},
  {Branchini}, {Brescia}, {Brinchmann}, {Camera}, {Ca{\~n}as-Herrera},
  {Capobianco}, {Carbone}, {Cardone}, {Carretero}, {Casas}, {Castander},
  {Castellano}, {Castignani}, {Cavuoti}, {Chambers}, {Cimatti},
  {Colodro-Conde}, {Congedo}, {Conselice}, {Conversi}, {Copin}, {Courbin},
  {Courtois}, {Cropper}, {Cuby}, {Da Silva}, {da Silva}, {Degaudenzi}, {de
  Jong}, {De Lucia}, {Di Giorgio}, {Dinis}, {Dolding}, {Dole}, {Douspis},
  {Dubath}, {Duncan}, {Dupac}, {Dusini}, {Ealet}, {Escoffier}, {Fabricius},
  {Farina}, {Farinelli}, {Faustini}, {Ferriol}, {Fotopoulou}, {Fourmanoit},
  {Frailis}, {Franceschi}, {Franzetti}, {Galeotta}, {George}, {Gillard},
  {Gillis}, {Giocoli}, {G{\'o}mez-Alvarez}, {Gracia-Carpio}, {Granett},
  {Grazian}, {Grupp}, {Guzzo}, {Gwyn}, {Haugan}, {Herent}, {Hoar}, {Hoekstra},
  {Holliman}, {Holmes}, {Hook}, {Hormuth}, {Hornstrup}, {Hudelot}, {Ili{\'c}},
  {Jhabvala}, {Joachimi}, {Keih{\"a}nen}, {Kermiche}, {Kiessling}, {Kubik},
  {Kuijken}, {K{\"u}mmel}, {Kunz}, {Kurki-Suonio}, {Lahav}, {Le Boulc'h}, {Le
  Brun}, {Le Mignant}, {Liebing}, {Ligori}, {Lilje}, {Lindholm}, {Lloro},
  {Mainetti}, {Maino}, {Maiorano}, {Mansutti}, {Marcin}, {Marggraf},
  {Markovic}, {Martinelli}, {Martinet}, {Marulli}, {Massey}, {Maurogordato},
  {McCracken}, {Medinaceli}, {Mei}, {Melchior}, {Mellier}, {Meneghetti},
  {Merlin}, {Meylan}, {Mora}, {Moresco}, {Morris}, {Moscardini}, {Mourre},
  {Nakajima}, {Neissner}, {Nichol}, {Niemi}, {Nightingale}, {Nutma}, {Padilla},
  {Paltani}, {Pasian}, {Peacock}, {Pedersen}, {Percival}, {Pettorino}, {Pires},
  {Polenta}, {Pollack}, {Poncet}, {Popa}, {Pozzetti}, {Racca}, {Raison},
  {Rebolo}, {Renzi}, {Rhodes}, {Riccio}, {Rix}, {Romelli}, {Roncarelli},
  {Rossetti}, {Rusholme}, {Saglia}, {Sakr}, {S{\'a}nchez}, {Sapone},
  {Sartoris}, {Sauvage}, {Schewtschenko}, {Schneider}, {Scodeggio}, {Secroun},
  {Sefusatti}, \& {Seidel}}]{Q1}
{Euclid Collaboration}, {Aussel}, H., {Tereno}, I., {et~al.} 2026,
  \bibinfo{title}{{Euclid Quick Data Release (Q1): I. Data release overview},}
  \aap, 711, A1, \dodoi{10.1051/0004-6361/202554610}

\bibitem[{{}{Euclid Collaboration: F. Passalacqua} {et~al.}(2026){Euclid
  Collaboration: F. Passalacqua}, {Anselmi}, {Monaco}, \&
  ohters}]{Passalacqua2026}
{Euclid Collaboration: F. Passalacqua}, {Anselmi}, S., {Monaco}, P., \& ohters.
  2026, \bibinfo{title}{{Euclid preparation. First investigation of the impact
  of cross-contamination on spectroscopic redshift measurements with
  pixel-level simulations},} arXiv e-prints, arXiv:2606.24697,
  \dodoi{10.48550/arXiv.2606.24697}

\bibitem[{{}{Euclid Collaboration: K. Jahnke} {et~al.}(2025){Euclid
  Collaboration: K. Jahnke}, {Gillard}, {Schirmer}, {et~al.}}]{EuclidSkyNISP}
{Euclid Collaboration: K. Jahnke}, {Gillard}, W., {Schirmer}, M., {et~al.}
  2025, \bibinfo{title}{Euclid - III. The NISP Instrument,} A\&A, 697, A3,
  \dodoi{10.1051/0004-6361/202450786}

\bibitem[{{}{Euclid Collaboration: M. Cropper} {et~al.}(2025){Euclid
  Collaboration: M. Cropper}, {Al-Bahlawan}, {Amiaux}, {et~al.}}]{EuclidSkyVIS}
{Euclid Collaboration: M. Cropper}, {Al-Bahlawan}, A., {Amiaux}, J., {et~al.}
  2025, \bibinfo{title}{Euclid - II. The VIS instrument,} A\&A, 697, A2,
  \dodoi{10.1051/0004-6361/202450996}

\bibitem[{{}{Euclid Collaboration: R. Scaramella} {et~al.}(2022){Euclid
  Collaboration: R. Scaramella}, {Amiaux}, {Mellier},
  {et~al.}}]{Scaramella-EP1}
{Euclid Collaboration: R. Scaramella}, {Amiaux}, J., {Mellier}, Y., {et~al.}
  2022, \bibinfo{title}{{Euclid preparation. I. The Euclid Wide Survey},} \aap,
  662, A112, \dodoi{10.1051/0004-6361/202141938}

\bibitem[{{}{Euclid Collaboration: V. Le Brun} {et~al.}(2026){Euclid
  Collaboration: V. Le Brun}, {Bethermin}, {Moresco}, \& others.}]{LeBrun2026}
{Euclid Collaboration: V. Le Brun}, {Bethermin}, M., {Moresco}, M., \& others.
  2026, \bibinfo{title}{{Euclid Quick Data Release (Q1): VII. Characteristics
  and limitations of the spectroscopic measurements},} \aap, 711, A7,
  \dodoi{10.1051/0004-6361/202554607}

\bibitem[{{}{Euclid Collaboration: Y. Copin} {et~al.}(2026){Euclid
  Collaboration: Y. Copin} {et~al.}}]{Copin2026}
{Euclid Collaboration: Y. Copin}, {et~al.} 2026, \bibinfo{title}{{Euclid Quick
  Data Release (Q1): VI. From spectrograms to spectra: The SIR spectroscopic
  processing function},} \aap, 711, A6, \dodoi{10.1051/0004-6361/202554627}

\bibitem[{{}{Euclid Collaboration: Y. Mellier} {et~al.}(2024){Euclid
  Collaboration: Y. Mellier} {et~al.}}]{EuclidSkyOverview}
{Euclid Collaboration: Y. Mellier}, {et~al.} 2024, \bibinfo{title}{{\Euclid. I.
  Overview of the \Euclid mission},} \aap, accepted, arXiv:2405.13491.
\newblock \doarXiv{2405.13491}

\bibitem[{{}{European Space Agency} {et~al.}(2025){European Space Agency},
  {European Space Agency}, \& { Euclid
  Consortium}}]{https://doi.org/10.57780/esa-2853f3b}
{European Space Agency}, {European Space Agency}, \& { Euclid Consortium}.
  2025, Euclid Quick Release (Q1), European Space Agency,
  \dodoi{10.57780/ESA-2853F3B}

\bibitem[{J. {Gagn{\'e}} {et~al.}(2026){Gagn{\'e}}, {Faherty}, {Ruiz Diaz},
  {Coulombe}, {Bickle}, {Schneider}, {Kirkpatrick}, {Kuchner}, {Meisner},
  {Caselden}, {Burgasser}, {Casewell}, {Honaker}, {Kiwy}, {Marocco}, {Bardalez
  Gagliuffi}, {Stevnbak Andersen}, {Ruiz Arroyo}, {Baller}, {Beaulieu}, {Bell},
  {Bilsing}, {Bohling}, {Colin}, {Colombo}, {Deen}, {Dereveanco}, {Dixon},
  {Durantini Luca}, {Flores}, {Franck}, {Fulvi}, {Gallmann}, {Gantier},
  {Glebov}, {Gramaize}, {Hamlet}, {Hinckley}, {Jablonski}, {Ja{\l}owiczor},
  {Kabatnik}, {Kasprowitz}, {Ly}, {Martin}, {Marzak}, {McColgan}, {McEwan},
  {Michaels}, {Pendrill}, {Perlin}, {Pumphrey}, {Rabe}, {Raway}, {Robledo},
  {Roser}, {Roy}, {Sainio}, {Schindler}, {Schonau}, {Sch{\"u}mann},
  {Selg-Mann}, {Serio}, {Smith}, {Stenner}, {Tanner}, {Th{\'e}venot}, {Thakur},
  {Torres Guerrero}, {Ventura}, {Voloshin}, {Walla}, {W{\c{e}}dracki},
  {Weyandt}, {Wilhite}, \& {Zitouni}}]{Gagne2026}
{Gagn{\'e}}, J., {Faherty}, J.~K., {Ruiz Diaz}, A., {et~al.} 2026,
  \bibinfo{title}{{A SPHEREx Pipeline and Spectral Library for Ultracool
  Dwarfs},} \apj, 1009, 15, \dodoi{10.3847/1538-4357/ae8aed}

\bibitem[{{}{Gaia Collaboration} {et~al.}(2023){Gaia Collaboration},
  {Vallenari}, {Brown}, {Prusti}, {de Bruijne}, {Arenou}, {Babusiaux},
  {Biermann}, {Creevey}, {Ducourant}, {Evans}, {Eyer}, {Guerra}, {Hutton},
  {Jordi}, {Klioner}, {Lammers}, {Lindegren}, {Luri}, {Mignard}, {Panem},
  {Pourbaix}, {Randich}, {Sartoretti}, {Soubiran}, {Tanga}, {Walton},
  {Bailer-Jones}, {Bastian}, {Drimmel}, {Jansen}, {Katz}, {Lattanzi}, {van
  Leeuwen}, {Bakker}, {Cacciari}, {Casta{\~n}eda}, {De Angeli}, {Fabricius},
  {Fouesneau}, {Fr{\'e}mat}, {Galluccio}, {Guerrier}, {Heiter}, {Masana},
  {Messineo}, {Mowlavi}, {Nicolas}, {Nienartowicz}, {Pailler}, {Panuzzo},
  {Riclet}, {Roux}, {Seabroke}, {Sordo}, {Th{\'e}venin}, {Gracia-Abril},
  {Portell}, {Teyssier}, {Altmann}, {Andrae}, {Audard}, {Bellas-Velidis},
  {Benson}, {Berthier}, {Blomme}, {Burgess}, {Busonero}, {Busso},
  {C{\'a}novas}, {Carry}, {Cellino}, {Cheek}, {Clementini}, {Damerdji},
  {Davidson}, {de Teodoro}, {Nu{\~n}ez Campos}, {Delchambre}, {Dell'Oro},
  {Esquej}, {Fern{\'a}ndez-Hern{\'a}ndez}, {Fraile}, {Garabato},
  {Garc{\'\i}a-Lario}, {Gosset}, {Haigron}, {Halbwachs}, {Hambly}, {Harrison},
  {Hern{\'a}ndez}, {Hestroffer}, {Hodgkin}, {Holl}, {Jan{\ss}en}, {Jevardat de
  Fombelle}, {Jordan}, {Krone-Martins}, {Lanzafame}, {L{\"o}ffler}, {Marchal},
  {Marrese}, {Moitinho}, {Muinonen}, {Osborne}, {Pancino}, {Pauwels},
  {Recio-Blanco}, {Reyl{\'e}}, {Riello}, {Rimoldini}, {Roegiers}, {Rybizki},
  {Sarro}, {Siopis}, {Smith}, {Sozzetti}, {Utrilla}, {van Leeuwen}, {Abbas},
  {{\'A}brah{\'a}m}, {Abreu Aramburu}, {Aerts}, {Aguado}, {Ajaj},
  {Aldea-Montero}, {Altavilla}, {{\'A}lvarez}, {Alves}, {Anders}, {Anderson},
  {Anglada Varela}, {Antoja}, {Baines}, {Baker}, {Balaguer-N{\'u}{\~n}ez},
  {Balbinot}, {Balog}, {Barache}, {Barbato}, {Barros}, {Barstow},
  {Bartolom{\'e}}, {Bassilana}, {Bauchet}, {Becciani}, {Bellazzini},
  {Berihuete}, {Bernet}, {Bertone}, {Bianchi}, {Binnenfeld}, {Blanco-Cuaresma},
  {Blazere}, {Boch}, {Bombrun}, {Bossini}, {Bouquillon}, {Bragaglia},
  {Bramante}, {Breedt}, {Bressan}, {Brouillet}, {Brugaletta}, {Bucciarelli},
  {Burlacu}, {Butkevich}, {Buzzi}, {Caffau}, {Cancelliere}, {Cantat-Gaudin},
  {Carballo}, {Carlucci}, {Carnerero}, {Carrasco}, {Casamiquela}, {Castellani},
  {Castro-Ginard}, {Chaoul}, {Charlot}, {Chemin}, {Chiaramida}, {Chiavassa},
  {Chornay}, {Comoretto}, {Contursi}, {Cooper}, {Cornez}, {Cowell}, {Crifo},
  {Cropper}, {Crosta}, {Crowley}, {Dafonte}, {Dapergolas}, {David}, {David},
  {de Laverny}, {De Luise}, \& {De March}}]{Gaia2023}
{Gaia Collaboration}, {Vallenari}, A., {Brown}, A.~G.~A., {et~al.} 2023,
  \bibinfo{title}{{Gaia Data Release 3. Summary of the content and survey
  properties},} \aap, 674, A1, \dodoi{10.1051/0004-6361/202243940}

\bibitem[{T.~R. {Geballe} {et~al.}(2002){Geballe}, {Knapp}, {Leggett}, {Fan},
  {Golimowski}, {Anderson}, {Brinkmann}, {Csabai}, {Gunn}, {Hawley},
  {Hennessy}, {Henry}, {Hill}, {Hindsley}, {Ivezi{\'c}}, {Lupton}, {McDaniel},
  {Munn}, {Narayanan}, {Peng}, {Pier}, {Rockosi}, {Schneider}, {Smith},
  {Strauss}, {Tsvetanov}, {Uomoto}, {York}, \& {Zheng}}]{Geballe02}
{Geballe}, T.~R., {Knapp}, G.~R., {Leggett}, S.~K., {et~al.} 2002,
  \bibinfo{title}{{Toward Spectral Classification of L and T Dwarfs: Infrared
  and Optical Spectroscopy and Analysis},} \apj, 564, 466,
  \dodoi{10.1086/324078}

\bibitem[{J. {Gizis}(2024){Gizis}}]{Gizis2024}
{Gizis}, J. 2024, {An Introduction to Brown Dwarfs; From very-low-mass stars to
  super-Jupiters} (IOP Publishing Ltd)

\bibitem[{J.~E. {Gizis}(1997){Gizis}}]{Gizis1997}
{Gizis}, J.~E. 1997, \bibinfo{title}{{M-Subdwarfs: Spectroscopic Classification
  and the Metallicity Scale},} \aj, 113, 806, \dodoi{10.1086/118302}

\bibitem[{J.~E. {Gizis} {et~al.}(2000){Gizis}, {Monet}, {Reid}, {Kirkpatrick},
  {Liebert}, \& {Williams}}]{2000AJ....120.1085G}
{Gizis}, J.~E., {Monet}, D.~G., {Reid}, I.~N., {et~al.} 2000,
  \bibinfo{title}{{New Neighbors from 2MASS: Activity and Kinematics at the
  Bottom of the Main Sequence},} \aj, 120, 1085, \dodoi{10.1086/301456}

\bibitem[{S.~L. {Hawley} {et~al.}(2002){Hawley}, {Covey}, {Knapp},
  {Golimowski}, {Fan}, {Anderson}, {Gunn}, {Harris}, {Ivezi{\'c}}, {Long},
  {Lupton}, {McGehee}, {Narayanan}, {Peng}, {Schlegel}, {Schneider}, {Spahn},
  {Strauss}, {Szkody}, {Tsvetanov}, {Walkowicz}, {Brinkmann}, {Harvanek},
  {Hennessy}, {Kleinman}, {Krzesinski}, {Long}, {Neilsen}, {Newman}, {Nitta},
  {Snedden}, \& {York}}]{2002AJ....123.3409H}
{Hawley}, S.~L., {Covey}, K.~R., {Knapp}, G.~R., {et~al.} 2002,
  \bibinfo{title}{{Characterization of M, L, and T Dwarfs in the Sloan Digital
  Sky Survey},} \aj, 123, 3409, \dodoi{10.1086/340697}

\bibitem[{J.~D. {Kirkpatrick} {et~al.}(1997){Kirkpatrick}, {Henry}, \&
  {Irwin}}]{Kirkpatrick97}
{Kirkpatrick}, J.~D., {Henry}, T.~J., \& {Irwin}, M.~J. 1997,
  \bibinfo{title}{{Ultra-Cool M Dwarfs Discovered by QSO Surveys.I: The APM
  Objects},} \aj, 113, 1421, \dodoi{10.1086/118357}

\bibitem[{J.~D. {Kirkpatrick} {et~al.}(1991){Kirkpatrick}, {Henry}, \&
  {McCarthy}}]{Kirkpatrick1991}
{Kirkpatrick}, J.~D., {Henry}, T.~J., \& {McCarthy}, Jr., D.~W. 1991,
  \bibinfo{title}{{A Standard Stellar Spectral Sequence in the
  Red/Near-Infrared: Classes K5 to M9},} \apjs, 77, 417, \dodoi{10.1086/191611}

\bibitem[{J.~D. {Kirkpatrick} {et~al.}(1995){Kirkpatrick}, {Henry}, \&
  {Simons}}]{Kirkpatrick95}
{Kirkpatrick}, J.~D., {Henry}, T.~J., \& {Simons}, D.~A. 1995,
  \bibinfo{title}{{The Solar Neighborhood. II. The First List of Dwarfs with
  Spectral Types of M7 and Cooler},} \aj, 109, 797, \dodoi{10.1086/117323}

\bibitem[{J.~D. {Kirkpatrick} {et~al.}(1999){Kirkpatrick}, {Reid}, {Liebert},
  {Cutri}, {Nelson}, {Beichman}, {Dahn}, {Monet}, {Gizis}, \&
  {Skrutskie}}]{Kirkpatrick99}
{Kirkpatrick}, J.~D., {Reid}, I.~N., {Liebert}, J., {et~al.} 1999,
  \bibinfo{title}{{Dwarfs Cooler than ``M``: The Definition of Spectral Type
  ``L'' Using Discoveries from the 2 Micron All-Sky Survey (2MASS)},} \apj,
  519, 802, \dodoi{10.1086/307414}

\bibitem[{J.~D. {Kirkpatrick} {et~al.}(2000){Kirkpatrick}, {Reid}, {Liebert},
  {Gizis}, {Burgasser}, {Monet}, {Dahn}, {Nelson}, \&
  {Williams}}]{Kirkpatrick00}
{Kirkpatrick}, J.~D., {Reid}, I.~N., {Liebert}, J., {et~al.} 2000,
  \bibinfo{title}{{67 Additional L Dwarfs Discovered by the Two Micron All Sky
  Survey},} \aj, 120, 447, \dodoi{10.1086/301427}

\bibitem[{J.~D. {Kirkpatrick} {et~al.}(2008){Kirkpatrick}, {Cruz}, {Barman},
  {Burgasser}, {Looper}, {Tinney}, {Gelino}, {Lowrance}, {Liebert},
  {Carpenter}, {Hillenbrand}, \& {Stauffer}}]{2008ApJ...689.1295K}
{Kirkpatrick}, J.~D., {Cruz}, K.~L., {Barman}, T.~S., {et~al.} 2008,
  \bibinfo{title}{{A Sample of Very Young Field L Dwarfs and Implications for
  the Brown Dwarf ``Lithium Test'' at Early Ages},} \apj, 689, 1295,
  \dodoi{10.1086/592768}

\bibitem[{J.~D. {Kirkpatrick} {et~al.}(2011){Kirkpatrick}, {Cushing}, {Gelino},
  {Griffith}, {Skrutskie}, {Marsh}, {Wright}, {Mainzer}, {Eisenhardt},
  {McLean}, {Thompson}, {Bauer}, {Benford}, {Bridge}, {Lake}, {Petty},
  {Stanford}, {Tsai}, {Bailey}, {Beichman}, {Bloom}, {Bochanski}, {Burgasser},
  {Capak}, {Cruz}, {Hinz}, {Kartaltepe}, {Knox}, {Manohar}, {Masters},
  {Morales-Calder{\'o}n}, {Prato}, {Rodigas}, {Salvato}, {Schurr}, {Scoville},
  {Simcoe}, {Stapelfeldt}, {Stern}, {Stock}, \& {Vacca}}]{Kirkpatrick11}
{Kirkpatrick}, J.~D., {Cushing}, M.~C., {Gelino}, C.~R., {et~al.} 2011,
  \bibinfo{title}{{The First Hundred Brown Dwarfs Discovered by the Wide-field
  Infrared Survey Explorer (WISE)},} \apjs, 197, 19,
  \dodoi{10.1088/0067-0049/197/2/19}

\bibitem[{J.~D. {Kirkpatrick} {et~al.}(2012){Kirkpatrick}, {Gelino}, {Cushing},
  {Mace}, {Griffith}, {Skrutskie}, {Marsh}, {Wright}, {Eisenhardt}, {McLean},
  {Mainzer}, {Burgasser}, {Tinney}, {Parker}, \& {Salter}}]{Kirkpatrick2012}
{Kirkpatrick}, J.~D., {Gelino}, C.~R., {Cushing}, M.~C., {et~al.} 2012,
  \bibinfo{title}{{Further Defining Spectral Type ``Y'' and Exploring the
  Low-mass End of the Field Brown Dwarf Mass Function},} \apj, 753, 156,
  \dodoi{10.1088/0004-637X/753/2/156}

\bibitem[{J.~D. {Kirkpatrick} {et~al.}(2021){Kirkpatrick}, {Gelino}, {Faherty},
  {Meisner}, {Caselden}, {Schneider}, {Marocco}, {Cayago}, {Smart},
  {Eisenhardt}, {Kuchner}, {Wright}, {Cushing}, {Allers}, {Bardalez Gagliuffi},
  {Burgasser}, {Gagn{\'e}}, {Logsdon}, {Martin}, {Ingalls}, {Lowrance},
  {Abrahams}, {Aganze}, {Gerasimov}, {Gonzales}, {Hsu}, {Kamraj}, {Kiman},
  {Rees}, {Theissen}, {Ammar}, {Andersen}, {Beaulieu}, {Colin}, {Elachi},
  {Goodman}, {Gramaize}, {Hamlet}, {Hong}, {Jonkeren}, {Khalil}, {Martin},
  {Pendrill}, {Pumphrey}, {Rothermich}, {Sainio}, {Stenner}, {Tanner},
  {Th{\'e}venot}, {Voloshin}, {Walla}, {W{\k{e}}dracki}, \& {Backyard Worlds:
  Planet 9 Collaboration}}]{Kirkpatrick2021}
{Kirkpatrick}, J.~D., {Gelino}, C.~R., {Faherty}, J.~K., {et~al.} 2021,
  \bibinfo{title}{{The Field Substellar Mass Function Based on the Full-sky 20
  pc Census of 525 L, T, and Y Dwarfs},} \apjs, 253, 7,
  \dodoi{10.3847/1538-4365/abd107}

\bibitem[{F. {Kiwy} {et~al.}(2026){Kiwy}, {Kirkpatrick}, {Schneider},
  {Meisner}, {Faherty}, {Kuchner}, {Bardalez Gagliuffi}, {Casewell}, {Bickle},
  \& {The Backyard Worlds: Planet 9 Collaboration}}]{Kiwy2026}
{Kiwy}, F., {Kirkpatrick}, J.~D., {Schneider}, A.~C., {et~al.} 2026,
  \bibinfo{title}{{A Search for Late-type Brown Dwarfs in the Euclid Quick Data
  Release 1},} \aj, 171, 108, \dodoi{10.3847/1538-3881/ae2fb7}

\bibitem[{A.~M. {Koekemoer} {et~al.}(2007){Koekemoer}, {Aussel}, {Calzetti},
  {Capak}, {Giavalisco}, {Kneib}, {Leauthaud}, {Le F{\`e}vre}, {McCracken},
  {Massey}, {Mobasher}, {Rhodes}, {Scoville}, \& {Shopbell}}]{COSMOS2007}
{Koekemoer}, A.~M., {Aussel}, H., {Calzetti}, D., {et~al.} 2007,
  \bibinfo{title}{{The COSMOS Survey: Hubble Space Telescope Advanced Camera
  for Surveys Observations and Data Processing},} \apjs, 172, 196,
  \dodoi{10.1086/520086}

\bibitem[{P.~M. {Korngut} {et~al.}(2018){Korngut}, {Bock}, {Akeson}, {Ashby},
  {Bleem}, {Boland}, {Bolton}, {Bradford}, {Braun}, {Bryan}, {Capak}, {Chang},
  {Coffey}, {Cooray}, {Crill}, {Dor{\'e}}, {Eifler}, {Feng}, {Habib},
  {Heitmann}, {Hemmati}, {Hirata}, {Jeong}, {Kim}, {Kirkpatrick},
  {Kowalkowski}, {Krause}, {Lisse}, {Mauskopf}, {Masters}, {McGuire},
  {Melnick}, {Nguyen}, {Nayyeri}, {Oberg}, {dePutter}, {Purcell}, {Rocca},
  {Runyan}, {Sandstrom}, {Smith}, {Song}, {Stickley}, {Stober}, {Susca},
  {Teplitz}, {Tolls}, {Unwin}, {Werner}, {Windhorst}, \& {Zemcov}}]{SPHEREx}
{Korngut}, P.~M., {Bock}, J.~J., {Akeson}, R., {et~al.} 2018,
  \bibinfo{title}{{SPHEREx: an all-sky NIR spectral survey},} in Society of
  Photo-Optical Instrumentation Engineers (SPIE) Conference Series, Vol. 10698,
  Space Telescopes and Instrumentation 2018: Optical, Infrared, and Millimeter
  Wave, ed. M.~{Lystrup}, H.~A. {MacEwen}, G.~G. {Fazio}, N.~{Batalha},
  N.~{Siegler}, \& E.~C. {Tong}, 106981U, \dodoi{10.1117/12.2312860}

\bibitem[{S.~K. {Leggett} {et~al.}(2025){Leggett}, {Phillips}, \&
  {Tremblin}}]{Leggett2025}
{Leggett}, S.~K., {Phillips}, M.~W., \& {Tremblin}, P. 2025,
  \bibinfo{title}{{Exploration of the Near-infrared Colors of Cold Y Dwarfs
  from the Ground and Space},} \apj, 991, 193, \dodoi{10.3847/1538-4357/adf21f}

\bibitem[{S. {L{\'e}pine} {et~al.}(2007){L{\'e}pine}, {Rich}, \&
  {Shara}}]{Lepine2007}
{L{\'e}pine}, S., {Rich}, R.~M., \& {Shara}, M.~M. 2007,
  \bibinfo{title}{{Revised Metallicity Classes for Low-Mass Stars: Dwarfs (dM),
  Subdwarfs (sdM), Extreme Subdwarfs (esdM), and Ultrasubdwarfs (usdM)},} \apj,
  669, 1235, \dodoi{10.1086/521614}

\bibitem[{M.~C. {Liu} {et~al.}(2011){Liu}, {Delorme}, {Dupuy}, {Bowler},
  {Albert}, {Artigau}, {Reyl{\'e}}, {Forveille}, \& {Delfosse}}]{Liu2011}
{Liu}, M.~C., {Delorme}, P., {Dupuy}, T.~J., {et~al.} 2011,
  \bibinfo{title}{{CFBDSIR J1458+1013B: A Very Cold (>T10) Brown Dwarf in a
  Binary System},} \apj, 740, 108, \dodoi{10.1088/0004-637X/740/2/108}

\bibitem[{D.~L. {Looper} {et~al.}(2007){Looper}, {Kirkpatrick}, \&
  {Burgasser}}]{2007AJ....134.1162L}
{Looper}, D.~L., {Kirkpatrick}, J.~D., \& {Burgasser}, A.~J. 2007,
  \bibinfo{title}{{Discovery of 11 New T Dwarfs in the Two Micron All Sky
  Survey, Including a Possible L/T Transition Binary},} \aj, 134, 1162,
  \dodoi{10.1086/520645}

\bibitem[{K.~L. {Luhman}(2014){Luhman}}]{Luhman14}
{Luhman}, K.~L. 2014, \bibinfo{title}{{Discovery of a
  \raisebox{-0.5ex}\textasciitilde250 K Brown Dwarf at 2 pc from the Sun},}
  \apjl, 786, L18, \dodoi{10.1088/2041-8205/786/2/L18}

\bibitem[{K.~L. {Luhman} {et~al.}(2011){Luhman}, {Burgasser}, \&
  {Bochanski}}]{Luhman2011}
{Luhman}, K.~L., {Burgasser}, A.~J., \& {Bochanski}, J.~J. 2011,
  \bibinfo{title}{{Discovery of a Candidate for the Coolest Known Brown
  Dwarf},} \apjl, 730, L9, \dodoi{10.1088/2041-8205/730/1/L9}

\bibitem[{K.~L. {Luhman} {et~al.}(2024){Luhman}, {Tremblin}, {Alves de
  Oliveira}, {Birkmann}, {Baraffe}, {Chabrier}, {Manjavacas}, {Parker}, \&
  {Valenti}}]{Luhman24}
{Luhman}, K.~L., {Tremblin}, P., {Alves de Oliveira}, C., {et~al.} 2024,
  \bibinfo{title}{{JWST/NIRSpec Observations of the Coldest Known Brown
  Dwarf},} \aj, 167, 5, \dodoi{10.3847/1538-3881/ad0b72}

\bibitem[{E. {Manjavacas} {et~al.}(2016){Manjavacas}, {Goldman}, {Alcal{\'a}},
  {Zapatero-Osorio}, {B{\'e}jar}, {Homeier}, {Bonnefoy}, {Smart}, {Henning}, \&
  {Allard}}]{Manjavacas2016}
{Manjavacas}, E., {Goldman}, B., {Alcal{\'a}}, J.~M., {et~al.} 2016,
  \bibinfo{title}{{Hunting for brown dwarf binaries and testing atmospheric
  models with X-Shooter},} \mnras, 455, 1341, \dodoi{10.1093/mnras/stv2048}

\bibitem[{E.~C. {Martin} {et~al.}(2017){Martin}, {Mace}, {McLean}, {Logsdon},
  {Rice}, {Kirkpatrick}, {Burgasser}, {McGovern}, \& {Prato}}]{Emily2017}
{Martin}, E.~C., {Mace}, G.~N., {McLean}, I.~S., {et~al.} 2017,
  \bibinfo{title}{{Surface Gravities for 228 M, L, and T Dwarfs in the NIRSPEC
  Brown Dwarf Spectroscopic Survey},} \apj, 838, 73,
  \dodoi{10.3847/1538-4357/aa6338}

\bibitem[{E.~L. {Martin} {et~al.}(1997){Martin}, {Basri}, {Delfosse}, \&
  {Forveille}}]{ege97}
{Martin}, E.~L., {Basri}, G., {Delfosse}, X., \& {Forveille}, T. 1997,
  \bibinfo{title}{{Keck HIRES spectra of the brown dwarf DENIS-P
  J1228.2-1547},} \aap, 327, L29

\bibitem[{E.~L. {Martin} {et~al.}(1998){Martin}, {Basri}, {Zapatero-Osorio},
  {Rebolo}, \& {L{\'o}pez}}]{ege98}
{Martin}, E.~L., {Basri}, G., {Zapatero-Osorio}, M.~R., {Rebolo}, R., \&
  {L{\'o}pez}, R.~J.~G. 1998, \bibinfo{title}{{The First L-Type Brown Dwarf in
  the Pleiades},} \apjl, 507, L41, \dodoi{10.1086/311675}

\bibitem[{E.~L. {Martin} {et~al.}(1999){Martin}, {Delfosse}, {Basri},
  {Goldman}, {Forveille}, \& {Zapatero Osorio}}]{ege99}
{Martin}, E.~L., {Delfosse}, X., {Basri}, G., {et~al.} 1999,
  \bibinfo{title}{{Spectroscopic Classification of Late-M and L Field Dwarfs},}
  \aj, 118, 2466, \dodoi{10.1086/301107}

\bibitem[{E.~L. {Mart{\'\i}n} {et~al.}(2021){Mart{\'\i}n}, {Zhang}, {Esparza},
  {Gracia}, {Rasilla}, {Masseron}, \& {Burgasser}}]{EGE2021amonia}
{Mart{\'\i}n}, E.~L., {Zhang}, J.~Y., {Esparza}, P., {et~al.} 2021,
  \bibinfo{title}{{Ammonia-methane ratios from H-band near-infrared spectra of
  late-T and Y dwarfs},} \aap, 655, L3, \dodoi{10.1051/0004-6361/202142470}

\bibitem[{A. {Mohandasan} {et~al.}(2026){Mohandasan}, {Smart}, {Reyl{\'e}}, {Le
  Brun}, {P{\'e}rez-Garrido}, {Ba{\~n}ados}, {Jones}, {Dupuy}, {Goldman},
  {Casewell}, {Zapatero Osorio}, {Rejkuba}, {Mart{\'\i}n}, {Mu{\~n}oz Torres},
  {Dom{\'\i}nguez-Tagle}, {{\v{Z}}erjal}, {Hu{\'e}lamo}, {Lodieu}, {Cruz},
  {Majidi}, {Rebolo}, {Phillips}, {Zhang}, {Aghanim}, {Altieri}, {Amara},
  {Andreon}, {Auricchio}, {Baccigalupi}, {Baldi}, {Balestra}, {Bardelli},
  {Battaglia}, {Biviano}, {Bonchi}, {Branchini}, {Brescia}, {Brinchmann},
  {Camera}, {Ca{\~n}as-Herrera}, {Capobianco}, {Carbone}, {Carretero}, {Casas},
  {Castellano}, {Castignani}, {Cavuoti}, {Chambers}, {Cimatti},
  {Colodro-Conde}, {Congedo}, {Conselice}, {Conversi}, {Copin}, {Costille},
  {Courbin}, {Courtois}, {Cropper}, {Da Silva}, {Degaudenzi}, {De Lucia},
  {Dole}, {Douspis}, {Dubath}, {Dupac}, {Dusini}, {Escoffier}, {Farina},
  {Faustini}, {Ferriol}, {Fotopoulou}, {Frailis}, {Franceschi}, {Galeotta},
  {George}, {Gillard}, {Gillis}, {Giocoli}, {Gracia-Carpio}, {Granett},
  {Grazian}, {Grupp}, {Haugan}, {Hoar}, {Holmes}, {Hook}, {Hormuth},
  {Hornstrup}, {Jahnke}, {Jhabvala}, {Keih{\"a}nen}, {Kermiche}, {Kiessling},
  {Kubik}, {Kuijken}, {K{\"u}mmel}, {Kunz}, {Kurki-Suonio}, {Le Boulc'h}, {Le
  Brun}, {Le Mignant}, {Ligori}, {Lilje}, {Lindholm}, {Lloro}, {Mainetti},
  {Maino}, {Maiorano}, {Mansutti}, {Marcin}, {Marggraf}, {Martinelli},
  {Martinet}, {Marulli}, {Massey}, {Medinaceli}, {Mei}, {Mellier},
  {Meneghetti}, {Merlin}, {Meylan}, {Mora}, {Moresco}, {Moscardini},
  {Nakajima}, {Neissner}, {Niemi}, {Padilla}, {Paltani}, {Pasian}, {Pedersen},
  {Percival}, {Pettorino}, {Pires}, {Polenta}, {Poncet}, {Popa}, {Pozzetti},
  {Raison}, {Renzi}, {Rhodes}, {Riccio}, {Romelli}, {Roncarelli}, {Saglia},
  {Sakr}, {Sapone}, {Sartoris}, {Schewtschenko}, {Schirmer}, {Schneider},
  {Schrabback}, {Secroun}, {Seidel}, {Serrano}, {Simon}, {Sirignano}, {Sirri},
  {Stanco}, {Steinwagner}, {Surace}, {Tallada-Cresp{\'\i}}, {Taylor}, {Tereno},
  {Toledo-Moreo}, {Torradeflot}, {Tsyganov}, {Tutusaus}, {Valenziano},
  {Valiviita}, {Vassallo}, {Kleijn}, {Veropalumbo}, {Vibert}, {Wang}, {Weller},
  {Zacchei}, {Zamorani}, {Zerbi}, {Zucca}, {Mart{\'\i}n-Fleitas}, \&
  {Scottez}}]{Anjana2026}
{Mohandasan}, A., {Smart}, R.~L., {Reyl{\'e}}, C., {et~al.} 2026,
  \bibinfo{title}{{Euclid: Quick Data Release (Q1) - Ultracool dwarfs in the
  Euclid Deep Fields},} \aap, 713, A245, \dodoi{10.1051/0004-6361/202660157}

\bibitem[{V. {Navarro} {et~al.}(2020){Navarro}, {{\'A}lvarez}, {Del R{\'\i}o},
  {Guerra}, {Mart{\'\i}n-Rubio}, {Diego}, {Ramos}, {Marques}, {Henriques},
  {Pereira}, {Basso}, {H{\'e}liou}, \& {Marseille}}]{DataLabs}
{Navarro}, V., {{\'A}lvarez}, R., {Del R{\'\i}o}, S., {et~al.} 2020,
  \bibinfo{title}{{ESA Datalabs: Multi Mission Science Exploitation and
  Preservation Platform},} in Astronomical Society of the Pacific Conference
  Series, Vol. 527, Astronomical Data Analysis Software and Systems XXIX, ed.
  R.~{Pizzo}, E.~R. {Deul}, J.~D. {Mol}, J.~{de Plaa}, \& H.~{Verkouter}, 287

\bibitem[{M.~W. {Phillips} {et~al.}(2020){Phillips}, {Tremblin}, {Baraffe},
  {Chabrier}, {Allard}, {Spiegelman}, {Goyal}, {Drummond}, \&
  {H{\'e}brard}}]{ATMO}
{Phillips}, M.~W., {Tremblin}, P., {Baraffe}, I., {et~al.} 2020,
  \bibinfo{title}{{A new set of atmosphere and evolution models for cool T-Y
  brown dwarfs and giant exoplanets},} \aap, 637, A38,
  \dodoi{10.1051/0004-6361/201937381}

\bibitem[{J.~T. {Rayner} {et~al.}(2003){Rayner}, {Toomey}, {Onaka}, {Denault},
  {Stahlberger}, {Vacca}, {Cushing}, \& {Wang}}]{spex}
{Rayner}, J.~T., {Toomey}, D.~W., {Onaka}, P.~M., {et~al.} 2003,
  \bibinfo{title}{{SpeX: A Medium-Resolution 0.8-5.5 Micron Spectrograph and
  Imager for the NASA Infrared Telescope Facility},} \pasp, 115, 362

\bibitem[{I.~N. {Reid} {et~al.}(2006){Reid}, {Lewitus}, {Allen}, {Cruz}, \&
  {Burgasser}}]{2006AJ....132..891R}
{Reid}, I.~N., {Lewitus}, E., {Allen}, P.~R., {Cruz}, K.~L., \& {Burgasser},
  A.~J. 2006, \bibinfo{title}{{A Search for Binary Systems among the Nearest L
  Dwarfs},} \aj, 132, 891, \dodoi{10.1086/505626}

\bibitem[{A. {Sanghi} {et~al.}(2024){Sanghi}, {Liu}, {Dupuy}, {Best}, {Siverd},
  \& {Zhang}}]{Sanghi2024}
{Sanghi}, A., {Liu}, M.~C., {Dupuy}, T.~J., {et~al.} 2024,
  \bibinfo{title}{{Ultracool Dwarf Absolute Magnitude Versus Spectral Type
  Relations for Euclid and Roman Near-infrared Filters},} Research Notes of the
  American Astronomical Society, 8, 137, \dodoi{10.3847/2515-5172/ad4cec}

\bibitem[{A.~C. {Schneider} {et~al.}(2017){Schneider}, {Windsor}, {Cushing},
  {Kirkpatrick}, \& {Shkolnik}}]{Schneider2017}
{Schneider}, A.~C., {Windsor}, J., {Cushing}, M.~C., {Kirkpatrick}, J.~D., \&
  {Shkolnik}, E.~L. 2017, \bibinfo{title}{{A 2MASS/AllWISE Search for Extremely
  Red L Dwarfs: The Discovery of Several Likely L Type Members of
  {\ensuremath{\beta}} Pic, AB Dor, Tuc-Hor, Argus, and the Hyades},} \aj, 153,
  196, \dodoi{10.3847/1538-3881/aa6624}

\bibitem[{A.~C. {Schneider} {et~al.}(2015){Schneider}, {Cushing},
  {Kirkpatrick}, {Gelino}, {Mace}, {Wright}, {Eisenhardt}, {Skrutskie},
  {Griffith}, \& {Marsh}}]{Schneider15}
{Schneider}, A.~C., {Cushing}, M.~C., {Kirkpatrick}, J.~D., {et~al.} 2015,
  \bibinfo{title}{{Hubble Space Telescope Spectroscopy of Brown Dwarfs
  Discovered with the Wide-field Infrared Survey Explorer},} \apj, 804, 92,
  \dodoi{10.1088/0004-637X/804/2/92}

\bibitem[{A.~C. {Schneider} {et~al.}(2026){Schneider}, {Kuchner}, {Faherty},
  {Meisner}, {Kirkpatrick}, {Burgasser}, {Bardalez Gagliuffi}, {Bickle},
  {Caselden}, {Casewell}, {Gagn{\'e}}, {Honaker}, {Kiwy}, {Marocco},
  {Rothermich}, {Stevnbak Andersen}, {Ruiz Arroyo}, {Baller}, {Beaulieu},
  {Bell}, {Bilsing}, {Bohling}, {Colin}, {Colombo}, {Deen}, {Dereveanco},
  {Dixon}, {Durantini Luca}, {Flores}, {Frank}, {Fulvi}, {Gallmann}, {Gantier},
  {Glebov}, {Gramaize}, {Hamlet}, {Hinckley}, {Jablonski}, {Ja{\l}owiczor},
  {Kabatnik}, {Kasprowitz}, {Ly}, {Martin}, {Marzak}, {McColgan}, {McEwan},
  {Michaels}, {Pendrill}, {Perlin}, {Pumphrey}, {Rabe}, {Raway}, {Robledo},
  {Roser}, {Roy}, {Sainio}, {Schindler}, {Schonau}, {Sch{\"u}mann},
  {Selg-Mann}, {Serio}, {Sirbescu-Stanley}, {Smith}, {Stenner}, {Sunjoto},
  {Tanner}, {Th{\'e}venot}, {Thakur}, {Torres Guerrero}, {Ventura}, {Voloshin},
  {Walla}, {W{\c{e}}dracki}, {Weyandt}, {Wilhite}, \&
  {Zitouni}}]{Schneider2026}
{Schneider}, A.~C., {Kuchner}, M.~J., {Faherty}, J.~K., {et~al.} 2026,
  \bibinfo{title}{{Three Thousand Motion-confirmed L and T Dwarf Candidates
  from the Backyard Worlds: Planet 9 Citizen Science Project},} \aj, 171, 341,
  \dodoi{10.3847/1538-3881/ae5b60}

\bibitem[{M.~B. {Taylor}(2005){Taylor}}]{2005ASPC..347...29T}
{Taylor}, M.~B. 2005, \bibinfo{title}{{TOPCAT \& STIL: Starlink Table/VOTable
  Processing Software},} in Astronomical Society of the Pacific Conference
  Series, Vol. 347, Astronomical Data Analysis Software and Systems XIV, ed.
  P.~{Shopbell}, M.~{Britton}, \& R.~{Ebert}, 29

\bibitem[{C.~G. {Tinney} {et~al.}(2018){Tinney}, {Kirkpatrick}, {Faherty},
  {Mace}, {Cushing}, {Gelino}, {Burgasser}, {Sheppard}, \&
  {Wright}}]{Tinney2018}
{Tinney}, C.~G., {Kirkpatrick}, J.~D., {Faherty}, J.~K., {et~al.} 2018,
  \bibinfo{title}{{New Y and T Dwarfs from WISE Identified by Methane
  Imaging},} \apjs, 236, 28, \dodoi{10.3847/1538-4365/aabad3}

\bibitem[{Z. {Tu} {et~al.}(2025){Tu}, {Wang}, {Chen}, \& {Liu}}]{Tu2025}
{Tu}, Z., {Wang}, S., {Chen}, X., \& {Liu}, J. 2025, \bibinfo{title}{{A Large
  Sample of JWST/NIRSpec Brown Dwarfs: New Distant Discoveries},} \apjs, 281,
  49, \dodoi{10.3847/1538-4365/ae112e}

\bibitem[{Z. {Tu} {et~al.}(2026){Tu}, {Wang}, {Huang}, {Chen}, \&
  {Liu}}]{Tu2026}
{Tu}, Z., {Wang}, S., {Huang}, H., {Chen}, X., \& {Liu}, J. 2026,
  \bibinfo{title}{{SPHEREx Ultracool Dwarf Spectrophotometric Atlas (SUDA):
  Atmospheric and Fundamental Parameters of Ultracool Dwarfs},} \apj, 1008,
  250, \dodoi{10.3847/1538-4357/ae9082}

\bibitem[{M. {{\v{Z}}erjal} {et~al.}(2026){{\v{Z}}erjal}, {Dominguez-Tagle},
  {Vitas}, {Sedighi}, {Mart{\'\i}n}, {Zapatero Osorio}, {Olivares}, {Mu{\~n}oz
  Torres}, {Tsilia}, {Zhang}, {Barrado}, {B{\'e}jar}, {Bouy}, {Burgasser},
  {Cruz}, {Lodieu}, {Mas Buitrago}, {Phan-Bao}, {Solano}, {Tata}, {Goldman},
  {Mohandasan}, {Reyl{\'e}}, {Smart}, {Aghanim}, {Altieri}, {Amara}, {Andreon},
  {Auricchio}, {Baccigalupi}, {Baldi}, {Balestra}, {Bardelli}, {Battaglia},
  {Biviano}, {Bonchi}, {Branchini}, {Brescia}, {Brinchmann}, {Camera},
  {Ca{\~n}as-Herrera}, {Capobianco}, {Carbone}, {Carretero}, {Casas},
  {Castellano}, {Castignani}, {Cavuoti}, {Chambers}, {Cimatti},
  {Colodro-Conde}, {Congedo}, {Conselice}, {Conversi}, {Copin}, {Courbin},
  {Courtois}, {Cropper}, {Cuby}, {Da Silva}, {Degaudenzi}, {De Lucia},
  {Dolding}, {Dole}, {Douspis}, {Dubath}, {Dupac}, {Dusini}, {Escoffier},
  {Farina}, {Faustini}, {Ferriol}, {Fotopoulou}, {Frailis}, {Franceschi},
  {Galeotta}, {George}, {Gillis}, {Giocoli}, {G{\'o}mez-Alvarez},
  {Gracia-Carpio}, {Granett}, {Grazian}, {Grupp}, {Haugan}, {Hoar}, {Holmes},
  {Hormuth}, {Hornstrup}, {Jahnke}, {Jhabvala}, {Keih{\"a}nen}, {Kermiche},
  {Kiessling}, {Kubik}, {Kuijken}, {K{\"u}mmel}, {Kunz}, {Kurki-Suonio}, {Le
  Boulc'h}, {Le Brun}, {Ligori}, {Lilje}, {Lindholm}, {Lloro}, {Mainetti},
  {Maino}, {Maiorano}, {Mansutti}, {Marggraf}, {Martinelli}, {Martinet},
  {Marulli}, {Massey}, {Medinaceli}, {Mei}, {Mellier}, {Meneghetti}, {Merlin},
  {Meylan}, {Mora}, {Moresco}, {Moscardini}, {Nakajima}, {Neissner}, {Niemi},
  {Padilla}, {Paltani}, {Pasian}, {Pedersen}, {Percival}, {Pettorino}, {Pires},
  {Polenta}, {Poncet}, {Popa}, {Pozzetti}, {Raison}, {Rebolo}, {Renzi},
  {Rhodes}, {Riccio}, {Romelli}, {Roncarelli}, {Saglia}, {Sakr}, {Sapone},
  {Sartoris}, {Schewtschenko}, {Schirmer}, {Schneider}, {Secroun}, {Seidel},
  {Seiffert}, {Serrano}, {Simon}, {Sirignano}, {Sirri}, {Stanco},
  {Steinwagner}, {Tallada-Cresp{\'\i}}, {Taylor}, {Tereno}, {Toft},
  {Toledo-Moreo}, {Torradeflot}, {Tsyganov}, {Tutusaus}, {Valenziano},
  {Valiviita}, {Vassallo}, {Verdoes Kleijn}, {Veropalumbo}, {Wang}, {Weller},
  {Zacchei}, {Zamorani}, {Zerbi}, {Zucca}, {Mart{\'\i}n-Fleitas}, \&
  {Scottez}}]{Marusa2025}
{{\v{Z}}erjal}, M., {Dominguez-Tagle}, C., {Vitas}, N., {et~al.} 2026,
  \bibinfo{title}{{Euclid: Quick Data Release (Q1) ─ A photometric search for
  ultracool dwarfs in the Euclid Deep Fields},} \aap, 711, A37,
  \dodoi{10.1051/0004-6361/202556980}

\bibitem[{Z. {Zhang} {\&} Y. {Li}(2025{\natexlab{a}}){Zhang} \& {Li}}]{USMILE}
{Zhang}, Z., \& {Li}, Y. 2025{\natexlab{a}}, \bibinfo{title}{{Ultracool dwarf
  Science with MachIne LEarning (USMILE). I. Scalable Tree-based Models for
  Photometric Spectral Classification and New Discoveries from LSST Data
  Preview 1 and Euclid Quick Data Release 1},} \aj, 170, 360,
  \dodoi{10.3847/1538-3881/ae100b}

\bibitem[{Z.~H. {Zhang} {et~al.}(2017){Zhang}, {Pinfield}, {G{\'a}lvez-Ortiz},
  {Burningham}, {Lodieu}, {Marocco}, {Burgasser}, {Day-Jones}, {Allard},
  {Jones}, {Homeier}, {Gomes}, \& {Smart}}]{Zhang2017}
{Zhang}, Z.~H., {Pinfield}, D.~J., {G{\'a}lvez-Ortiz}, M.~C., {et~al.} 2017,
  \bibinfo{title}{{Primeval very low-mass stars and brown dwarfs - I. Six new L
  subdwarfs, classification and atmospheric properties},} \mnras, 464, 3040,
  \dodoi{10.1093/mnras/stw2438}

\end{thebibliography}

\bibliographystyle{aasjournalv7.1}



\end{document}